\documentclass[12pt,notitlepage,a4paper]{article}

\pdfoutput=1

\usepackage{a4wide}
\usepackage{epsfig}
\usepackage{color,graphicx}
\usepackage{cite}
\usepackage{amssymb,amsmath}

\newcommand{\be}{\begin{equation}}
\newcommand{\ee}{\end{equation}}
\newcommand{\bea}{\begin{eqnarray}}
\newcommand{\eea}{\end{eqnarray}}

\begin{document}

\begin{center}  

\vskip 2cm 

\centerline{\Large {\bf Compactifications of $5d$ SCFTs with a twist}}

\vskip 1cm

\renewcommand{\thefootnote}{\fnsymbol{footnote}}

   \centerline{
    {\large \bf Gabi Zafrir${}^{a}$} \footnote{gabizaf@techunix.technion.ac.il}}

\vspace{1cm}
\centerline{{\it ${}^a$ Department of Physics, Technion, Israel Institute of Technology}} \centerline{{\it Haifa, 32000, Israel}}
\vspace{1cm}

\end{center}

\vskip 0.3 cm

\setcounter{footnote}{0}
\renewcommand{\thefootnote}{\arabic{footnote}}   
   
\begin{abstract}

We study the compactification of $5d$ SCFTs to $4d$ on a circle with a twist in a discrete global symmetry element of these SCFTs. We present evidence that this leads to various $4d$ $\mathcal{N}=2$ isolated SCFTs. These include many known theories as well as seemingly new ones. The known theories include the recently discovered rank $1$ $SU(4)$ SCFT and its mass deformations. One application of the new SCFTs is in the dual descriptions of the $4d$ gauge theory $SU(N)+1S+(N-2)F$. Also interesting is the appearance of a theory with rank $1$ and $F_4$ global symmetry. 

\end{abstract}
 
 \newpage
 
\tableofcontents

\section{Introduction}

In recent years there has been an increased interest in the compactification of higher dimensional field theories in order to better understand lower dimensional ones. The most notable case being compactification of the $6d$ $(2,0)$ theory on a Riemann surface initiated in \cite{Gai}. This leads to many $4d$ $\mathcal{N}=2$ SCFTs and can be used to uncover various properties of these theories. A nice feature of this construction is that it naturally leads to Argyres-Seiberg type dualities\cite{AS} which are manifested as different pair of pants decomposition of the same Riemann surface. 

 This motivate the studying of compactification of other higher dimensional field theories. One possibility is to study the compactification of $6d$ $(1,0)$ SCFTs with its richer selection of possible theories. Indeed this has been recently studied for selected types of $6d$ $(1,0)$ SCFTs\cite{OSTY1,ZVX,OSTY2,Zaf2,OS}. Instead in this article we wish to concentrate on a different route, the compactification of $5d$ SCFTs on a circle. 

The existence of $5d$ SCFTs with $8$ supercharges has first been noted in \cite{SEI}. These provide UV completions to various $5d$ gauge theories which are non-renormalizable as the inverse gauge coupling squared has dimension of mass. One can interpret the gauge theory as the low-energy description of the SCFT perturbed by a mass deformation identified with the inverse gauge coupling squared. Interestingly, the gauge theory seems to contain considerable information about the UV SCFT such as its BPS spectrum, where the massive states are realized as instantons in the gauge theory which are particles in $5d$. Therefore we shall sometimes drop the "low-energy" term and simply refer to these as gauge theory descriptions of the UV SCFT.   

In general the dynamics of instanton particles play an important role in the UV completion of the gauge theory. A nice example for this is given by the phenomenon of enhancement of symmetry where the fixed point has a larger global symmetry than the low-energy gauge theory. From the SCFT point of view the extra symmetry is broken by the mass deformation. The mass deformation itself can be identified as a vev to a scalar in a background vector multiplet associated with a global symmetry whose Cartan remains as a symmetry of the gauge theory. This symmetry is the topological symmetry whose conserved currents is the instanton number, $j_T = *\mbox{Tr}(F\wedge F)$. The broken symmetry is manifested in the gauge theory by the appearance of additional conserved currents whose origin is these instantonic particles.  

\begin{figure}
\center
\includegraphics[width=0.75\textwidth]{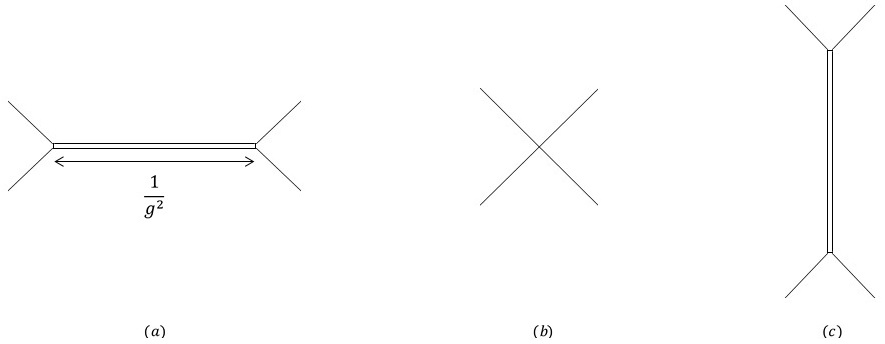} 
\caption{(a) The brane web for an $SU_0(2)$ gauge theory. The arrow shows the distance corresponding in the gauge theory to the inverse gauge coupling squared. (b) Taking the $\frac{1}{g^2}\rightarrow 0$ limit leads to the web describing the $5d$ SCFT. (c) The web for $\frac{1}{g^2}< 0$. Note that performing S-duality leads us back to the original theory so this limit also has a low-energy description as an $SU_0(2)$ gauge theory.}
\label{Ils0}
\end{figure}

Five dimensional SCFTs can in turn be studied by embedding them in string theory. This can be conveniently achieved using brane webs\cite{HA,HAK} where the SCFT is realized as the low-energy theory on a group of $5$-branes in type IIB string theory. For example consider the web shown in figure \ref{Ils0} (a). The low-energy theory living on the two D$5$-branes is an $SU_0(2)$ gauge theory\footnote{The subscript denotes the value of the $SU(2)$ $\theta$ angle\cite{SM}. We shall employ this to denote the $\theta$ angle for $USp$ group or Chern-Simons level for $SU$ groups. When denoting gauge theories we shall also use $F$ for matter in the fundamental representation, $AS$ for matter in the antisymmetric representation and $S$ for matter in the symmetric representation. When $SO$ groups are involved we use $V$ for matter in the vector representation. When writing quiver theories, we use the notation $G_1\times G_2\times...$ where it is understood that there is a single bifundamental hyper associated with every $\times$.}. The mass deformation associated with the $SU(2)$ coupling constant is visible as the distance between the two pairs of $(1,-1)$ and $(1,1)$ $5$-branes. Also visible are the BPS spectrum of the theory. For example F-strings represent the W-boson while D-strings represent the instanton particles. 

We can consider taking the $\frac{1}{g^2}\rightarrow 0$ limit in the web. This leads to the web in figure \ref{Ils0} (b) where all the $5$-branes intersect at a point. Now there is no mass scale in the problem so the theory living on the $5$-branes is an SCFT. All mass parameters in $5d$ are real and can be both positive and negative. Particularly this means that we can deform the SCFT in a different way corresponding to the negative of the deformation associated with $\frac{1}{g^2}$. This is shown in figure \ref{Ils0} (c). Note that the resulting low-energy theory is identical to the original as can be seen by performing an S-duality. So we see that deforming this $5d$ SCFT by a positive deformation leads to an $SU_0(2)$ gauge theory with coupling $\frac{1}{g^2}$, while doing a negative deformation leads to an $SU_0(2)$ gauge theory with coupling $-\frac{1}{g^2}$. This phenomenon, where different mass deformations of the same $5d$ SCFT can lead to different low-energy gauge theory descriptions is called a $5d$ duality. For the $SU_0(2)$ case, this is a self-duality yet there are many other known examples where different gauge theories are related in this way\cite{HA,BPTY,BGZ,Zaf,BZ,BZ1,GC}.   

One can now study the compactification of $5d$ SCFTs on a circle to $4d$. This has been previously explored for various SCFTs in \cite{BB,OSTY2,ZafD}. It can be used to realize various isolated non-Lagrangian $4d$ theories of the type considered in \cite{Gai}. The $5d$ SCFT lift of these theories generally have a low-energy gauge theory description and so can be studied by conventional means. Also in some cases, the $4d$ Argyres-Seiberg dualities lift to $5d$ dualities between two low-energy gauge theory descriptions of the same $5d$ SCFT\cite{BZ}. This then allows studying these dualities via conventional techniques. 

When compactifying a theory on a circle one can impose various twists under symmetries of the theory. For example, holonomies under continuous global symmetries are generally incorporated where in supersymmetric theories they lead to various mass parameters. In the case of compactification of $5d$ SCFTs these complete the real mass parameters of $\mathcal{N}=1$ $5d$ theories to the complex mass parameters of $\mathcal{N}=2$ $4d$ theories. Here we wish to study the compactification where we perform a twist by a discrete symmetry. That is we consider compactification of $5d$ SCFTs on a circle imposing that upon traversing the circle the theory is transformed by a discrete element of its global symmetry group.

We shall concentrate on $5d$ SCFTs with a brane web description, particularly the ones whose non-twisted reductions were discussed in \cite{BB,OSTY2}. The $5d$ SCFT is a strongly coupled non-perturbative beast so direct evaluation is usually not possible. Instead, as common in this field, we shall start by examining various simple cases, and by studying their properties, conjecture the resulting $4d$ theories, which in the cases at hand turns out to be isolated $4d$ SCFTs. This is then subjected to a variety of consistency checks. 

Once the simpler cases are understood, we can use them to study more general cases where we do not have a candidate $4d$ theory. This then suggests that this compactification leads to a variety of unknown $4d$ theories. We further study some of their properties and perform various consistency checks on our conjectures.

The structure of this article is as follows. In section $2$ we discuss twisted compactification of the $5d$ SCFT represented in string theory by the intersection of $N$ coincident D$5$-branes and $k$ coincident NS$5$-branes. We start with the simpler case of $N=2$ where we propose identifications for the resulting theories among known class S theories. We then test these identifications by studying dualities and mass deformations of these theories. We then move on to the general case where we conjecture the resulting theories to be new isolated SCFTs. We employ various dualities to study their properties and to serve as consistency checks. One application for this is to study the duality frames for $4d$ $SU(N)$ gauge theory with symmetric matter and $N-2$ fundamentals.

In section $3$ we move on to study the twisted compactification of the $5d$ SCFT represented in string theory by $N$ coincident $5$-brane junctions. We consider two different twists one under a $Z_3$ discrete element and one under a $Z_2$ one. In the $Z_3$ case we first examine various low $N$ cases identifying these with various known $4d$ SCFTs. Interestingly one of these is the recently discovered rank $1$ $SU(4)$ SCFT found in \cite{CDTn}. Besides providing an additional string theory construction for this theory, by examining its mass deformation, we get string theory constructions also for other rank $1$ SCFT generated by mass deforming the $SU(4)$ SCFT, originally introduced in \cite{ALLM1}. We suspect the general case to lead to unknown isolated $4d$ SCFTs and we comment on some of their properties. The $Z_2$ twist is more mysterious with a variety of seemingly unknown $4d$ theories including one with rank $1$ and $F_4$ global symmetry. 

In section $4$ we use the known Hall-Littlewood index for class S theories\cite{GRRY,GR,GRR} and properties of the compactification to conjecture an expression for the Hall-Littlewood index for the theories we presented. This is then checked for the cases where the conjectured $4d$ theory is known. We end with some conclusion. Appendix A gives a short review of the Hall-Littlewood index. Appendix B discuses aspects of $5d$ index calculations for the $5d$ $T_4$ theory and related theories that have interesting application to the rank $1$ $SU(4)$ SCFT and related theories.   



\section{$Z_2$ twist on the $SU(N)^2\times SU(k)^2\times U(1)$ $5d$ SCFT}

\begin{figure}
\center
\includegraphics[width=0.5\textwidth]{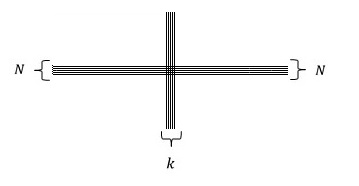} 
\caption{A $5d$ SCFT with an $SU(N)^2\times SU(k)^2\times U(1)$ global symmetry.}
\label{Ils14}
\end{figure}

 We start by considering the $5d$ SCFT engineered in string theory by the intersection of $N$ D$5$-branes and $k$ NS$5$-branes, shown in figure \ref{Ils14}. This $5d$ SCFT has an $SU(N)^2\times SU(k)^2\times U(1)$ global symmetry (enhanced to $SU(2N)\times SU(2)^2$ when $k=2$ or $SU(2k)\times SU(2)^2$ when $N=2$). Note that exchanging $N$ and $k$ leads to the same SCFT as can be seen by performing S-duality on the web in figure \ref{Ils14}. It has two convenient low-energy gauge theory descriptions. One is given by an $NF+SU(N)^{k-1}+NF$ quiver and is generated via a mass deformation breaking the $SU(k)^2$ global symmetry. Alternatively performing a mass deformation breaking the $SU(N)^2$ global symmetry leads to the low-energy quiver gauge theory $kF+SU(k)^{N-1}+kF$\cite{BPTY}. 

This theory has a $Z_2 \times Z_2$ discrete symmetry (enhanced to the dihedral group $D_4$ when $N=k$) given by exchanging the two $SU(N)$ or $SU(k)$ groups. We shall be particularly interested in the element which simultaneously exchanges the two $SU(N)$ and $SU(k)$ groups, given in the web by a $\pi$ rotation in the plane. In both gauge theory descriptions it is given by a combination of charge conjugation and quiver reflection. In this section we shall investigate the $4d$ theory resulting from circle compactification of this $5d$ SCFT with a twist in this discrete element. In other words we reduce the $5d$ SCFT on a circle where we enforce that upon traversing the circle the theory return to itself acted by this discrete element.

 This is an interesting twist to consider as it can be naturally implemented in a brane construction of the SCFT. For this let's consider what a $\pi$ rotation of the web entails in string theory. First this includes a $\pi$ rotation of the spacetime plane where the web lives. This can also be interpreted as a reflection of the two coordinates spanning the plane. We shall call this operation $I_{45}$. 

In addition the rotation of the web changes the charges of the $5$-branes. Particularly, a $(p,q)$ $5$-branes is mapped under this operation to a $(-p,-q)$ $5$-branes. Thus, in addition to the spacetime reflection $I_{45}$, we must also perform the $SL(2,Z)$ transformation $-I$: 

\be
-I = \left(
\begin{array}{cc}
-1 & 0 \\
0 & -1
\end{array}
\right) \,.
\ee

This in turn is equal to $\Omega (-1)^{F_L}$ where $\Omega$ is worldsheet parity and $F_L$ the left moving spacetime fermion number. Therefore a $\pi$ rotation of the web can be implemented in string theory by the operation $I_{45} \Omega (-1)^{F_L}$. The twisted compactification we consider then can be implemented by compactifying a direction common to all branes and enforce that upon traversing the circle we return to the system acted by $I_{45} \Omega (-1)^{F_L}$. 

In fact this type of compactification is just a generalization of the Dabholkar-Park background\cite{DP}. This specific twisted compactification was actually studied in \cite{PU} which considered the T-dual configuration (see also \cite{Ke,HKLTY1}). They found that performing T-duality on the circle leads to type IIB on the dual circle in the presence of an $O6^+$ and $O6^-$-planes. In particular this also shows that this compactification preserves the same supersymmetry as an $O7$-plane, and so applying this twisted compactification on a brane web should lead to a $4d$ system with $\mathcal{N}$$=2$ supersymmetry.

\subsection{The $N=2$ case and related theories}

We wish to begin by presenting some simple examples before discussing the general case.

\subsubsection{The $N=2, k=3$ case and related theories}

\begin{figure}
\center
\includegraphics[width=0.5\textwidth]{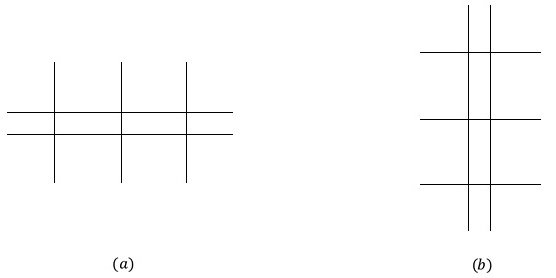} 
\caption{The brane web representation of a $5d$ SCFT, described by the collapsed web. (a) A deformation of the SCFT illustrating the $2F+SU(2)\times SU(2)+2F$ gauge theory description. (b) The S-dual web now illustrating the $SU_0(3)+6F$ gauge theory description.}
\label{Ils1}
\end{figure}

 Consider the $5d$ SCFT shown in figure \ref{Ils1}, which is the $N=2, k=3$ case of the general SCFT in figure \ref{Ils14}. It has an $SU(2)^2\times SU(6)$ global symmetry and two convenient gauge theory descriptions, one being $2F+SU(2)\times SU(2)+2F$ and the other $SU_0(3)+6F$. Like the other cases, it also has a $Z_2$ discrete symmetry given in the web by a $\pi$ rotation in the plane. This is identified with quiver reflection in the $2F+SU(2)\times SU(2)+2F$ theory and charge conjugation in the $SU_0(3)+6F$ theory. 

Now we want to consider compactifying it on a circle with a twist involving this $Z_2$ discrete symmetry. We inquire as to what theory we get in $4d$. This theory should have a $1$-dimensional Coulomb branch as only one of the two Coulomb branch dimensions of the $5d$ SCFT is symmetric under this $Z_2$ discrete symmetry. The two $SU(2)^2$ are mapped to one another so only the symmetric combination survives. 

The $SU(6)$ global symmetry should be broken to the $Z_2$ invariant part which is $USp(6)$. This can be seen as follows: we consider the action of the $Z_2$ as mapping the two $SU(3)$ subgroups of $SU(6)$ with charge conjugation, as suggested by the web. Under the $U(1)\times SU(3)\times SU(3)\subset SU(6)$ subgroup the adjoint of $SU(6)$ decomposes as: $\bold{35}=(\bold{8},\bold{1})^0 + (\bold{1},\bold{8})^0 + (\bold{3},\bar{\bold{3}})^2 + (\bar{\bold{3}},\bold{3})^{-2} + (\bold{1},\bold{1})^0$. Under the $Z_2$ action the adjoints of both $SU(3)$ groups are mapped to one another so we get only one $SU(3)$. The $(\bold{3},\bar{\bold{3}})$ is projected to the $\bold{3}\times \bold{3} = \bold{6} + \bar{\bold{3}}$ where only the symmetric combination, $\bold{6}$, is invariant. So we get the conserved currents $\bold{8}^0 + \bold{6}^2 + \bold{6}^{-2} + \bold{1}^0$ which builds the adjoint of $USp(6)$. Thus the resulting $4d$ SCFT should have an $SU(2)\times USp(6)$ global symmetry. 

Further we can use the web to calculate the Higgs branch dimension of the resulting SCFT. The Higgs branch dimension is given by the number of possible motions of the $5$-branes along the $7$-branes. The $5d$ SCFT has a $12$ dimensional Higgs branch. This is manifested in the web by the directions given by: breaking the $2$ D$5$-branes on a D$7$-brane both on the left and the right of the web (this gives $1$ direction for each side), breaking the $3$ NS$5$-branes on the $(0,1)$ $7$-branes both on the top and the bottom of the web (this gives $3$ directions for each side), and finally separating the remaining $2$ D$5$-branes and $3$ NS$5$-branes along the $7$-branes (this gives $4$ directions, one for each brane modulo a global translation).  

To find the Higgs branch dimension for the $4d$ theory resulting from the twisted compactification we must limit the counting to those motions invariant under the $Z_2$ discrete symmetry. As the two sides are mapped to one another, the motion on both sides are identified. This leaves determining whether there are constraints when separating the $5$-branes along the $7$-branes which is the direction fixed by the orbifolding. In other words, we need to determine if it is consistent to have a brane mapped to itself. For this we performing T-duality which maps this configuration to a group of NS$5$-branes, D$4$-branes and D$6$-branes in the presence of an $O6^+$ and $O6^-$-planes. There is no impediment to separating the NS$5$-branes along the $O6$-planes. Also we can tune parameters so as to have the D$4$-branes sit on top of the $O6^+$-plane where they can be separated along it. Thus, we conclude that we can separate $5$-branes along this orbifold.



We can now count all the possible breakings consistent with the $Z_2$ discrete symmetry, where we find: $1$ direction from breaking the $2$ D$5$-branes on a D$7$-brane simultaneously on both sides of the web, $3$ directions from breaking the $3$ NS$5$-branes on the $(0,1)$ $7$-branes simultaneously on the top and bottom of the web and $4$ directions from separating the remaining $2$ D$5$-branes and $3$ NS$5$-branes along the $7$-branes. This gives an $8$ dimensional Higgs branch. There is an isolated rank $1$ $4d$ SCFT with an $SU(2)\times USp(6)$ global symmetry and an $8$ dimensional Higgs branch\cite{AW}. Thus, the natural conjecture is that the preceding compactification leads to this theory. Next we wish to test this conjecture.

\subsubsection*{Dualities}

\begin{figure}
\center
\includegraphics[width=1.0\textwidth]{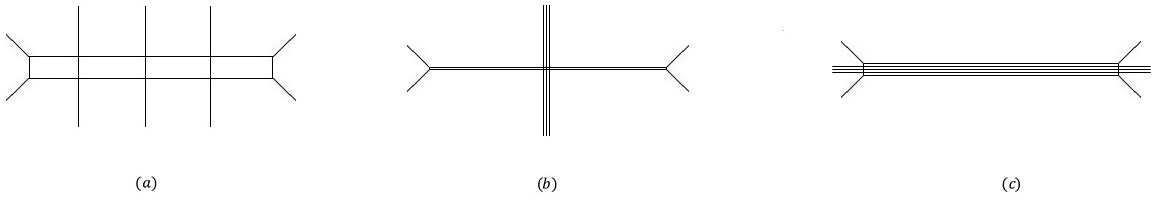} 
\caption{(a) The brane web representation of a $5d$ SCFT, generated from the one in figure \ref{Ils1} by gauging both $SU(2)$ global symmetries. (b) A mass deformation of the SCFT corresponding to the limit $g^{-2}_{SU(2)}\rightarrow \infty$ where $g_{SU(2)}$ is the coupling constant of both edge $SU(2)$ gauge groups. (c) The mass deformation corresponding to the limit $g^{-2}_{SU(2)}\rightarrow -\infty$ where we have also performed S-duality on the web. That this deformation is a continuation of the previous one is apparent as it preserves the $U(6)$ global symmetry. In this limit the $SU(2)$ quiver description is inadequate, but there is a different description as an $SU_0(5)+6F$ gauge theory where this limit corresponds to $g^{-2}_{SU(5)}\rightarrow \infty$ for $g_{SU(5)}$ being the coupling constant of $SU(5)$.}
\label{Ils2}
\end{figure}

As one piece of evidence for this conjecture, we shall show that we can recover $4d$ dualities involving the $SU(2)\times USp(6)$ SCFT from the $5d$ construction. Consider gauging both $SU(2)$ global symmetries of the $5d$ SCFT in figure \ref{Ils1} with the same coupling $g_{5d}$ so as to preserve the $Z_2$ discrete symmetry. This leads to the SCFT shown in figure \ref{Ils2} (a). Now let's reduce this SCFT with the $Z_2$ twist while taking the limit $R\rightarrow 0$, $g_{5d}\rightarrow 0$ keeping $\frac{g^2_{5d}}{R}$ fixed. 

Let's consider first doing the reduction in the $R>>g^2_{5d}>0$ limit. At energy scales of $g^{-2}_{5d}>>E>>\frac{1}{R}$ the $5d$ theory is effectively described by weakly gauging the $SU(2)^2$ global symmetry of the $5d$ SCFT in figure \ref{Ils1} by two $SU(2)$ gauge groups with identical couplings $g^2_{5d}$. At energies of $\frac{1}{R}>>E$ we get a $4d$ theory. Under the twist the two $SU(2)$ gauge theories are identified so we get just one gauge group with coupling $g^2_{4d} \sim \frac{g^2_{5d}}{R}$. The $5d$ SCFT should reduce to the proposed $4d$ $SU(2)\times USp(6)$ SCFT. So we see that in this limit the resulting $4d$ theory is an $SU(2)$ gauging of the $SU(2)$ global symmetry of the proposed $4d$ $SU(2)\times USp(6)$ SCFT.

 Consider approaching this limit from a different direction given by $g^2_{5d}<0$, shown in figure \ref{Ils2} (c). Now the $SU(2)$ description is inadequate and we should switch to a different description of the SCFT given by performing S-duality on the brane web. In this description we have an $SU_0(5)+6F$ gauge theory with coupling $-g^2_{5d}>0$. We now ask what happens to this theory under the twisted reduction. The twist should project the $SU(5)$ to $SO(5)$ and the $6F$ to the $3V$. This follows from the global symmetry as well as the Higgs branch analysis, both agreeing with the $4d$ gauge theory $SO(5)+3V$. 

\begin{figure}
\center
\includegraphics[width=1.0\textwidth]{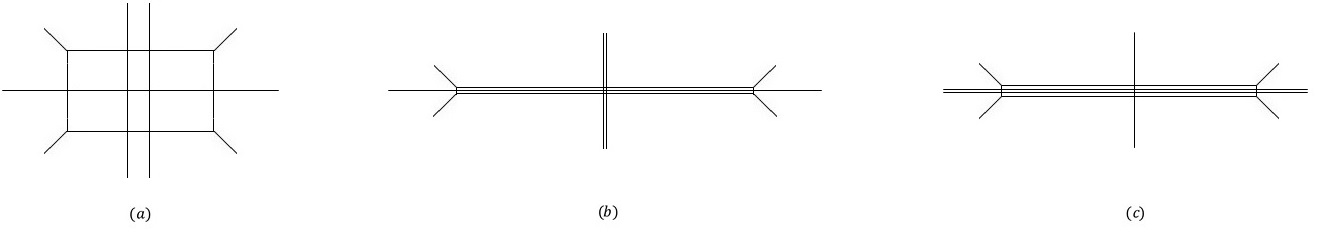} 
\caption{(a) The brane web representation of a $5d$ SCFT, generated from the one in figure \ref{Ils1} by double gauging an $SU(3)+1F$ into the $SU(6)$ global symmetry. (b) A mass deformation of the SCFT corresponding to the limit $g^{-2}_{SU(3)}\rightarrow \infty$ where $g_{SU(3)}$ is the coupling constant of both edge $SU(3)$ gauge groups. (b) The mass deformation corresponding to the limit $g^{-2}_{SU(3)}\rightarrow -\infty$ where we have also performed S-duality on the web. That this deformation is a continuation of the previous one is apparent as it preserves the $U(2)^2\times U(1)^2$ global symmetries associated with the semi-infinite $5$-branes. In this limit the $SU(3)$ quiver description is inadequate, but there is a different description as an $2F+SU_0(4)\times SU_0(4)+2F$ gauge theory where this limit corresponds to $g^{-2}_{SU(4)}\rightarrow \infty$ for $g_{SU(4)}$ the coupling constant of both $SU(4)$ gauge groups.}
\label{Ils3}
\end{figure}

Thus, we arrive at the $4d$ gauge theory $SO(5)+3V$ with $g^2_{4d} \sim \frac{g^2_{5d}}{R}$ which is weakly coupled. In fact this is a conformal theory with a marginal parameter $g^2_{4d}$. So we see that we can compactify the same $5d$ SCFT in the same limit getting different weakly coupled descriptions in different ranges of the marginal parameter $\frac{g^2_{5d}}{R}$. This implies that these two theories are dual. That the $4d$ $SU(2)\times USp(6)$ SCFT obeys such a duality is indeed known\cite{AW}. 

There is another way to generate a $4d$ conformal theory by gauging part of the global symmetry of the $SU(2)\times USp(6)$ SCFT which we can directly implement in the web. This is done by gauging the $USp(6)$ global symmetry with an $SU(3)+1F$ gauge theory. On the $5d$ SCFT this should lift to a double $SU(3)+1F$ gauging of the $SU(6)$ global symmetry. The resulting $5d$ SCFT is shown in figure \ref{Ils3} (a) where we have also shown the SCFT in the two limits of $g_{5d}\rightarrow 0$ for $g^2_{5d}>0$ in figure \ref{Ils3} (b) and $g^2_{5d}<0$ in figure \ref{Ils3} (c). When $g^2_{5d}>0$ the $SU(3)$ group is weakly coupled and we expect the $4d$ theory to be a weakly coupled $SU(3)+1F$ gauging the $SU(2)\times USp(6)$ SCFT. However when $g^2_{5d}<0$ the $5d$ SCFT is more appropriately described by an $2F+SU_0(4)\times SU_0(4)+2F$ gauge theory with equal couplings $\sim -g^2_{5d}$. The two groups are identified under the $Z_2$ symmetry. The identification and charge conjugation implies that the bifundamental matter should decompose to a symmetric and an antisymmetric of $SU(4)$ of which only the symmetric is $Z_2$ invariant. Therefore, when reduced to $4d$, we expect to get a weakly coupled $SU(4)+1S+2F$ gauge theory. Again these are two description of the same theory in different limits of a marginal operator and so describe a duality. This duality indeed appears in \cite{AW}. 

\subsubsection*{Mass deformations}

\begin{figure}
\center
\includegraphics[width=1.0\textwidth]{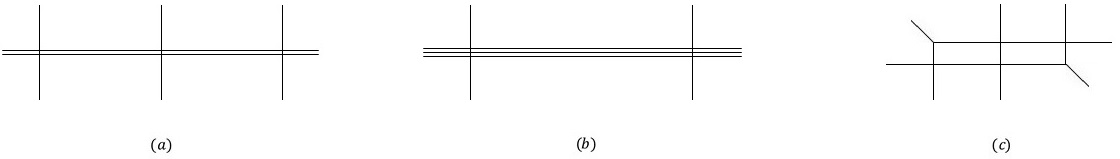} 
\caption{Mass deformations of the $5d$ SCFT in figure \ref{Ils1}. (a) A mass deformation corresponding to the limit $g^{-2}_{SU(2)}\rightarrow \infty$. (b)  A different mass deformation, shown in the S-dual frame, corresponding to the limit $g^{-2}_{SU(3)}\rightarrow \infty$. (c) The $5d$ SCFT we get after a mass deformation corresponding to integrating out a flavor on both sides.}
\label{Ils4}
\end{figure}

As a final piece of evidence we can consider mass deformations of this SCFT. For example, consider the deformation of the $5d$ SCFT shown in figure \ref{Ils4} (a). This mass deformation breaks the $SU(6)$ part of the global symmetry to $SU(2)^3$, but leaves the $SU(2)^2$ part unbroken. It is most convenient to describe this using the $2F+SU(2)\times SU(2)+2F$ gauge theory description where it corresponds to taking $g^{-2}_{SU(2)}\rightarrow \infty$. Reducing with the $Z_2$ we see that the resulting $4d$ theory should be $SU(2)+1S+2F$ which is an IR free gauge theory. Therefore the $4d$ $SU(2)\times USp(6)$ SCFT, resulting from the $Z_2$ twisted compactification of the $5d$ SCFT in figure \ref{Ils1}, should have a mass deformation leading to the $4d$ gauge theory $SU(2)+1S+2F$. Recently the mass deformations of the $SU(2)\times USp(6)$ SCFT were analyzed using the Seiberg-Witten curve in \cite{ALLM} who found that it indeed possess such a mass deformation. 

\begin{figure}
\center
\includegraphics[width=1.0\textwidth]{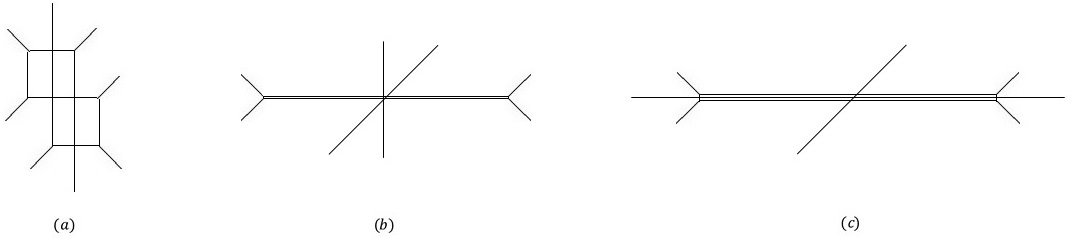} 
\caption{(a) The brane web representation of a $5d$ SCFT, generated from the one in figure \ref{Ils4} (c) by double gauging both $SU(2)$ subgroups of its $SU(4)$ global symmetry. Note that we have performed S-duality compared to the web shown in \ref{Ils4} (c). (b) A mass deformation of the SCFT corresponding to the limit $g^{-2}_{SU(2)}\rightarrow \infty$ where $g_{SU(2)}$ is the coupling constant of both edge $SU(2)$ gauge groups. (c) The mass deformation corresponding to the limit $g^{-2}_{SU(2)}\rightarrow -\infty$ where we have also performed S-duality on the web. That this deformation is a continuation of the previous one is apparent as it preserves the $U(1)^4$ global symmetry associated with the semi-infinite $5$-branes while breaking the $SU(2)^2$ associated with the semi-infinite $(1,1)$ $5$-branes. In this limit the previous description is inadequate, but there is a different description as an $1F+SU_{1}(3)\times SU_{-1}(3)+1F$ gauge theory where this limit corresponds to $g^{-2}_{SU(3)}\rightarrow \infty$ for $g_{SU(3)}$ the coupling constant of both $SU(3)$ gauge groups.}
\label{Ils5}
\end{figure}

We can also consider another mass deformation, shown in figure \ref{Ils4} (b), now breaking the $SU(2)^2$ global symmetry while preserving the $SU(6)$. This one is most conveniently addressed from the $SU_0(3)+6F$ description where it corresponds to the limit $g^{-2}_{SU(3)}\rightarrow \infty$. Again the previous discussion leads use to conclude that the resulting $4d$ theory is $SO(3)+3V$ which is again IR free. Such a mass deformation of the $SU(2)\times USp(6)$ SCFT was also found in \cite{ALLM}. 

There is an additional mass deformation we can consider, given in the web by integrating a flavor on both sides. This leads to a new $5d$ SCFT, shown in figure \ref{Ils4} (c), with gauge theory descriptions of $1F+SU(2)\times SU(2)+1F$ and $SU_0(3)+4F$. It has an $SU(4)\times U(1)^2$ global symmetry so we expect the $4d$ theory to have $USp(4)\times U(1)$ global symmetry. There is indeed a mass deformation of the $SU(2)\times USp(6)$ SCFT with that pattern of symmetry breaking leading to an isolated $4d$ SCFT with $USp(4)\times U(1)$ global symmetry\cite{ALLM}. It is natural to identify the resulting $4d$ SCFT with this theory.

\begin{figure}
\center
\includegraphics[width=0.65\textwidth]{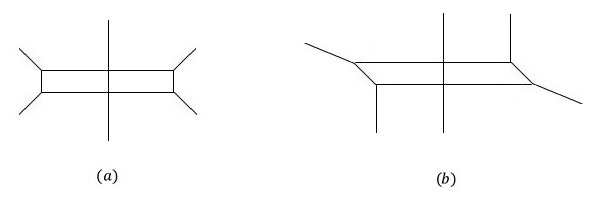} 
\caption{(a) The brane web representation of a $5d$ SCFT, generated from the one in figure \ref{Ils4} (c) by a mass deformation. It has two gauge theory descriptions as an $SU_{\pi}(2)\times SU_{\pi}(2)$ quiver gauge theory and an $SU_0(3)+2F$ one. (b) A different brane web of another $5d$ SCFT generated from the one in figure \ref{Ils4} by the opposite mass deformation. It has a gauge description has an $SU_{0}(2)\times SU_{0}(2)$ quiver gauge theory.}
\label{Ils505}
\end{figure}

We can test this in the same spirit as the previous tests. First we can compute the Higgs branch dimension finding $d_H=4$, which indeed agrees with the known Higgs branch dimension of the $USp(4)\times U(1)$ SCFT. Second we can gauge various global symmetries and study the resulting dualities. In this case we can gauge an $SU(2)\subset USp(4)$ which indeed gives a $4d$ conformal theory. The two interesting limits of this gauging are shown in figures \ref{Ils5} (b)+(c). The limit of figure \ref{Ils5} (b) describes a double weak gauging of the $USp(4)\times U(1)$ SCFT by an $SU(2)$ gauge group, while figure \ref{Ils5} (c) describes an $SU(3)+1S+1F$ gauge theory. This suggests that these are dual as was discovered in \cite{AW,CDT'}.

We can consider taking an additional mass deformation given by integrating an additional flavor. We can get to two different $5d$ SCFTs depending on the sign of the mass deformation. The first shown in figure \ref{Ils505} (a) has an $SU(2)\times U(1)^2$ global symmetry and gauge theory descriptions of $SU_{\pi}(2)\times SU_{\pi}(2)$ and $SU_0(3)+2F$. The second, shown in figure \ref{Ils505} (b), has an $SU(4)$ global symmetry and gauge theory description of $SU_{0}(2)\times SU_{0}(2)$. When reduced to $4d$ with a twist these should lead to $4d$ theories with global symmetries of $SU(2)\times U(1)$ and $USp(4)$ respectively. Examining mass deformations of the $USp(4)\times U(1)$ SCFT we find two natural candidates for these theories: $SU(2)+1S+1F$ for the $SU(2)\times U(1)$ theory and $SU(2)+2S$ for the $USp(4)$ theory. These are IR free gauge theories. We can further test this by comparing the dimension of the Higgs branch finding complete agreement.

\begin{figure}
\center
\includegraphics[width=0.5\textwidth]{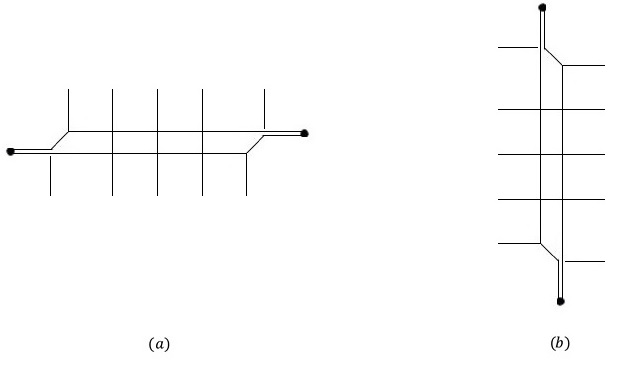} 
\caption{The brane web representation of a $5d$ SCFT, described by the collapsed web. (a) A deformation of the SCFT illustrating the $3F+SU(2)\times SU(2)+3F$ gauge theory description. (b) The S-dual web now illustrating the $SU_0(3)+8F$ gauge theory description.}
\label{Ils6}
\end{figure}

It is also interesting to consider the $5d$ SCFT we can get by adding flavors to the $5d$ gauge theories in figure \ref{Ils1}. When reduced with a twist this should lead to a $4d$ theory with a mass deformation leading to the $SU(2)\times USp(6)$ SCFT. Particularly consider the $5d$ SCFT shown in figure \ref{Ils6}. It has an $SU(10)$ global symmetry (see \cite{HKLTY,Yon,GC}) and two convenient gauge theory descriptions, one being $3F+SU(2)\times SU(2)+3F$ and the other $SU_0(3)+8F$. It also has the $Z_2$ discrete symmetry so we can compactify it on a circle with a twist under it. We expect this to lead to a rank $1$ $4d$ theory with $USp(10)$ global symmetry. There is indeed a rank $1$ isolated SCFT with $USp(10)$ global symmetry, first found in \cite{AW}. Furthermore this SCFT indeed has a mass deformation leading to the $SU(2)\times USp(6)$ SCFT\cite{ALLM}. We can also compute the Higgs branch dimension finding $d_H=16$ which agrees with the known Higgs branch dimension of the $USp(10)$ SCFT\cite{ALLM}.

\subsubsection{The $N=2$, general $k$ case and related theories}

In this subsection we generalize the previous discussion by the addition of NS$5$-branes. Like in the previous case, we can propose a known $4d$ SCFT as the result of the twisted compactification and test this using dualities. 

\begin{figure}
\center
\includegraphics[width=0.5\textwidth]{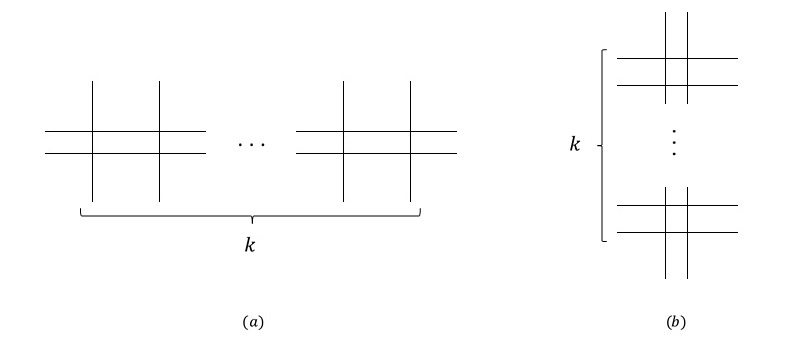} 
\caption{The brane web representation of a $5d$ SCFT, described by the collapsed web. (a) A deformation of the SCFT illustrating the $2F+SU(2)^{k-1}+2F$ gauge theory description. (b) The S-dual web now illustrating the $SU_0(k)+2kF$ gauge theory description.}
\label{Ils7}
\end{figure}

\begin{figure}
\center
\includegraphics[width=1.0\textwidth]{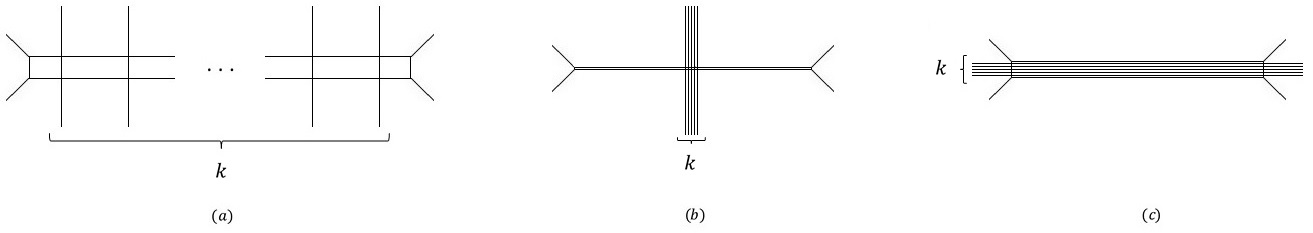} 
\caption{(a) The brane web representation of a $5d$ SCFT, generated from the one in figure \ref{Ils7} by gauging both $SU(2)$ global symmetries. (b) A mass deformation of the SCFT corresponding to the $g^{-2}_{SU(2)}\rightarrow \infty$ where $g_{SU(2)}$ is the coupling constant of both edge $SU(2)$ gauge groups. (b) The mass deformation corresponding to $g^{-2}_{SU(2)}\rightarrow -\infty$ where we have also performed S-duality on the web. That this deformation is a continuation of the previous one is apparent as it preserves the $U(2k)$ global symmetry. In this limit the $SU(2)$ quiver description is inadequate, but there is a different description as $SU_0(k+2)+2kF$ where this limit corresponds to $g^{-2}_{SU(k+2)}\rightarrow \infty$ for $g_{SU(k+2)}$ the coupling constant of $SU(k+2)$.}
\label{Ils8}
\end{figure}

Consider the $5d$ SCFT shown in figure \ref{Ils7}. It has an $SU(2)^2\times SU(2k)$ global symmetry and two convenient gauge theory descriptions given by $2F+SU(2)^{k-1}+2F$ and $SU_0(k)+2kF$. Reducing this $5d$ SCFT to $4d$ with a twist, we expect a $4d$ SCFT with an $SU(2)\times USp(2k)$ global symmetry. Further, we can gauge the $SU(2)$ global symmetry and consider reducing the theory in the $g^2\rightarrow 0, \frac{g^2}{R}$ fixed limit, in two different regimes of $\frac{g^2}{R}$. We find one describes a weak $SU(2)$ gauging of the aforementioned SCFT (see figure \ref{Ils8} (b)) while the other describing a weak $SO(k+2)+kV$ gauge theory (see figure \ref{Ils8} (c)). There is indeed a known duality of this form\cite{CDTDN}, leading us to identify the $SU(2)\times USp(2k)$ SCFT appearing in these dualities with the one resulting from the twisted compactification.

 In the $k$ even case, this theory can be constructed by the compactification of a D type $6d$ $(2,0)$ theory with twist\cite{CDTDN} which allows determining its properties. We can perform some consistency checks on this identification. First we can calculate the Higgs branch dimension of these SCFTs from the web. These can be compared against the class S result for even $k$ and against what is expected from the duality for general $k$ finding complete agreement.

\begin{figure}
\center
\includegraphics[width=0.6\textwidth]{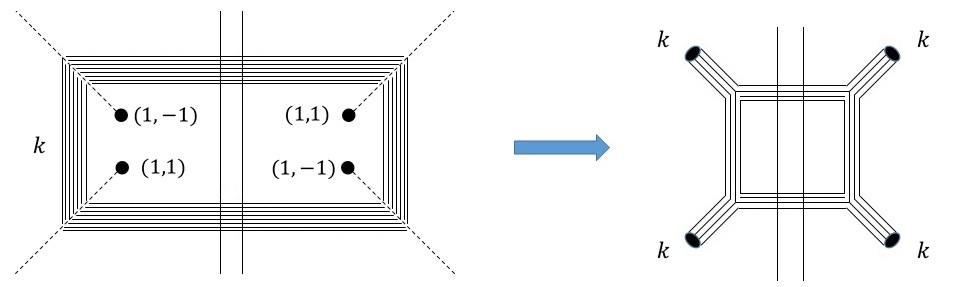} 
\caption{Gauging a $USp(2k)$ subgroup of an $SU(2k)$ global symmetry. The gauging is done by adding an $O7^-$ plane here shown after it has been resolved to a $(1,1)$ and $(1,-1)$ $7$-branes. We can proceed by pulling out the $7$-branes arriving at the configuration on the right, where a number next to a $7$-brane stands for the number of $5$-branes ending on that $7$-brane.}
\label{Ils10}
\end{figure}

Another check we can do is to consider gauging a part of the $USp(2k)$ global symmetry and so consider a different duality. One option is to gauge it with an $SU(k)+(k-2)F$ gauge group which is a conformal gauging. We shall consider this duality in the next section when we discuss the general case. When $k$ is even we can also consider a double gauging of the $USp(2k)$ global symmetry by a $USp(k)$ group which leads to an interesting duality\footnote{There is a generalization of this that works for every $k$ given by gauging with an $SU(k)+1AS$ gauge group. We will consider this in the next section.}. In the web this can be performed by adding an $O7^-$ plane and then resolving it as shown in figure \ref{Ils10}. 

\begin{figure}
\center
\includegraphics[width=1.0\textwidth]{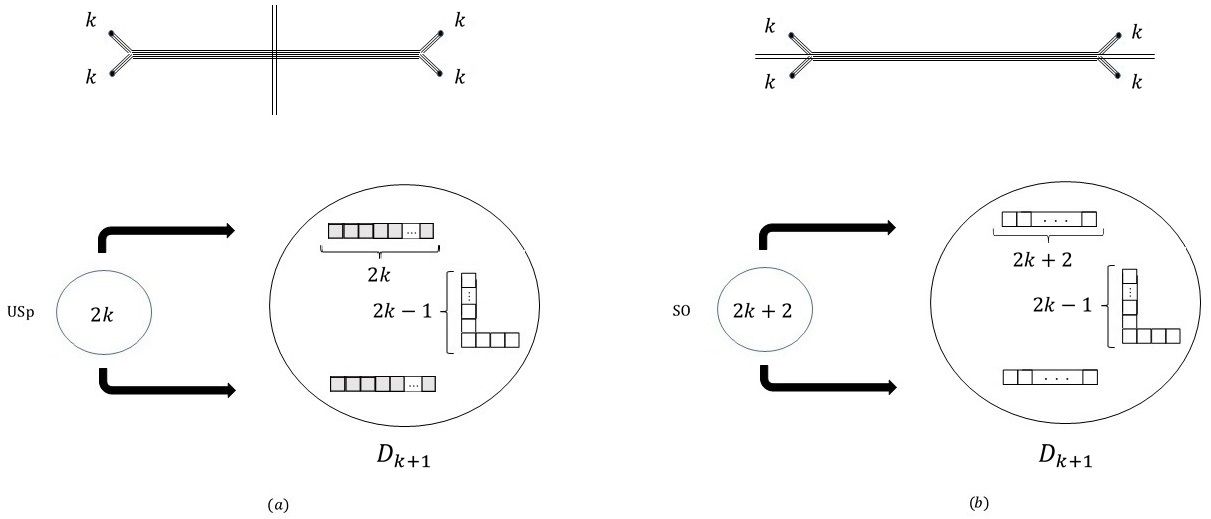} 
\caption{Two limits of the $USp(2k)$ gauging introduced in figure \ref{Ils10}. (a) This limit corresponds to $g^{-2}_{USp(2k)}\rightarrow \infty$. Reducing with a twist leads to the theory shown below where we use grayed Young diagrams to represent twisted punctures, and black arrows to represent gauging the appropriate symmetry of the shown class S theory. Further we use two arrows as both $USp(2k)$ groups, associated to the two punctures, are gauged. (b) This limit corresponds to $g^{-2}_{USp(2k)}\rightarrow -\infty$ and reducing it with a twist leads to the theory shown below.}
\label{Ils9}
\end{figure}

\begin{figure}
\center
\includegraphics[width=1.0\textwidth]{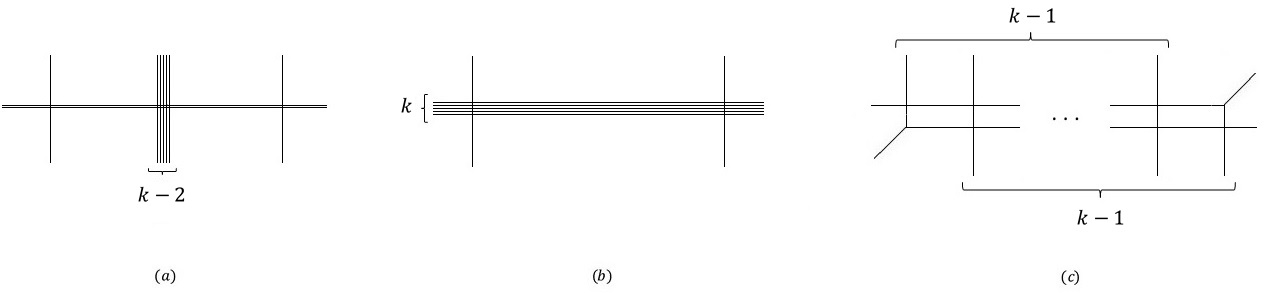} 
\caption{Two mass deformations of the $5d$ SCFT of figure \ref{Ils7}. A mass deformation corresponding to the limit $g^{-2}_{SU(2)}\rightarrow \infty$. (b)  A different mass deformation, shown in the S-dual frame, corresponding to the limit $g^{-2}_{SU(k)}\rightarrow \infty$. (c) A $5d$ SCFT we get after a mass deformation corresponding to integrating out a flavor on both sides.}
\label{Ils11}
\end{figure}

This leads to the duality shown in figure \ref{Ils9}. We can now use the known properties of the $SU(2)\times USp(2k)$ SCFT to check this duality by comparing the conformal anomalies, dimensions of Coulomb branch operators and global symmetries and their associated central charges finding complete agreement. Furthermore we can argue this duality from a class S construction, where we reduce the $6d$ $D_{k+1}$ $(2,0)$ theory on a torus with a single puncture whose associated Young diagram is shown in figure \ref{Ils9}, being the ungrayed one on the bottom left theory. In addition we add a $Z_2$ twist in the outer automorphism of $D_{k+1}$ on one of the cycles of the torus. We then get both theories in the bottom of figure \ref{Ils9} as different pair of pants decompositions of this Riemann surface (see also \cite{CDT'} for an example of this type of dualities for the twisted $A$ $(2,0)$ theory).

Finally we can also use this to study mass deformations of these SCFT. For example, figure \ref{Ils11} suggests that they should have a mass deformation, breaking the $SU(2)$ global symmetry, that leads to the IR free $SO(k)+kV$ gauge theory. There should also be another mass deformation, now breaking the $USp(2k)$ global symmetry, that leads to an IR free $SU(2)+2F$ gauging the $SU(2)$ global symmetry of the $SU(2)\times USp(2k-4)$ SCFT. It will be interesting to see if this can be verified by alternative means. 

We can also consider a mass deformation interpreted in the web by integrating a flavor. This leads to the $5d$ SCFT shown in figure \ref{Ils11} (c) having a $U(1)^2\times SU(2k-2)$ global symmetry. We can consider the result of reducing this SCFT to $4d$ with the twist where based on the previous example we expect a $4d$ SCFT with a $U(1)\times USp(2k-2)$ global symmetry. When $N$ is odd we can identify this theory with the class S theories introduced in \cite{CDT'}. One evidence for this is that the dimension of the Higgs branch agree. We shall give an additional piece of evidence in section $2.2.1$.

\begin{figure}
\center
\includegraphics[width=0.95\textwidth]{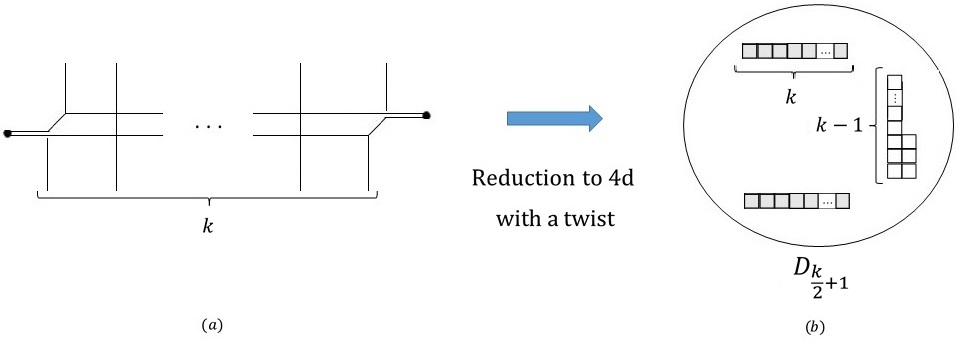} 
\caption{(a) The brane web for a $5d$ SCFT. (b) A $4d$ class S theory where $N$ is even. It is the theory we conjecture results from reducing the $5d$ SCFT in (a) with a $Z_2$ twist when $k$ is even.}
\label{Ils13}
\end{figure}

Finally we can consider the generalization of the $USp(10)$ theory by the addition of NS$5$-branes. The $5d$ SCFT, shown in figure \ref{Ils13} (a), has an $SU(2k)$ global symmetry so reducing it with a twist leads to a $4d$ SCFT with $USp(2k)$ global symmetry. When $k$ is even we can naturally identify it with the class S theory shown in \ref{Ils13} (b). This is supported as the global symmetry, dimension of the Coulomb branch and dimension of the Higgs branch all agree. Further we can again consider double gauging the $USp(2k)$ global group with a $USp(k)$ gauge group leading to the duality similar to this of figure \ref{Ils9} after forcing the two $5$-branes to end on the same $7$-brane. In the class S description this corresponds to changing the puncture with $SO(3)$ symmetry to the minimal puncture. The rest works out exactly as in the duality of figure \ref{Ils9} so we won't elaborate on it.


\subsection{The general case}

In this section we turn to analyzing the general case. Particularly, we consider the compactification of the $5d$ SCFT whose brane description is given in figure \ref{Ils14}. We wish to study its compactification to $4d$ with the $Z_2$ twist. We shall argue that this leads to an isolated $4d$ SCFT. The basic tool we use to study this is the dualities of the type considered in the previous subsection. 

\begin{figure}
\center
\includegraphics[width=1.0\textwidth]{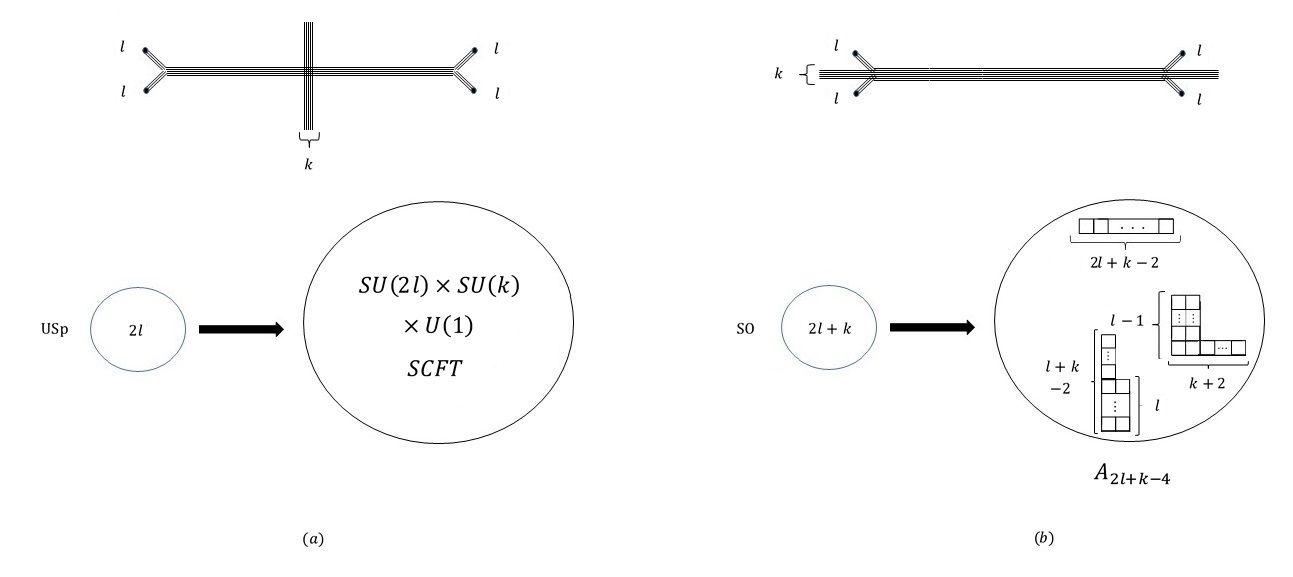} 
\caption{Two limits of a $USp(2l)$ gauging of an $SU(2l)$ global symmetry of the SCFT in figure \ref{Ils14}. (a) This limit corresponds to $g^{-2}_{USp(2l)}\rightarrow \infty$. Reducing with a twist leads to the theory shown below. (b) This limit corresponds to $g^{-2}_{USp(2l)}\rightarrow -\infty$ and reducing it with a twist leads to the theory shown below.}
\label{Ils15}
\end{figure}

\begin{figure}
\center
\includegraphics[width=1.1\textwidth]{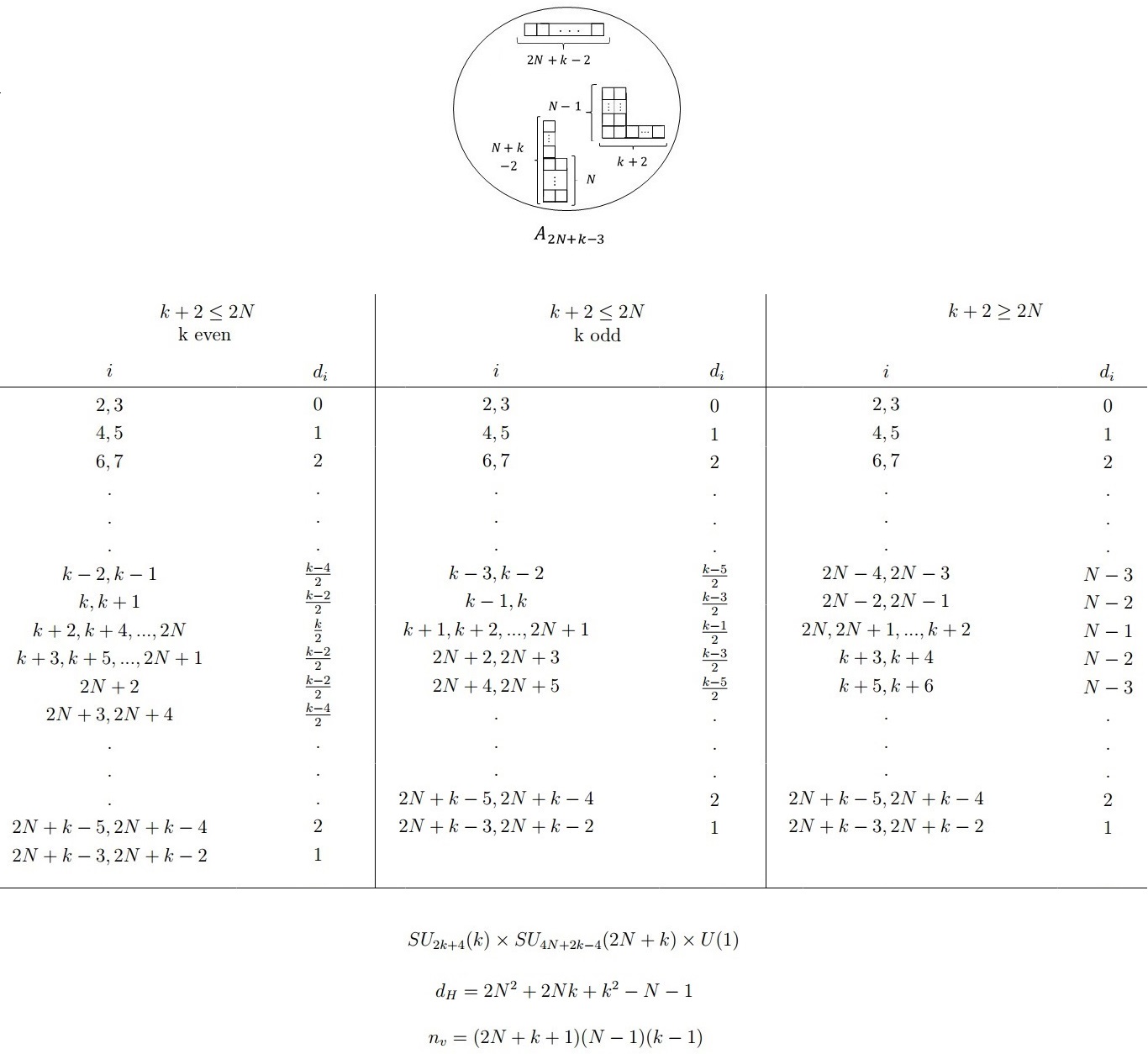} 
\caption{Properties of the $4d$ class S theory shown in the top of the picture. The table in the middle summarizes the spectrum and dimensions of Coulomb branch operators. These are somewhat different depending on whether $2N\geq k+2$ or $2N\leq k+2$ and whether $k$ is even or odd. In the table $i$ stands for the dimension of the operator and $d_i$ for the number of such operators present in the SCFT. Also written are the global symmetry with the central charges, Higgs branch dimension and effective number of vector multiplets. The global symmetry written is for the $N,k>2$ case, and is further enhanced to $SU(2)\times SO(4N+4)$ for $k=2$ and $SU(2k+4)$ for $N=2$. Note that for $N=k=2$ the theory becomes the rank $1$ $E_7$ theory.}
\label{SCFTP1}
\end{figure}

We start by considering the case of $N=2l$ where we can consider gauging the $SU(N)$ global symmetry by $USp(2l)$. The two interesting limits of this gauging are shown in figures \ref{Ils15} (a)+(b). These suggest the duality shown in the lower part of figure \ref{Ils15}. An important feature here is that the right side of the duality is given in terms of known theories allowing us to deduce the properties of the unknown theory. For this we rely on the properties of the class S theory appearing in the duality. Particularly we require the spectrum and dimensions of Coulomb branch operators, global symmetry and central charges. These can be evaluated using the methods of \cite{CD}, and as these play a prominent rule in the proceeding discussion we have summarized them in figure \ref{SCFTP1}. For the central charges we use the Higgs branch dimension and effective number of vector multiplets. These can be readily converted to the $a$ and $c$ conformal anomalies using: $d_H = 24(c-a), n_v = 4(2a-c)$.

\begin{figure}
\center
\includegraphics[width=1.1\textwidth]{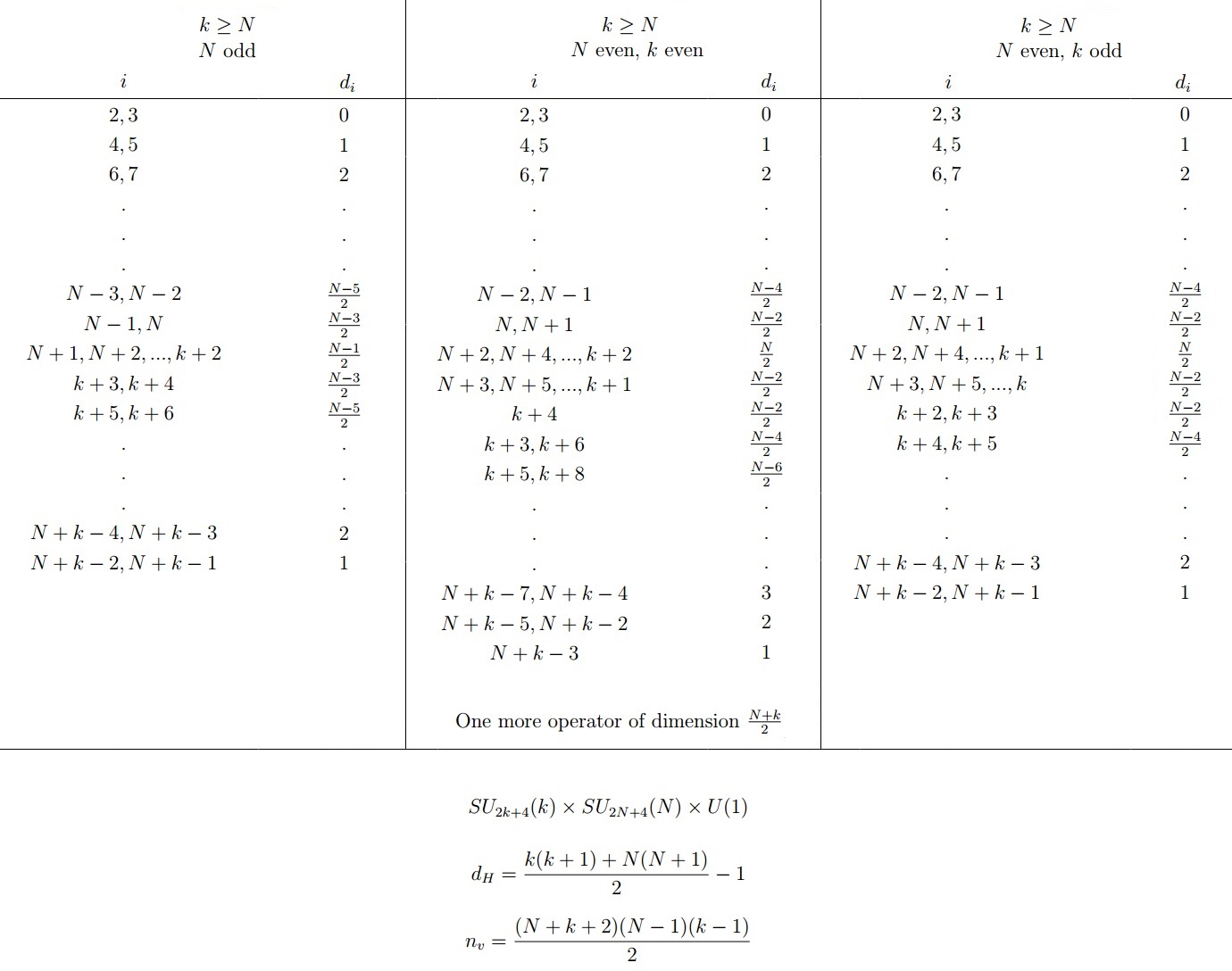} 
\caption{Properties of the conjectured $SU(k)\times SU(N)\times U(1)$ SCFT resulting from the twisted compactification of the $5d$ SCFT in figure \ref{Ils14}. These can be determined from the dualities in figures \ref{Ils15} and \ref{Ils16}. The table summarizes the spectrum and dimensions of Coulomb branch operators where we have assumed that $k\geq N$, the other case given by exchanging $N$ and $k$. The last entry in the middle table refers to the existence of one more Coulomb branch operator, in addition to the other ones appearing in the table. Also written are the global symmetry with the central charges, Higgs branch dimension and effective number of vector multiplets. In the global symmetry we have assumed that $N,k>2$, the cases of $N=2$ or $k=2$ being covered in the previous subsection.}
\label{SCFTP2}
\end{figure}

From these we see that the theory on the right hand side is an SCFT with a single marginal parameter. The duality suggests the theory on the left side should also be of this type. Therefore the $USp$ gauging should be conformal and the $SU(k)\times SU(N)\times U(1)$ theory should be an isolated SCFT. From the duality we can determine its properties, at least when $N$ is even, which are summarized in figure \ref{SCFTP2}.

\begin{figure}
\center
\includegraphics[width=1.0\textwidth]{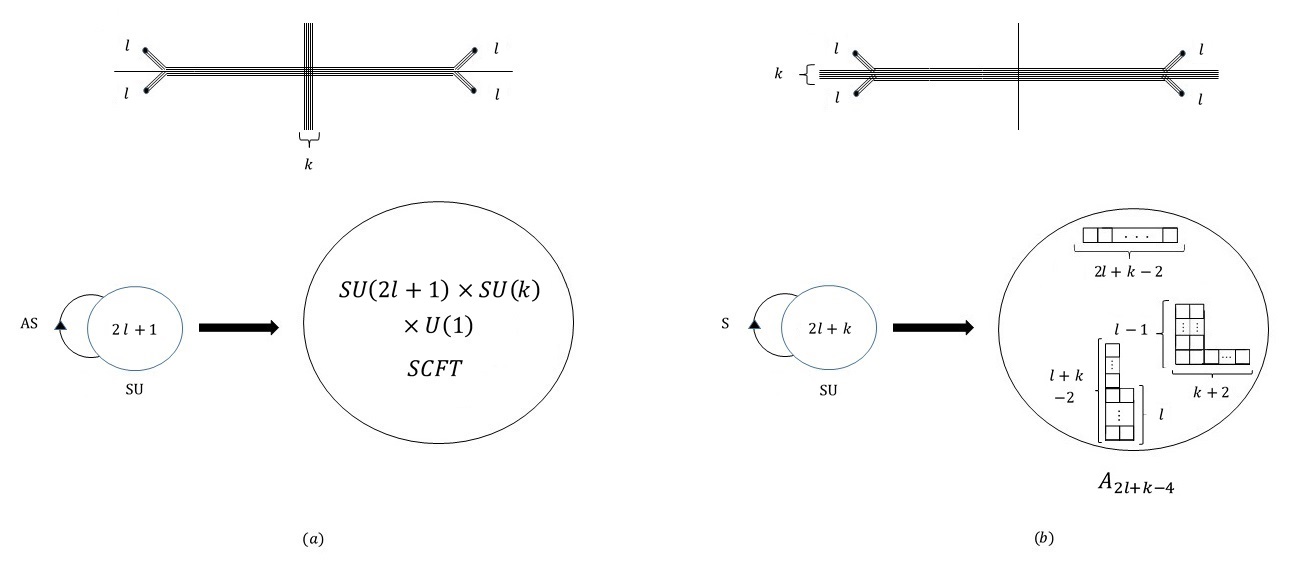} 
\caption{Two limits of a $SU(2N+1)+1AS$ gauging of an $SU(2N+1)$ global symmetry of the SCFT in figure \ref{Ils14}. (a) This limit corresponds to $g^{-2}_{SU(2N+1)}\rightarrow \infty$. Reducing with a twist leads to the theory shown below. (b) This limit corresponds to $g^{-2}_{SU(2N+1)}\rightarrow -\infty$ and reducing it with a twist leads to the theory shown below.}
\label{Ils16}
\end{figure}

\begin{figure}
\center
\includegraphics[width=1.0\textwidth]{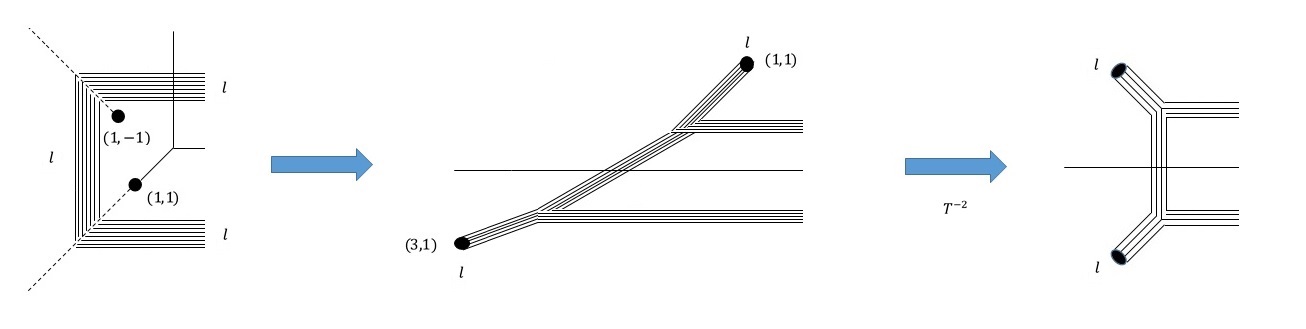} 
\caption{Starting from the left, depicting a resolved $O7^-$ plane with a stuck D$5$-brane and NS$5$-branes, we arrive to the right configuration.}
\label{Ils17}
\end{figure}

\begin{figure}
\center
\includegraphics[width=1.0\textwidth]{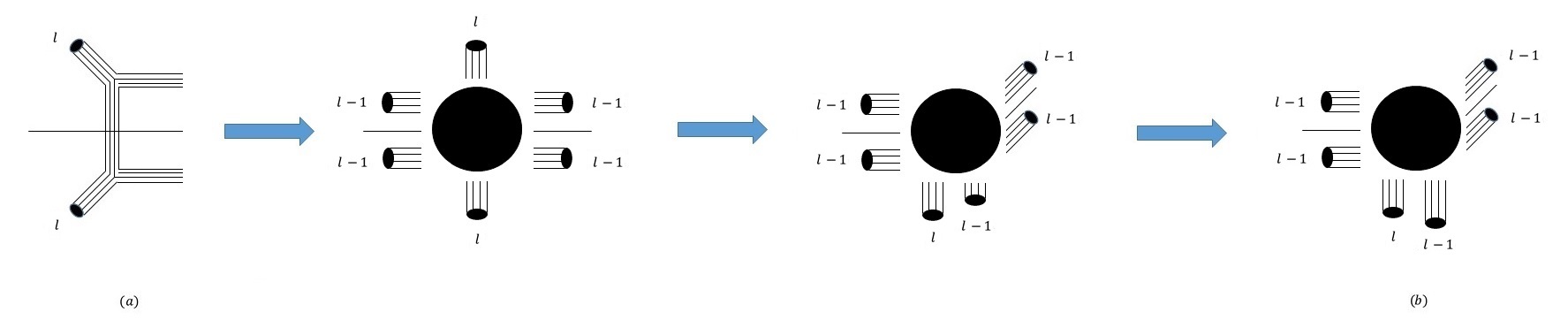} 
\caption{(a) A web for a $5d$ SCFT gauged by $SU(2l+1)$ in figure \ref{Ils16}. Performing a series of $7$-brane motions we arrive at the configuration in (b) which is of the form of \cite{BB}.}
\label{Ils18}
\end{figure}

In the $N=2l+1$ we can consider gauging the $SU(N)$ global symmetry by $SU(N)+1AS$. This and the resulting duality are shown in figure \ref{Ils16}. That this describes an $SU(2l+1)+1AS$ gauging can be reasoned by resolving an $O7^-$-plane with a stuck NS$5$-brane (see figure \ref{Ils17}). It is instructive to argue this also in an alternative way. We can interpret the system in figure \ref{Ils16} as an $SU(2l+1)$ gauging of, on one side the $SU(2l+1)$ of the $SU(2l+1)\times SU(k)\times U(1)$ SCFT, and on the other the $5d$ SCFT shown in figure \ref{Ils18} (a). Perfuming a series of $7$-brane motions we can map it to the one in figure \ref{Ils18} (b) which is of the form considered in \cite{BB}. Thus, there is a class S theory associated with this SCFT which describes an antisymmetric hyper and two fundamentals under the $SU(2l-1)$ global symmetry manifested in the punctures. Also note that we performed a transition of the type considered in \cite{ZafD} so there is an additional hyper in the theory of figure \ref{Ils18} (a).

Therefore the theory in figure \ref{Ils18} (a) is a collection of $l(2l+1)$ free hypers that transform as the $(\bold{1},\bold{(l-1)(2l-1)}) + (\bold{2},\bold{2l-1}) + (\bold{1},\bold{1})$ under the $SU(2)\times SU(2l-1)$ subgroup of $SU(2l+1)$. This can only be consistent with this theory describing a single hyper in the antisymmetric of $SU(2l+1)$.

We can now use the duality in figure \ref{Ils16} to study the $SU(N)\times SU(k)\times U(1)$ SCFT when both $N$ and $k$ are odd. This is again summarized in figure \ref{SCFTP2}. We can perform several consistency checks on the properties we find. First we find that these are indeed invariant under the interchange of $N$ and $k$ as suggested by the web. This is also necessary as we could have performed the same dualities by gauging the $SU(k)$ group instead and this structure guaranties that this as well is consistent. Another consistency check we can perform is to compare the Higgs branch dimension evaluated from the web against the one expected from the duality using $d_H = 24(a-c)$ where we again find agreement. We can also compare the dimension of the Coulomb branch required from the duality against the one expected from the web where again we find agreement.

\begin{figure}
\center
\includegraphics[width=1.0\textwidth]{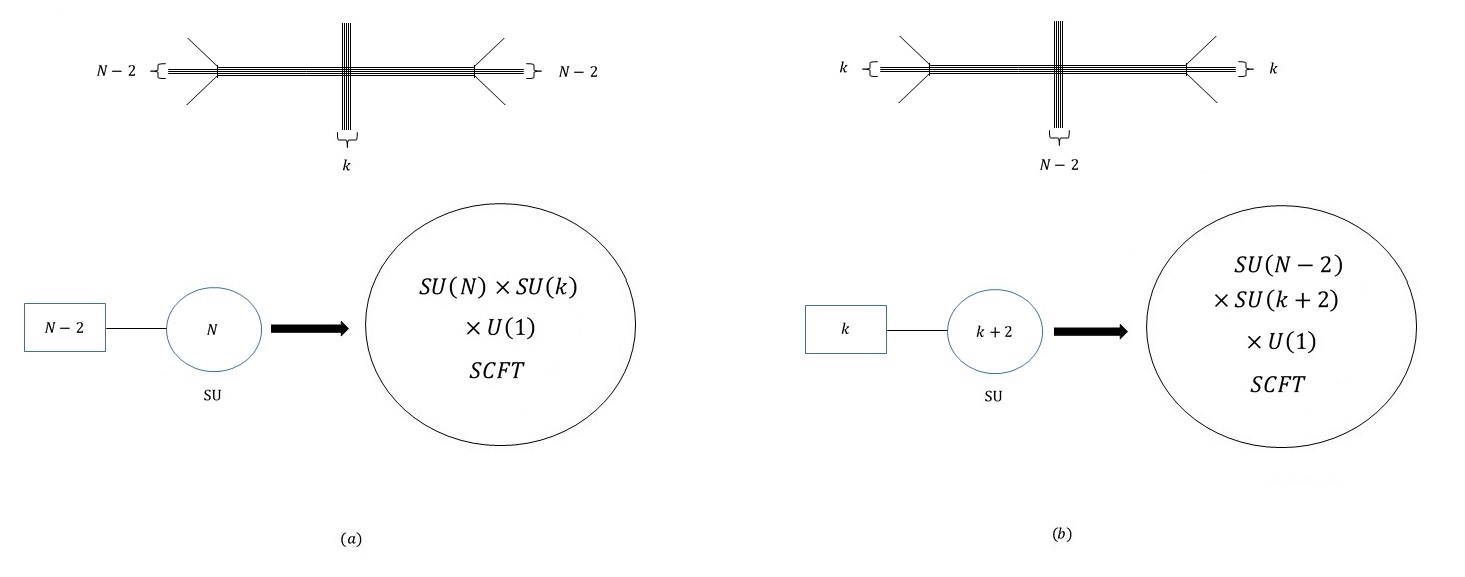} 
\caption{Two limits of a $SU(N)+(N-2)F$ gauging of an $SU(N)$ global symmetry of the SCFT in figure \ref{Ils14}. (a) This limit corresponds to $g^{-2}_{SU(N)}\rightarrow \infty$. Reducing with a twist leads to the theory shown below. (b) This limit corresponds to $g^{-2}_{SU(N)}\rightarrow -\infty$ and reducing it with a twist leads to the theory shown below.}
\label{Ils19}
\end{figure}

In order to perform additional consistency checks we consider other dualities. For example we can gauge the $SU(N)$ group with an $SU(N)+(N-2)F$ gauge theory which leads to a $4d$ conformal theory. This gives to the duality shown in figure \ref{Ils19}. We can now test this duality by matching central charges using the properties of the $SU(N)\times SU(k)\times U(1)$ SCFT we determined from the previous dualities, and consistency now necessitates that these agree. We indeed find that they agree. 

\begin{figure}
\center
\includegraphics[width=1.1\textwidth]{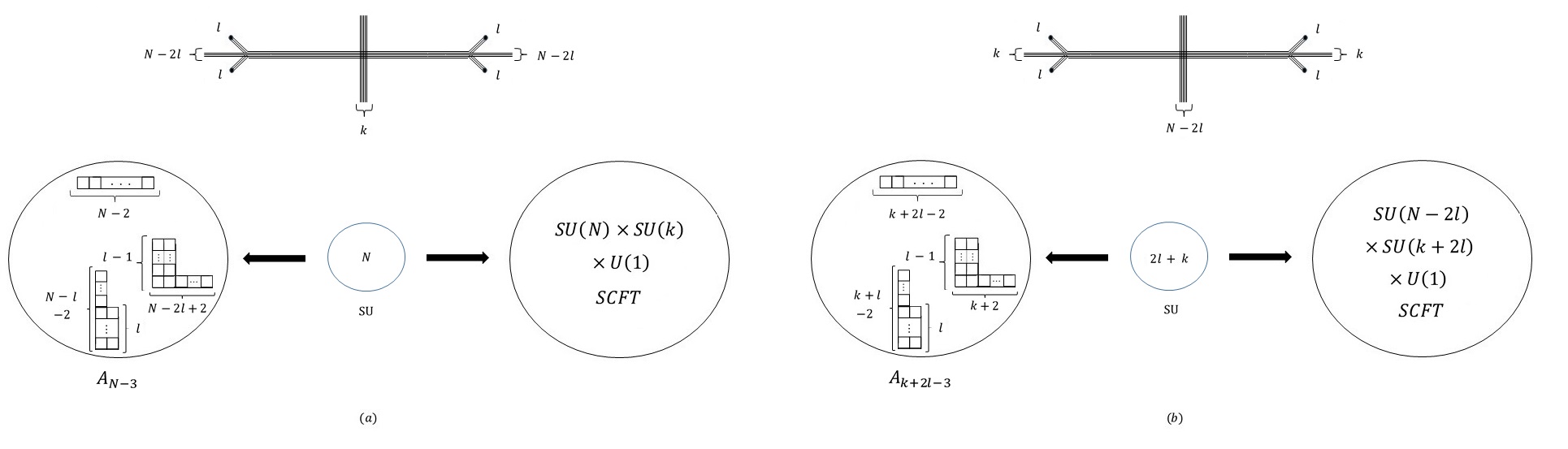} 
\caption{Two limits of a $SU(N)$ gauging of an $SU(N)$ global symmetry of the SCFT in figure \ref{Ils14} on one side and the SCFT of the type in \ref{SCFTP1} on the other side. (a) This limit corresponds to $g^{-2}_{SU(N)}\rightarrow \infty$. Reducing with a twist leads to the theory shown below. (b) This limit corresponds to $g^{-2}_{SU(N)}\rightarrow -\infty$ and reducing it with a twist leads to the theory shown below.}
\label{Ils20}
\end{figure}

We can in fact generate a number of dualities by considering the $5d$ SCFT shown in figure \ref{Ils20} for $l=1,2,...[\frac{N}{2}]$, where $[\frac{N}{2}]$ stands for the integer part of $\frac{N}{2}$. The figure shows two extreme limits of a particular mass deformation where the $5d$ SCFT can be described by a double $SU$ gauging, each connecting the $5d$ SCFT in figure \ref{Ils14} to the one in figure \ref{SCFTP1}. When reduced to $4d$ with the twist this naturally leads to the $4d$ dualities shown in figure \ref{Ils20}. Note that the two previous cases are just the $l=1$ and $l=[\frac{N}{2}]$ limits of this duality. We can now test this duality by comparing the various central charges finding complete agreement. We can also compare the dimension of Coulomb branch operators finding complete agreement.

\subsubsection{Additional theories}

From the $SU(N)\times SU(k)\times U(1)$ family of $4d$ SCFT we introduced we can generate additional theories by Higgsing and mass deformations. Like in class S theories, we can generate additional SCFTs by taking various Higgs branch limits of the starting SCFT. We can see the various possible Higgs branch limits from the brane web description. First, starting from a specific $SU(N)\times SU(k)\times U(1)$ theory, we can flow to ones with lower $N$ or $k$. This is described in the web by separating a group of D$5$ and NS$5$-branes from the web. 

Another Higgs branch limit is given by forcing a group of D$5$ or NS$5$-branes to end on the same $7$-brane. This limit is well known in the $5d$ class S theories of \cite{BB} as being the Higgs branch associate with partial closure of the punctures. Indeed specifying the distribution of $N$ $5$-branes on a group of $7$-branes constitute a partition of $N$ and so a Young diagram. Therefore we can describe theories generated by Higgsing down a $SU(N)\times SU(k)\times U(1)$ SCFT by introducing two Young diagrams, one with $N$ boxes while the other with $k$ boxes. The $SU(N)\times SU(k)\times U(1)$ SCFT itself is given by the one row diagram while other Young diagrams give other SCFT's generated by Higgsing.

 It is straightforward to generalize the methods we used in this section also to these cases. An example for this was given in section $2.1$ where we studied the $USp(2N)$ SCFTs. Thus, we can essentially study dualities and mass deformations of these SCFTs.

\begin{figure}
\center
\includegraphics[width=0.5\textwidth]{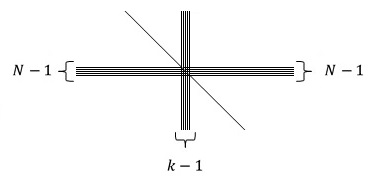} 
\caption{A $5d$ SCFT with an $SU(N-1)^2\times SU(k-1)^2\times U(1)^3$ global symmetry.}
\label{Ils27}
\end{figure}

Another way to generate theories is using mass deformations. These of course may not lead to an SCFT. Yet we can find one mass deformation which we can argue indeed gives an SCFT. We can consider the mass deformation which can be interpreted as integrating out a flavor on both sides in the gauge theory description of the $5d$ SCFT. This leads to the $5d$ SCFT shown in figure \ref{Ils27}, and reducing it to $4d$ with the $Z_2$ twist is expected to lead to a $4d$ theory. We now argue that this $4d$ theory is an isolated SCFT.


\begin{figure}
\center
\includegraphics[width=1.0\textwidth]{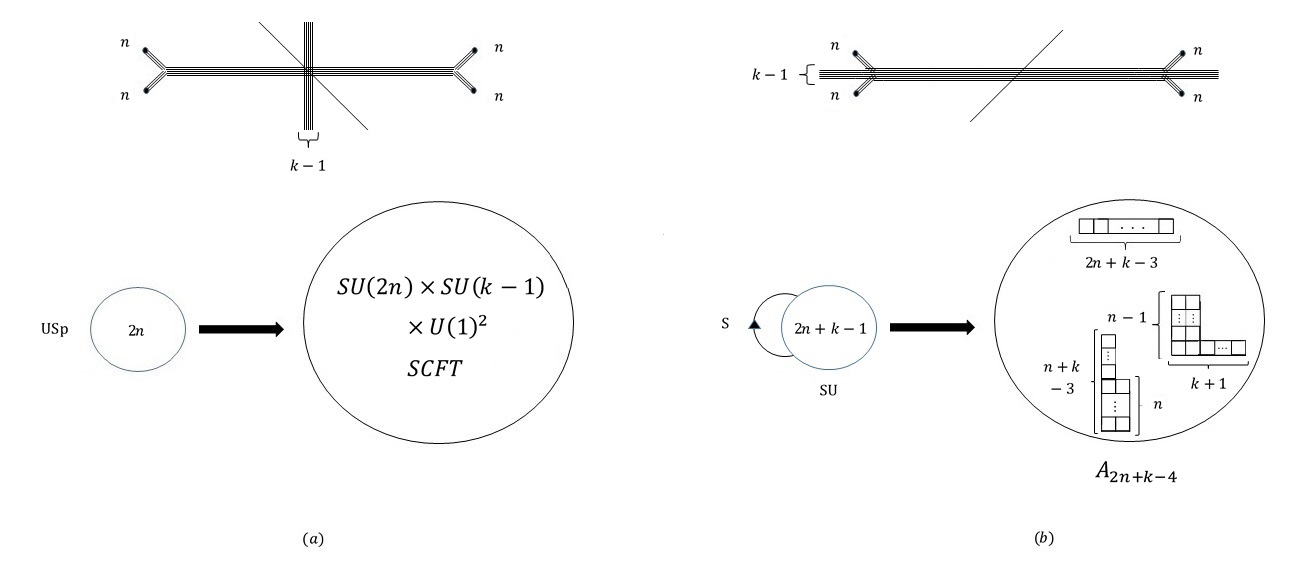} 
\caption{Two limits of a $USp(2n)$ gauging of an $SU(2n)$ global symmetry of the SCFT in figure \ref{Ils27}. (a) This limit corresponds to $g^{-2}_{USp(2n)}\rightarrow \infty$. Reducing with a twist leads to the theory shown below. (b) This limit corresponds to $g^{-2}_{USp(2n)}\rightarrow -\infty$ and reducing it with a twist leads to the theory shown below.}
\label{Ils28}
\end{figure}

\begin{figure}
\center
\includegraphics[width=1.0\textwidth]{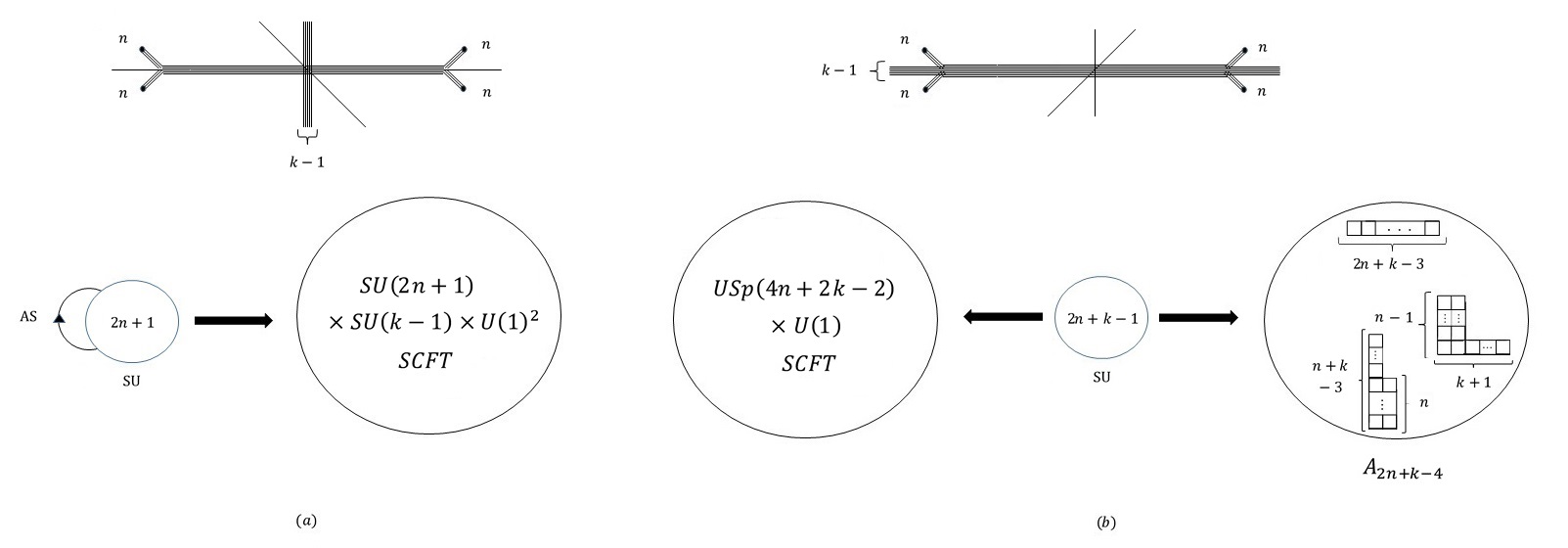} 
\caption{Two limits of a $SU(2n+1)+1AS$ gauging of an $SU(2n+1)$ global symmetry of the SCFT in figure \ref{Ils27}. (a) This limit corresponds to $g^{-2}_{SU(2n+1)}\rightarrow \infty$. Reducing with a twist leads to the theory shown below. (b) This limit corresponds to $g^{-2}_{SU(2n+1)}\rightarrow -\infty$ and reducing it with a twist leads to the theory shown below.}
\label{Ils29}
\end{figure}

Our method is the one we used previously: we consider dualities. Particularly, we concentrate on the $N=2n+1$ case and consider gauging a $USp(2n)$ gauge theory into the $SU(N-1)$ global symmetry. We can now reduce to $4d$ with the twist taking the scaling limit with $\frac{g^2}{R}$ fixed. Examining two limits of this reduction we arrive at the duality in figure \ref{Ils28}. Note that in the $k=2$ case this reduces to the $4d$ duality of \cite{CDT'}. This further supports identifying the $USp(4n)\times U(1)$ SCFT introduced there with the $4d$ theory resulting from the twisted compactification of the $5d$ SCFT in figure \ref{Ils27} for $k=2, N=2n+1$.

\begin{figure}
\center
\includegraphics[width=1.1\textwidth]{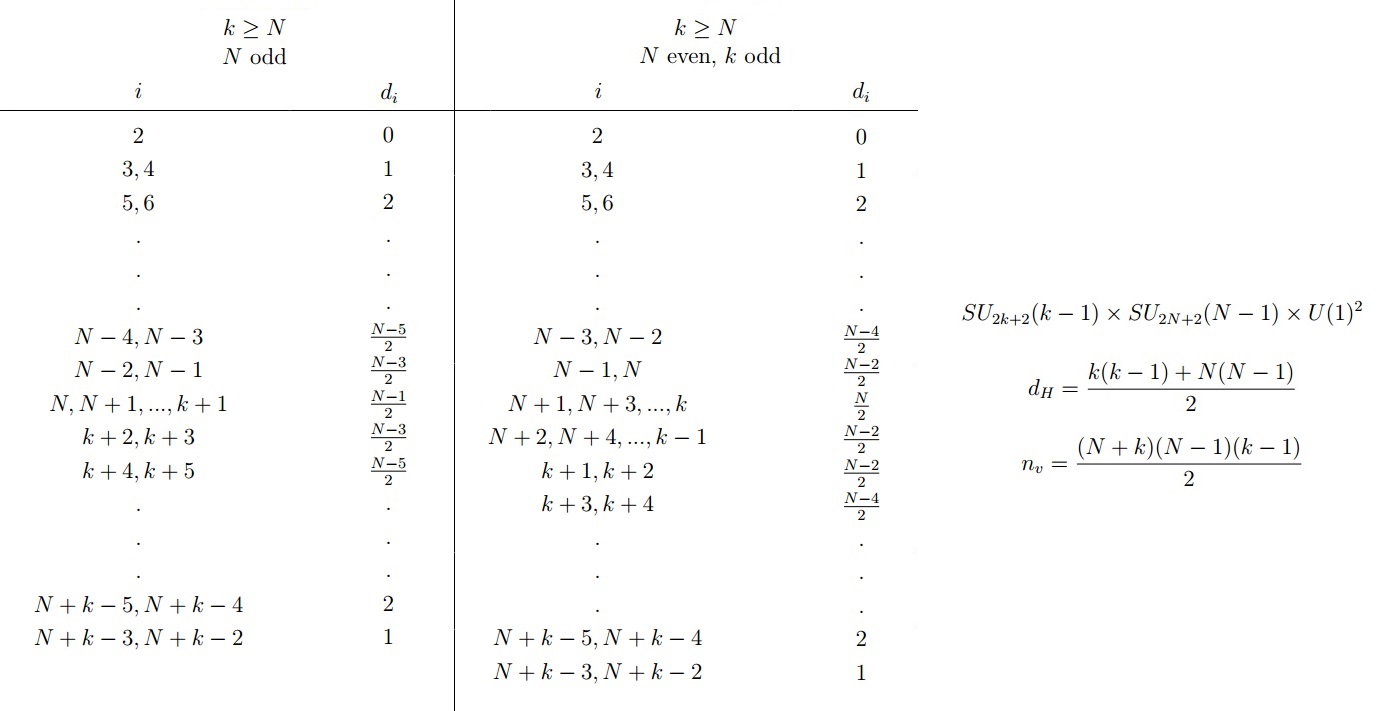} 
\caption{Properties of the conjectured $SU(k-1)\times SU(N-1)\times U(1)^2$ SCFT resulting from the twisted compactification of the $5d$ SCFT in figure \ref{Ils27}. These can be determined from the duality in figures \ref{Ils28}. The table on the left summarize the spectrum and dimensions of Coulomb branch operators where we have assumed that $k\geq N$. The other case is given by exchanging $N$ and $k$. Also written are the global symmetry with the central charges, Higgs branch dimension and effective number of vector multiplets. In the global symmetry we have assumed that $N,k>2$ having already discussed the $k=2$ and $N=2$ cases in section $2.1$. Note that the $N, k$ even case is not given since there is no duality which we case use to uncover them.}
\label{SCFTP3}
\end{figure}

\begin{figure}
\center
\includegraphics[width=1.0\textwidth]{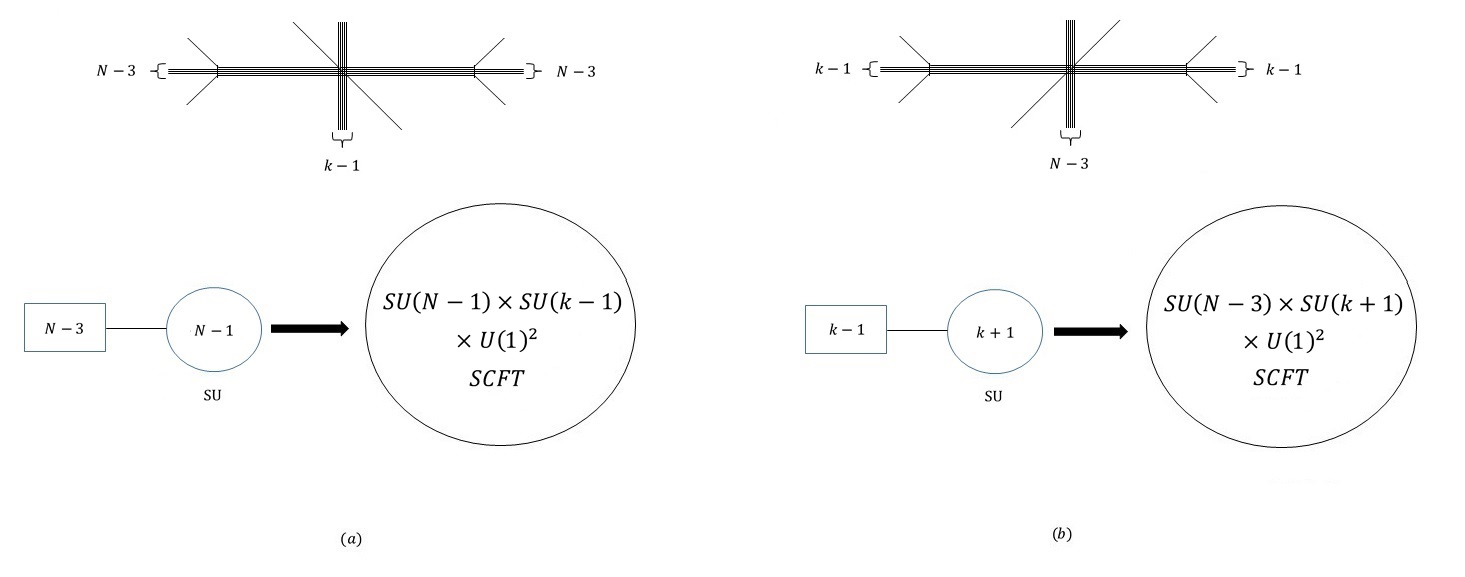} 
\caption{Two limits of a $SU(N-1)+(N-3)F$ gauging of an $SU(N-1)$ global symmetry of the SCFT in figure \ref{Ils27}. (a) This limit corresponds to $g^{-2}_{SU(N-1)}\rightarrow \infty$. Reducing with a twist leads to the theory shown below. (b) This limit corresponds to $g^{-2}_{SU(N-1)}\rightarrow -\infty$ and reducing it with a twist leads to the theory shown below.}
\label{Ils30}
\end{figure}

Like in the previous case, in the $N=2n+2$ case we can still carry this by gauging an $SU(2n+1)+1AS$ which gives the duality in figure \ref{Ils29}. Note that one side of the duality now involves the $USp(2N)\times U(1)$ SCFT that we introduced in section $2.1$. With the exception of the $N$ and $k$ even case, one side in it is made of known theories so we know that it is a conformal theory with a single marginal operators. Thus, the other side must also be an SCFT with the marginal operator being the $USp(2n)$ or $SU(2n+1)$ gauge coupling. This implies that the SCFT generated from the reduction of the $5d$ SCFT in figure \ref{Ils27} is an isolated SCFT. 

We can now perform the same consistency checks as before. First we can use the duality in figure \ref{Ils28} to determine the properties of the SCFT when either $N$ or $k$ are odd. We can in principle use the duality in figure \ref{Ils29} to study the duality when both $N$ and $k$ are even, but this requires understanding the properties of the $USp(2n)\times U(1)$ SCFT. These are known when $n$ is even using the results of \cite{CDT'}, but not when $n$ is odd. We have summarized the properties of this expected SCFT in figure \ref{SCFTP3} except in the $N, k$ even case. This can be evaluated directly from the duality in figure \ref{Ils28}. The duality in figure \ref{Ils29}, when applicable, can then be used as a consistency check. As other consistency checks we have verified that all properties are invariant under interchange of $N$ and $k$ and that the Coulomb branch and Higgs branch dimension, evaluated using $d_H = 24(c-a)$, agrees with what is expected from the web.  

We can also consider other dualities. For example figure \ref{Ils30} shows the duality when gauging an $SU(N-1)+(N-3)F$. We can now compare all the quantities in figure \ref{SCFTP3} finding that they agree. It is straightforward to generalize the other dualities also to this case. We shall not carry this out here. 

\subsubsection{Dualities for $SU$ with symmetric matter}

\begin{figure}
\center
\includegraphics[width=1.0\textwidth]{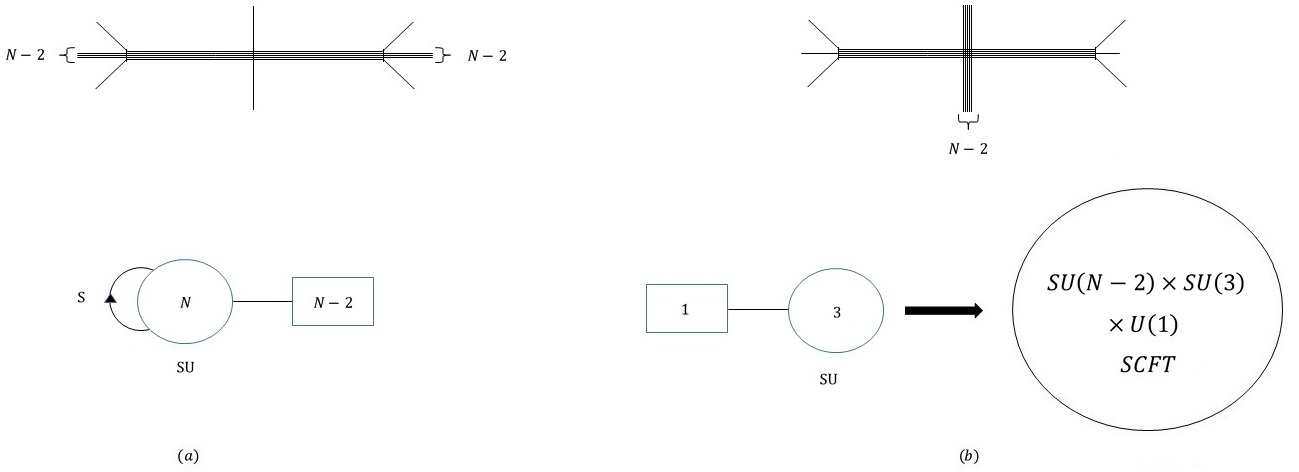} 
\caption{Two limits of a $SU(N)+(N-2)F$ gauging of an $SU(N)$ global symmetry of the $SU(N)\times SU(1)\times U(1)$ SCFT. (a) This limit corresponds to $g^{-2}_{SU(N)}\rightarrow \infty$. Reducing with a twist leads to the theory shown below. (b) This limit corresponds to $g^{-2}_{SU(N)}\rightarrow -\infty$ and reducing it with a twist leads to the theory shown below.}
\label{Ils23}
\end{figure}

We can apply this to study $4d$ theories involving $SU$ with symmetric matter which, to our knowledge, have no known class S construction. For example consider the $4d$ gauge theory $SU(N)+1S+(N-2)F$. The beta function vanishes so this is a conformal theory. We can engineer it from the $5d$ description as follows. Consider the $5d$ SCFT shown in figure \ref{Ils23} (a). Reducing it to $4d$ with a twist and taking a scaling limit leads to the $4d$ $SU(N)+1S+(N-2)F$ gauge theory. We can now use this to study dualities of this theory. For example figure \ref{Ils23} (b) shows the same reduction, but with a different weak coupling description. This leads to the duality shown in figure \ref{Ils23}.

\begin{figure}
\center
\includegraphics[width=1.0\textwidth]{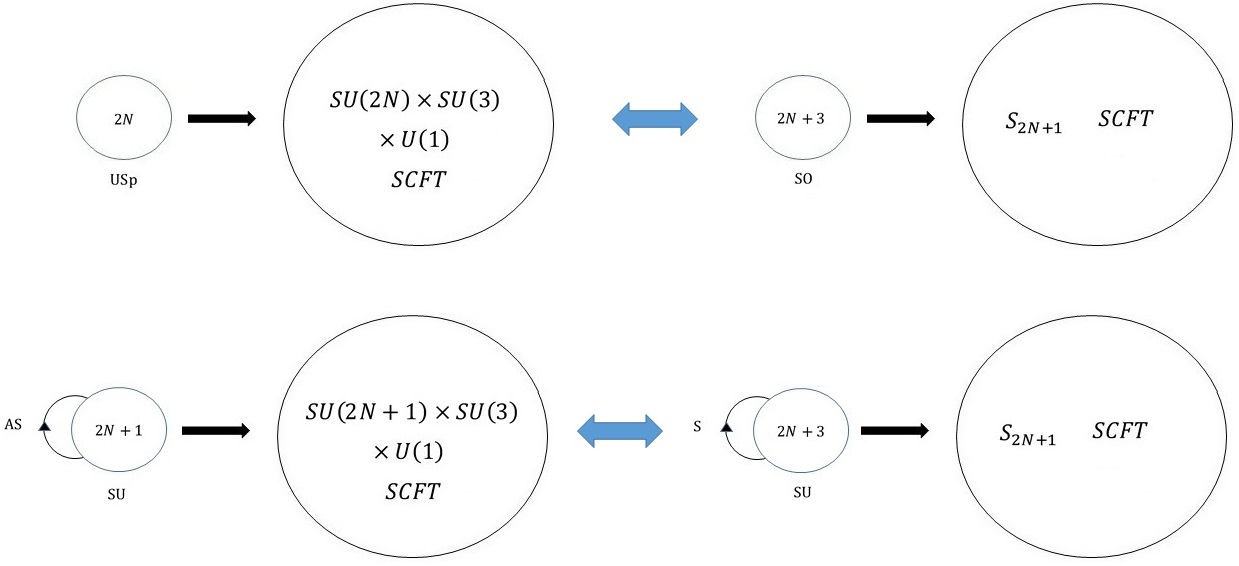} 
\caption{Two dualities of the $SU(N)\times SU(3)\times U(1)$ for $N$ even (upper) and odd (lower) which we get by setting $k=3$ in the dualities of figures \ref{Ils15} and \ref{Ils16}.}
\label{Ils24}
\end{figure}

It is interesting to also study dualities resulting from gauging the $SU(N-2)$ global symmetry of the SCFT appearing in figure \ref{Ils23}. Using the results in figures \ref{Ils15} and \ref{Ils16} we find the dualities in figure \ref{Ils24} where we have used the naming conventions of \cite{CD}. The $S_N$ theory that appears in these dualities is dual to an $SU(N)+1AS+(N+2)F$ gauge theory when its $SU(3)$ subgroup is gauged by $SU(3)+1F$.

\begin{figure}
\center
\includegraphics[width=1.0\textwidth]{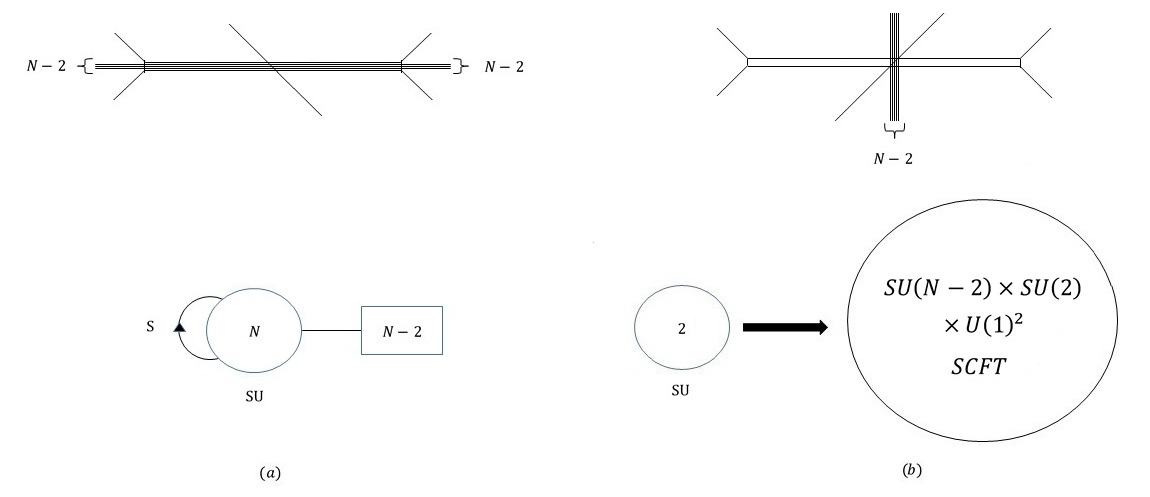} 
\caption{Two limits of a $SU(N)+(N-2)F$ gauging of an $SU(N)$ global symmetry of the $SU(N)\times U(1)^2$ SCFT. (a) This limit corresponds to $g^{-2}_{SU(N)}\rightarrow \infty$. Reducing with a twist leads to the theory shown below. (b) This limit corresponds to $g^{-2}_{SU(N)}\rightarrow -\infty$ and reducing it with a twist leads to the theory shown below.}
\label{Ils31}
\end{figure}

\begin{figure}
\center
\includegraphics[width=1.0\textwidth]{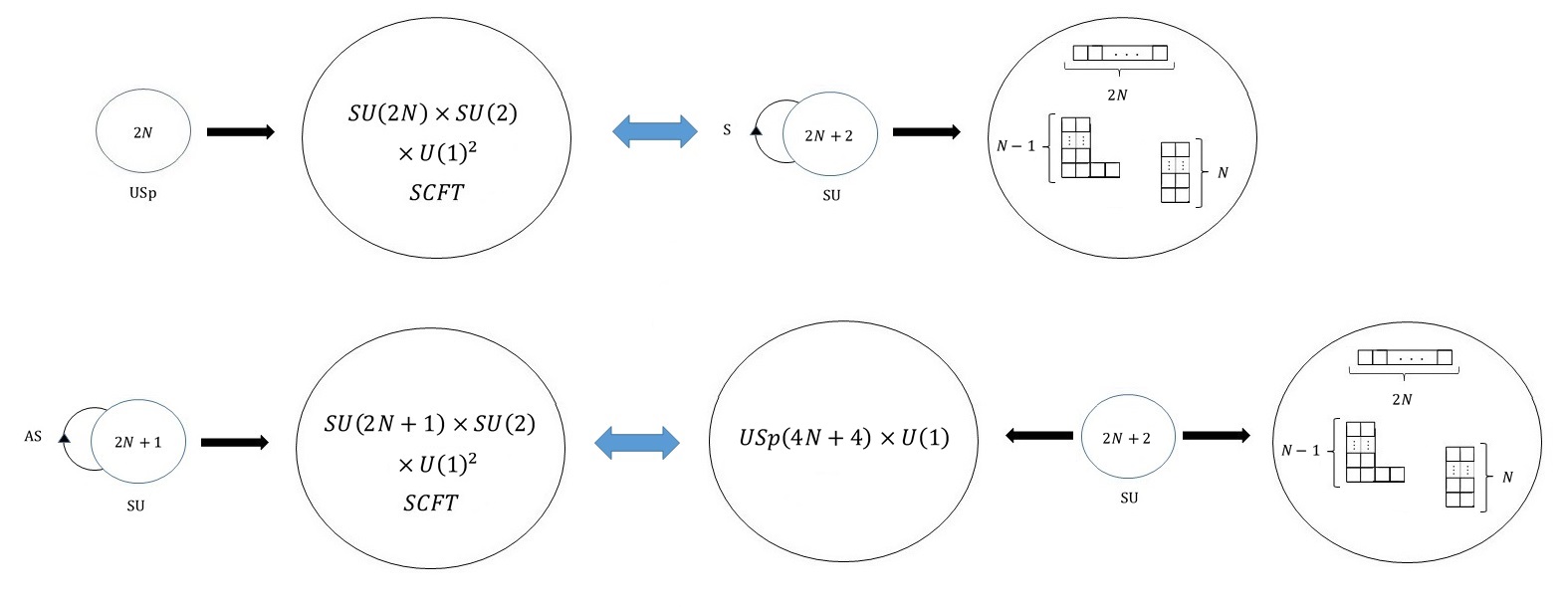} 
\caption{Two dualities of the $SU(N)\times SU(2)\times U(1)^2$ for $N$ even (upper) and odd (lower) which we get by setting $k=3$ in the dualities of figures \ref{Ils28} and \ref{Ils29}.}
\label{Ils32}
\end{figure}

We can get another dual frame of the $4d$ $SU(N)+1S+(N-2)F$ SCFT by using the family of SCFTs introduced in the previous subsection. For this we consider reducing with a twist the $5d$ SCFT of figure \ref{Ils31} (a) while taking a scaling limit leading to the $4d$ gauge theory $SU(N)+1S+(N-2)F$. Again by considering the same reduction in a different range of the parameters we get the dual description shown in figure \ref{Ils31} (b). We can again consider gauging the $SU(N-2)$ global symmetry. Using the results in figures \ref{Ils28} and \ref{Ils29} we find the dualities in figure \ref{Ils32}. The class S theory appearing in the figure as the property that gauging its $SU_8(2)$ global symmetry leads to a $USp(2N)+(2N+2)F$ gauge theory.

\begin{figure}
\center
\includegraphics[width=1.1\textwidth]{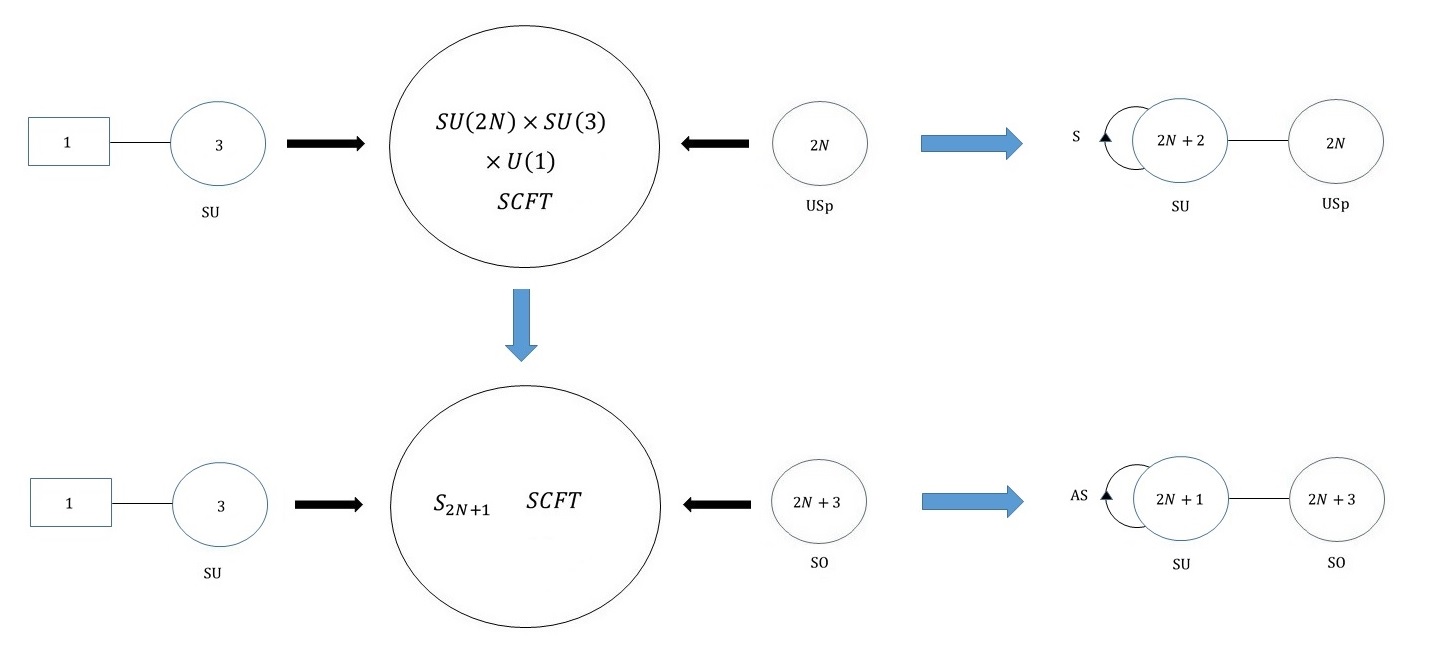} 
\caption{Combining the two dualities in figures \ref{Ils23} and \ref{Ils24} we get that the two theories on the right are dual.}
\label{Ils25}
\end{figure}

\begin{figure}
\center
\includegraphics[width=1.1\textwidth]{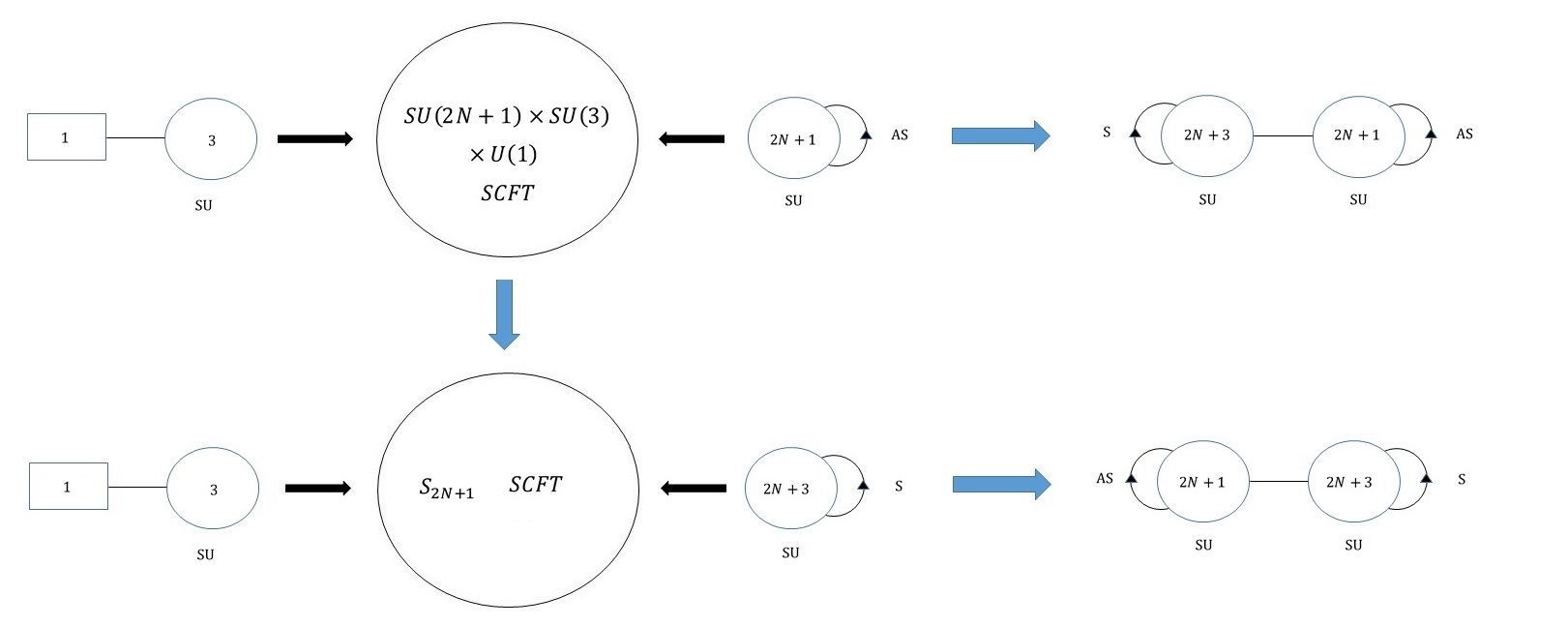} 
\caption{Combining the two dualities in figures \ref{Ils23} and \ref{Ils24} we get that the two theories on the right are dual.}
\label{Ils26}
\end{figure}

We can use this to motivate a completely perturbative duality by performing both gauging simultaneously. Consider the theory shown in figure \ref{Ils25} where we have gauged the $SU(2N)\times SU(3)\times U(1)$ SCFT by an $SU(3)+1F$ and $USp(2N)$ gauge theories. Then by applying the dualities of figures \ref{Ils23} and \ref{Ils24} we get a duality between two gauge theories. We can play the same game also on the $SU(2N+1)\times SU(3)\times U(1)$ SCFT, now gauging by an $SU(3)+1F$ and $SU(2N+1)+1AS$ gauge theories. This is shown in figure \ref{Ils26} and results in a self duality.


\begin{figure}
\center
\includegraphics[width=1.1\textwidth]{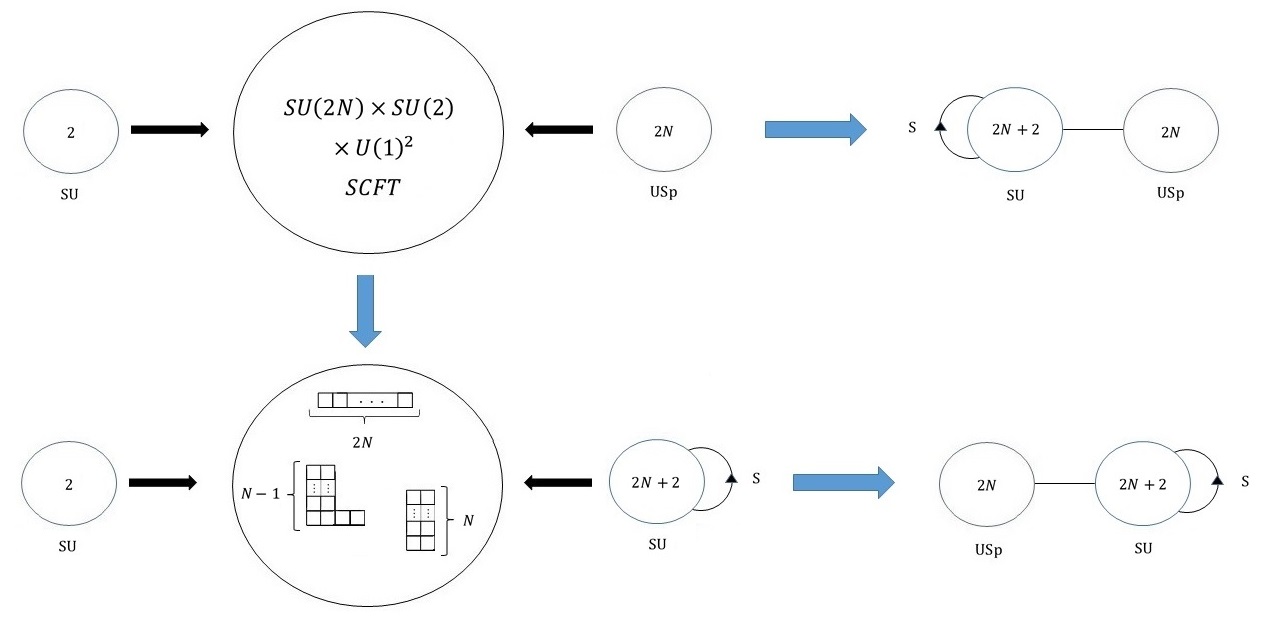} 
\caption{Combining the two dualities in figures \ref{Ils31} and \ref{Ils32} we get that the two theories on the right are dual.}
\label{Ils330}
\end{figure}

\begin{figure}
\center
\includegraphics[width=1.1\textwidth]{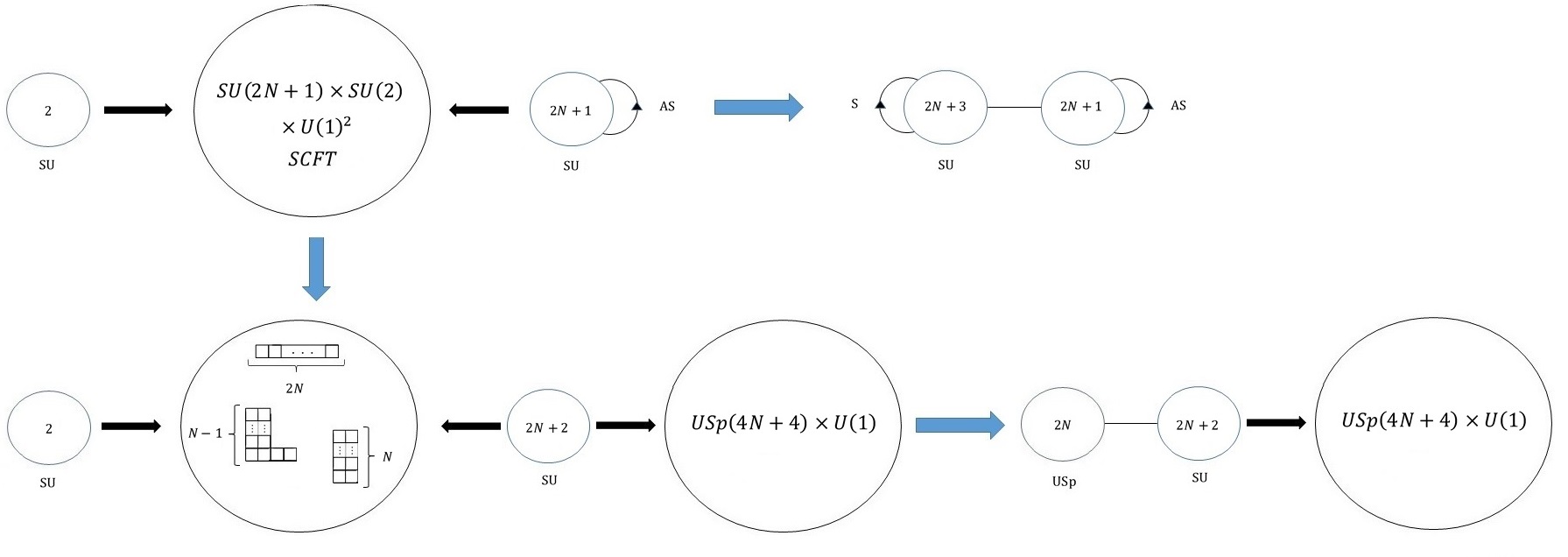} 
\caption{Combining the two dualities in figures \ref{Ils31} and \ref{Ils32} we get that the two theories on the right are dual.}
\label{Ils331}
\end{figure}

We can also consider simultaneously gauging the $SU(2)$ and $SU(N-2)$ global symmetries of the $SU(N-2)\times SU(2)\times U(1)^2$ SCFT. This leads to the dualities in figures \ref{Ils330} and \ref{Ils331}, all involving known theories. The case of figure \ref{Ils330} is a self duality while the one in figure \ref{Ils331} is not. It is interesting that this is the exact opposite of the previous case. 

Since these dualities all involve known theories they can be tested by ordinary means. For example consider the duality in figure \ref{Ils25}, which is between two gauge theories. As we have a perturbative description on both sides, it can be tested by more intricate means like comparing the superconformal index. Nevertheless evaluating the superconfrmal index in the general case is technically challenging. Therefore, as a simple test, we shall compromise on examining the $N=2$ case, which is the first new case\footnote{The $N=1$ case follows from the dualities in \cite{AW}.}. Also we shall compute the index in the simplified Hall-Littlewood limit\cite{GRRY} (see also appendix A).

We expand the index in a power series in $\tau$, evaluating to order $\tau^5$. We indeed find the indices agree being given by:

\be
I^{HL}_{AS/S+SU\times SO/USp} = 1 + 2\tau^2 + 4\tau^4 -(\frac{p^2}{a}+\frac{a}{p^2})\tau^5 + O(\tau^6)
\ee
where we use $p$ for the bifundamental $U(1)$ fugacity and $a$ for the $U(1)$ fugacity associated with the symmetric or antisymmetric matter, depending on the theory.

We can do the same also for the duality in figure \ref{Ils331} for the case of $N=1$. Besides technical issues we also need the Hall-Littlewood index of the $U(1)\times USp(8)$ SCFT. We can calculate it using the conjectured formula for the index given in \cite{CDT'}. We find:

\be
I^{HL}_{U(1)\times USp(8)} = 1 + \tau^2 (\chi[\bold{36}]+1) + \tau^4(\chi[\bold{330}]+\chi[\bold{308}]+\chi[\bold{36}]+1) +(h^5+\frac{1}{h^5})\chi[\bold{42}]\tau^5 + O(\tau^6)
\ee
where we used $\chi[\bold{d}]$ for the character of the $d$ dimensional representation under the non-abelian global symmetry, in this case being $USp(8)$. Also we used $h$ as the fugacity for the $U(1)$ global symmetry. 

We can now use this to evaluate the indices for the two proposed dual theories. For the $1S+SU(5)\times SU(3)+1F$ theory we find:

\be
I^{HL}_{1S+SU(5)\times SU(3)+1F} = 1 + 3\tau^2 + 6\tau^4 +(c f^2 p+\frac{1}{c f^2 p})(\frac{f^3}{c p}+\frac{c p}{f^3})\tau^5 + O(\tau^6)
\ee
where we again use $p$ for the bifundamental fugacity, $f$ for the symmetric fugacity and $c$ for the fundamental one. 

For the $SU(2)\times SU(4)$ gauging of the $U(1)\times USp(8)$ SCFT, we find:

\be
I^{HL}_{SU(2)\times SU(4)\hookrightarrow U(1)\times USp(8)} = 1 + 3\tau^2 + 6\tau^4 +(h^5+\frac{1}{h^5})(x^4+\frac{1}{x^4})\tau^5 + O(\tau^6)
\ee
where $x$ is the fugacity for the $U(1)$ commutant of $SU(4)$ inside $USp(8)$. We now see that the two indices indeed match, to the evaluated order, if we identify $h^5, x^4$ with $c f^2 p, \frac{f^3}{c p}$.  

\section{Twisted compactification of $T_N$ theories}

\begin{figure}
\center
\includegraphics[width=0.31\textwidth]{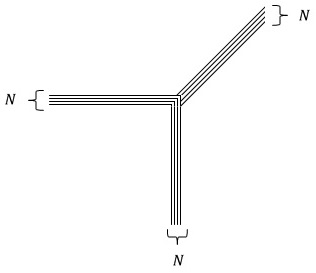} 
\caption{The brane web for the $5d$ $T_N$ theory.}
\label{Ils34}
\end{figure}

In this section we move on to investigate twisted compactification of the $5d$ $T_N$ theory whose web description is given in figure \ref{Ils34} \cite{BB}. This SCFT has an $S_3$ discrete symmetry which acts by permutating the three $SU(N)$ global symmetries. In the analogous $4d$ theory this is seen as permutating the three maximal punctures. We can now consider compactifying this theory to $4d$ with a twist in an abelian subgroup of $S_3$. There are two cases to consider, $Z_2$ and $Z_3$. We shall first start with the $Z_3$ case and then move on to discuss the $Z_2$ case. 

\subsection{$Z_3$ twist}

In this subsection we discuss compactification of the $5d$ $T_N$ theory on a circle with a $Z_3 \subset S_3$ twist. The $Z_3$ element we twist by can be conveniently represented by the $SL(2,Z)$ element $TS$:

\be
TS = \left(
\begin{array}{cc}
-1 & 1 \\
-1 & 0
\end{array}
\right) \,.
\ee

Like in the previous case, we shall start by studying some simple low $N$ cases where we can readily identify the resulting theory with a known $4d$ SCFT. We then test this by performing various consistency checks. In the general $N$ case we cannot readily identify them with any known theory leading us to believe these are new. We discuss some of their expected properties.  

\subsubsection*{$N=2$ case}

We begin with the $N=2$ case where the $5d$ SCFT reduces to eight free half-hypermultiplets in the $(\bold{2},\bold{2},\bold{2})$ of the $SU(2)^3$ global symmetry. In this case we have a perturbative description of the SCFT so we can easily determine the resulting $4d$ theory. The twist project the free half-hypermultiplets to the $Z_3$ symmetric part which is four free half-hypermultiplets in the $\otimes^3_{Sym} \bold{2} = \bold{4}$ of the $SU(2)$ global symmetry. 

We can also support this from the Higgs branch dimension. This can be counted in the web by looking at the directions compatible with the $Z_3$ twist. There are two such directions agreeing with what is expected from four free half-hypermultiplets. The first is given by forcing the two $(1,0), (0,1)$ and $(1,1)$ $5$-branes to end on the same $7$-brane while the second is given by separating the two $5$-brane junctions\footnote{There is a question as to whether this direction indeed survives the projection. Unlike the previous case we do not have a T-dual description to use in order to answer this. We can however use knowledge of the theory to determine that such a direction exists.}. 

\subsubsection*{$N=3$ case}

Next we move on to the $N=3$ case. The theory we now consider is the rank $1$ $E_6$ theory. Compactifying with a twist we get a $4d$ theory with rank $1$ and a $5$ dimensional Higgs branch. We can also consider the global symmetry of the $4d$ theory. The $5d$ theory has an $E_6$ global symmetry and so possesses moment map operators in the adjoint of $E_6$ which decomposes to $(\bold{8},\bold{1},\bold{1})+(\bold{1},\bold{8},\bold{1})+(\bold{1},\bold{1},\bold{8})+(\bold{3},\bold{3},\bold{3})+(\bar{\bold{3}},\bar{\bold{3}},\bar{\bold{3}})$ under the $SU(3)^3$ subgroup of $E_6$. Projecting to the $Z_3$ invariant part we get moment map operators in the $\bold{8} + \bold{10}+ \bar{\bold{10}}$ of the symmetric $SU(3)$ global symmetry. This builds the adjoint of $SO(8)$. Thus, we conjecture that the resulting $4d$ theory should be $SU(2)+4F$ which has an $SO(8)$ global symmetry, rank $1$ and a $5$ dimensional Higgs branch. 

\subsubsection*{$N=4$ case}

\begin{figure}
\center
\includegraphics[width=0.9\textwidth]{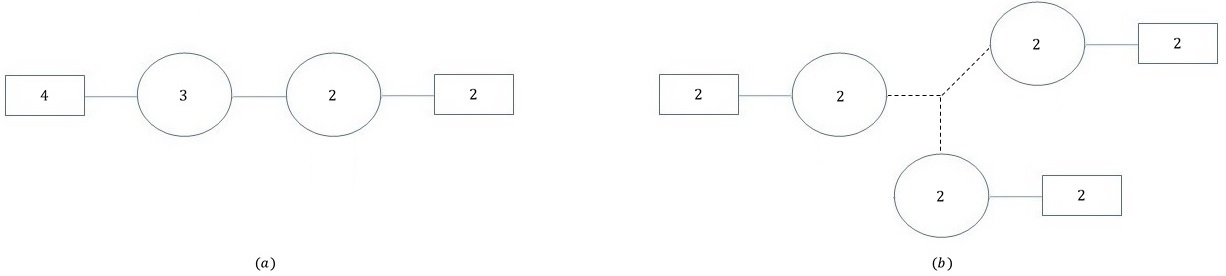} 
\caption{(a) The gauge theory description of the $5d$ $T_4$ theory. (b) Dualizing the $SU_0(3)+6F$ part to $2F+SU(2) \times SU(2)+2F$ we get this theory where the dashed line stands for an half-hyper in the $(\bold{2},\bold{2},\bold{2})$.}
\label{Ils35}
\end{figure}

Now we move to the $N=4$ case that is the $5d$ $T_4$ theory. It will be instructive to consider a gauge theory description of the SCFT where the $Z_3$ symmetry is manifest. For this we start with the gauge theory description given in \cite{BZ} shown in figure \ref{Ils35} (a). We dualize the $SU_0(3)+6F$ part to $2F+SU(2) \times SU(2)+2F$ which gives the gauge theory in figure \ref{Ils35} (b). This theory has an $S_3$ symmetry given by permutating the three $SU(2)$ groups which is the descendent of the $S_3$ symmetry of the $T_4$ SCFT. 

Now we inquire about $4d$ theory resulting from twisted compactification of the $5d$ $T_4$ SCFT. Both the web and the gauge theory description suggests it should have rank $1$, and we also expect an $SU(4)$ global symmetry. There is indeed a known theory with these properties which is the recently discovered rank $1$ $SU(4)$ SCFT in \cite{CDTn}. It is natural to expect that we get this theory. We next provide further evidence for this. First we can compare the dimension of the Higgs branch where the web suggests a $9$ dimensional Higgs branch agreeing with the Higgs branch dimension of the rank $1$ $SU(4)$ SCFT\cite{CDTn}. We can also ask what is the theory we get on the Higgs branch. The $T_4$ theory has a $Z_3$ symmetric direction along which it is reduced to $T_3$. Thus the previous discussion suggests that our theory should have a Higgs branch direction leading to $SU(2)+4F$. This also agrees with properties of the rank $1$ $SU(4)$ SCFT\cite{CDTn}.

As a more intricate check we can consider the Hall-Littlewood chiral ring\cite{BPRR}. In \cite{CDTn} it was determined to be comprised from the following operators: $\tau^2 \chi[\bold{15}] + \tau^3 (\chi[\bold{20}''] +  \chi[\bar{\bold{20}}'']) + \tau^4 \chi[\bold{50}] + O(\tau^5)$ ($\tau$ is the same fugacity appearing in the Hall-Littlewood index, see appendix A). As we now argue this naturally follows from this construction. Let's consider the $5d$ $T_4$ theory compactified with a twist. As the Higgs branch is invariant under quantum corrections the $4d$ Higgs branch should be identical to a subspace of the original invariant under the twist. For the theory at hand, the Hall-Littlewood index, which is made from operators belonging to the Hall-Littlewood chiral ring, should be identical to the Hilbert series of the Higgs branch\cite{GRRY,BHM}. Thus symmetric combinations of Higgs branch operators of the $5d$ theory should descend to Higgs branch operators of the $4d$ theory, and thus to part of the Hall-Littlewood chiral ring. So we expect to be able to identify the Hall-Littlewood chiral ring of the $4d$ theory in operators of the $5d$ theory. Note that there could exist operators acting on the subspace without a corresponding operator on the full space, or alternatively constraints acting on the subspace with no analogue on the full space. Thus, this procedure may not generate the full Hall-Littlewood chiral ring.

We can determine the basic Higgs branch operators of the $5d$ $T_4$ theory directly from $5d$ as done in the appendix. However we can use a shortcut utilizing the fact that compactifying the $5d$ $T_4$ theory leads to the $4d$ one and that the Higgs branch of the two is identical. It is well known that the Hall-Littlewood chiral ring of the $4d$ $T_4$ theory is comprised of the basic operators: $\tau^2 (\chi[\bold{15},\bold{1},\bold{1}] + \chi[\bold{1},\bold{15},\bold{1}] + \chi[\bold{1},\bold{1},\bold{15}] ), \tau^3 ( \chi[\bold{4},\bold{4},\bold{4}] + \chi[\bar{\bold{4}},\bar{\bold{4}},\bar{\bold{4}}] ), \tau^4 \chi[\bold{6},\bold{6},\bold{6}]$. We next project these operators to their $Z_3$ invariants by identifying the three $SU(4)$ groups giving:

\be
\chi[\bold{15},\bold{1},\bold{1}] + \chi[\bold{1},\bold{15},\bold{1}] + \chi[\bold{1},\bold{1},\bold{15}] \rightarrow \chi[\bold{15}]
\ee

These are the moment map operators and this just tells us that the three $SU(4)$ groups are projected to the diagonal.

\be
\chi[\bold{4},\bold{4},\bold{4}] \rightarrow  \chi[\otimes^3_{Sym}\bold{4}] = \chi[\bold{20}'']
\ee
and likewise for the conjugate. 

\be
\chi[\bold{6},\bold{6},\bold{6}] \rightarrow  \chi[\otimes^3_{Sym}\bold{6}] = \chi[\bold{50}] + \chi[\bold{6}]
\ee
where the three index symmetric tensor of $SO(6)$ decomposes to the trace and traceless part (the $\bold{50}$).

Doing the projection on the Higgs branch operators of the $T_4$ theory we see that we indeed get the ones of the $SU(4)$ SCFT. Note that the $\tau^4 \chi[\bold{6}]$ operators corresponding to the trace of the three index symmetric tensor is apparently projected out. This is quite reasonable as this imply an additional constraint, a tracelessness condition, that only makes sense on the subspace. 

It is interesting if one can make this construction more precise. We will explore this further in the next section.

We can also consider mass deformations. The gauge theory description suggests at least two interesting ones. First There is the mass deformation leading to the low energy gauge theory. Since we have a perturbative description it is straightforward to carry out the reduction. The three $SU(2)+2F$ parts are identified leading to just one $SU(2)+2F$ gauge theory. The half trifundamentals should be projected to the symmetric part being the $\bold{4}$ of $SU(2)$. Thus we conclude that the resulting $4d$ SCFT should have a mass deformation leading to the IR gauge theory $SU(2)+\frac{1}{2}\bold{4}+2F$. This is consistent with the results of \cite{ALLM1} who examined the Seiberg-Witten curve of this theory. 

\begin{figure}
\center
\includegraphics[width=1.1\textwidth]{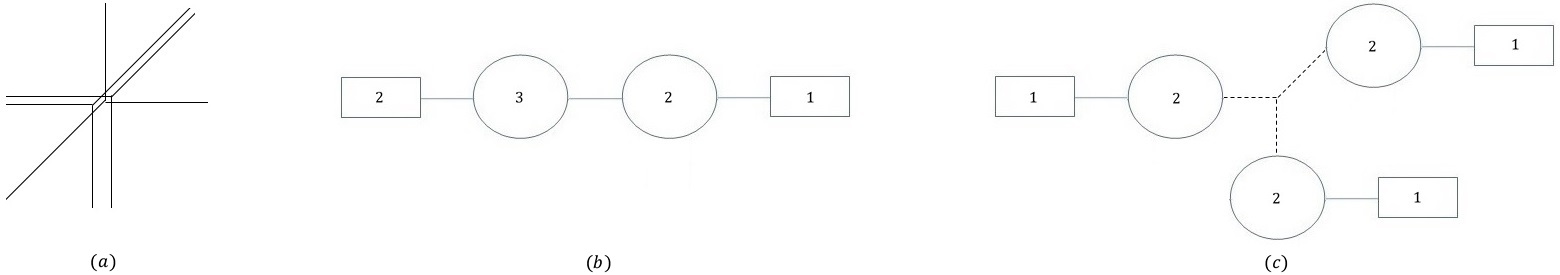} 
\caption{(a) The brane web for a $5d$ SCFT given by a mass deformation of the $5d$ $T_4$ theory. (b) A gauge theory description of the SCFT. (c) Dualizing the $SU_0(3)+4F$ part to $1F+SU(2) \times SU(2)+1F$ we get this theory where the dashed line standing for an half-hyper in the $(\bold{2},\bold{2},\bold{2})$.}
\label{Ils36}
\end{figure}

Another mass deformation is given by integrating out a flavor leading to the gauge theory in figure \ref{Ils36} (c). This gauge theory should originate from a $5d$ SCFT which can be generated from the $T_4$ SCFT via a mass deformation. It can be given a brane web description shown in figure \ref{Ils36} (a). As can be seen from the web and confirmed using the $5d$ superconformal index (see the appendix) this $5d$ SCFT should have an $SU(2)^3\times U(1)^3$ global symmetry leading us to suspect the $4d$ theory should have an $SU(2)\times U(1)$ global symmetry. In \cite{ALLM1} it was found that there is indeed a mass deformation leading to a $4d$ theory with an $SU(2)\times U(1)$ global symmetry which is expected to be yet another rank $1$ SCFT. It is natural to identify the resulting $4d$ theory with this SCFT.

There are two additional pieces of evidence for this identification. First it has a mass deformation leading to an $SU(2)+\frac{1}{2}\bold{4}+1F$ IR free gauge theory as expected from the $SU(2)\times U(1)$ SCFT\cite{ALLM1}. Second the web suggests it has a $3$ dimensional Higgs branch, which agrees with the results of \cite{ALLM1} if one uses $d_H = 24(c-a)$. We can also consider the Hall-Littlewood chiral ring of this SCFT from the $5d$ index similarly to the previous case. We carry this out in appendix B. 

\begin{figure}
\center
\includegraphics[width=0.7\textwidth]{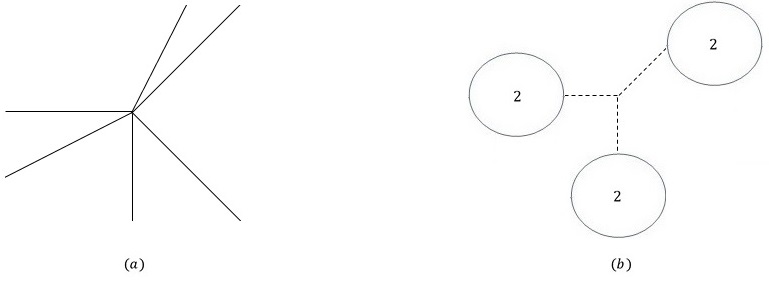} 
\caption{(a) The brane web for a $5d$ SCFT given by a mass deformation of the $5d$ SCFT in figure \ref{Ils36}. (b) A gauge theory description of the SCFT.}
\label{Ils37}
\end{figure}

Finally we can consider another mass deformation now leading to the gauge theory in figure \ref{Ils37} (b)\footnote{Naively there should be two different gauge theories depending on whether the $SU(2)$ $\theta$ angles are all $0$ or $\pi$. Yet we seem to find only one $Z_3$ symmetric brane web leading us to suspect that these are identical. In fact, due to the presence of the trifundamental, exchanging two $SU(2)$ gauge groups is effectively seen by the third as reversing the mass of one flavor and so changes its $\theta$ angle.}. Again this gauge theory should originate from a $5d$ SCFT which we identify with the web in figure \ref{Ils37} (a). This should have a $U(1)^3$ global symmetry leading us to expect the $4d$ SCFT to have a $U(1)$ global symmetry. The resulting theory should also have a mass deformation leading to an $SU(2)+\frac{1}{2}\bold{4}$ IR free gauge theory. Indeed in \cite{ALLM1} it was found that the $SU(2)\times U(1)$ SCFT has such a mass deformation leading to an SCFT with these characteristic. In fact this SCFT is suspected to have an enhanced $\mathcal{N}$$=3$ supersymmetry\cite{TN}. As a supporting evidence we note that the Higgs branch dimension is $1$ which agrees with the field theory analysis of \cite{TN}. 

 
\subsubsection*{Higher $N$ and related theories}

We can now continue to theories with higher $N$. For example for $N=5$ we expect a $4d$ theory with an $SU(5)$ global symmetry and rank $2$ Coulomb branch. It should also have a $14$ dimensional Higgs branch, along which it can be reduced to the rank $1$ $SU(4)$ SCFT. We also expect the Hall-Littlewood chiral to contain the operators: $\tau^2 \chi[\bold{24}], \tau^4 (\chi[\bold{35}] +  \chi[\bar{\bold{35}}]), \tau^6 ( \chi[\bold{175}''] +  \chi[\bar{\bold{175}}''] )$. To our awareness, no such $4d$ theory is known. The preceding thus suggests that there is host of possibly unknown $4d$ SCFTs given by the twisted compactification of the $5d$ $T_N$ theories with a $Z_3$ twist. 

We can further generate additional theories by Higgs branch flows and mass deformations. Higgs branch flows are readily visible from the brane webs. First we can pullout a group of $5$-brane junctions. This initiate a flow one $T_N$ to another one with lower $N$. Alternatively we can break some of the $5$-branes on the $7$-branes. As previously discussed this can be naturally implemented by associating to the SCFT a Young diagram with $N$ boxes. The theory given by the compactification of the $T_N$ theory is then represented by the one row Young diagram, while other choices giving different theories.

\begin{figure}
\center
\includegraphics[width=0.285\textwidth]{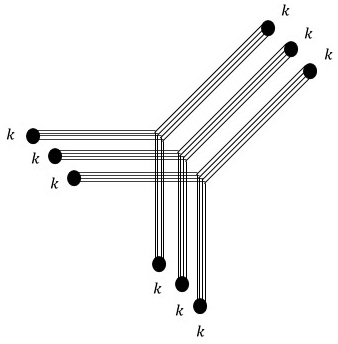} 
\caption{The brane web for the $5d$ rank $k$ $E_6$ theory.}
\label{Ils38}
\end{figure}

We can identify some of these theories with known $4d$ theories. As an example of a theory in this class that we can identify, consider the rank $k$ $E_6$ theory whose web is shown in figure \ref{Ils38}. It is natural to conjecture that reducing it with a twist leads to the $4d$ rank $k$ $SO(8)$ theory which is just the gauge theory $USp(2k)+1AS+4F$. Indeed this theory has a $5k-1$ dimensional Higgs branch agreeing with the web. In particular the one associated with the antisymmetric breaks the theory to $k$ copies of $SU(2)+4F$ agreeing with what is expected from the web. We shall see another example of a theory in this class in the next section.

We can also consider mass deformations. As seen in the $T_4$ example, this may lead to new SCFT as well as non-conformal theories. We suspect that this will be true also for cases with higher $N$ so besides the theories introduced so far there should be many additional theories that are mass deformations of these. As we shall now argue some of them are related to the ones introduced and to themselves via dualities. 

\subsubsection{Dualities}

\begin{figure}
\center
\includegraphics[width=1.0\textwidth]{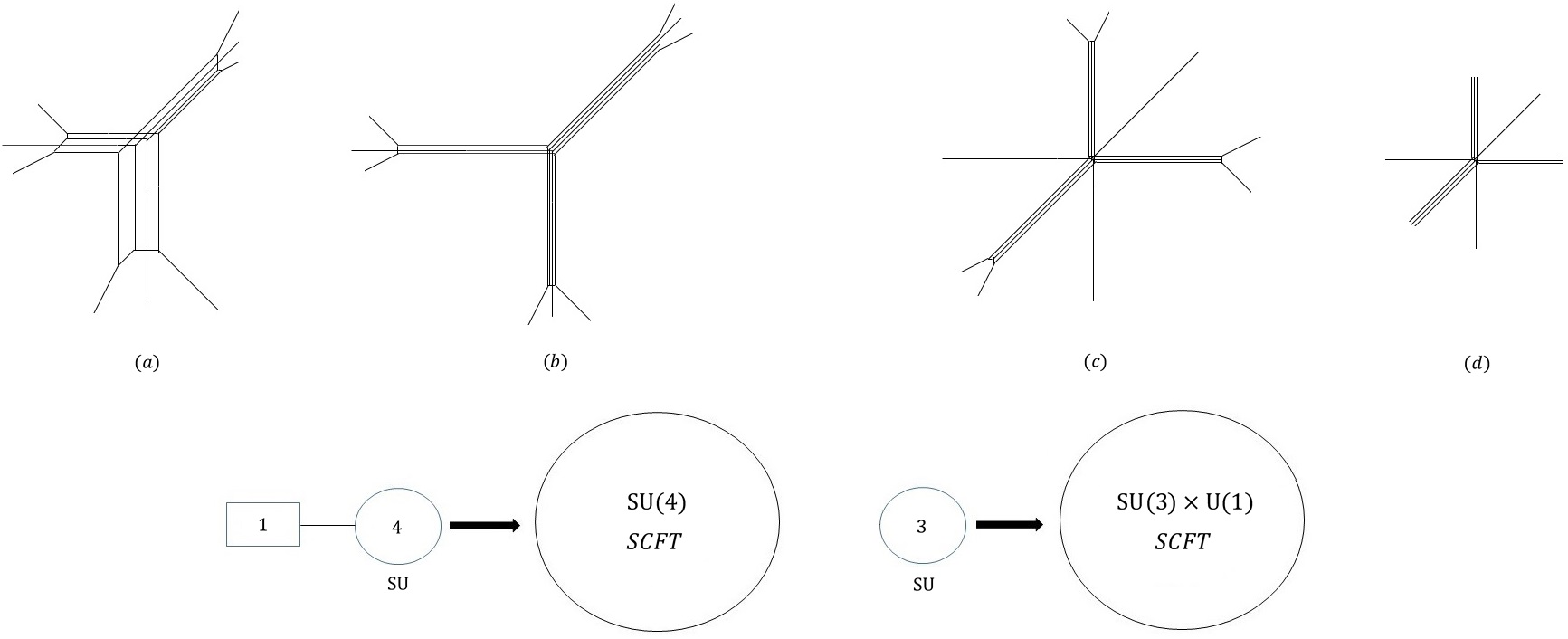} 
\caption{ (a) The brane web for the $5d$ SCFT resulting from a $Z_3$ symmetric $SU(4)+1F$ gauging of the $T_4$ theory. (b) The $5d$ SCFT in the limit of $g^{-2}_{SU(4)}\rightarrow \infty$. Bellow is the $4d$ theory resulting from a twisted compactification in this limit. (c) The $5d$ SCFT in the limit of $g^{-2}_{SU(4)}\rightarrow -\infty$. In this limit there is a better description as an $SU(3)$ gauging of the $5d$ SCFT in (d). Bellow is the $4d$ theory resulting from a twisted compactification in this limit.}
\label{Ils444}
\end{figure}

We can consider dualities of the class of theories we introduced in the same spirit as performed in section $2$. As a simple example consider the $Z_3$ symmetric gauging of the three $SU(4)$ global symmetry groups of the $T_4$ theory by an $SU(4)+1F$ gauge theory. The brane web for the resulting $5d$ SCFT is shown in figure \ref{Ils444} (a). That it is $Z_3$ symmetric is most readily visible by noting it is invariant under $TS$. We can consider reducing this theory to $4d$ with the $Z_3$ twist and taking the scaling limit $g^2_{SU(4)}\rightarrow 0, R\rightarrow 0$ keeping the ratio $\frac{g^2_{SU(4)}}{R}$ fixed.

We can examine the reduction in two different limits. First we can consider the limit $R >> g^2_{SU(4)}$ shown in figure \ref{Ils444} (b). In this limit the $SU(4)+1F$ gauge theory is weakly coupled and should reduce to an $SU(4)+1F$ gauging of the rank $1$ $SU(4)$ SCFT. This is a conformal gauging so the result is a $4d$ SCFT with a single marginal operator. 

Now consider the opposite limit when $g^2_{SU(4)}<0$, shown in figure \ref{Ils444} (c). Now the description as an $SU(4)+1F$ gauging is inadequate, but there is an alternative description given by a weakly coupled $SU(3)$ gauging of the SCFT in figure \ref{Ils444} (d). This SCFT in turn is given by a $Z_3$ symmetric mass deformation of the $T_5$ theory and should have an $SU(3)^3\times U(1)^3$ global symmetry and a $6$ dimensional Higgs branch. Thus, when compactified with a $Z_3$ twist should give a $4d$ theory with an $SU(3)\times U(1)$ global symmetry.

The result of the twisted compactification in the limit of figure \ref{Ils444} (c) should therefore be an $SU(3)$ gauging of this theory. Since we know from the opposite limit that this theory is an SCFT with a single marginal operator, it is quite reasonable that the $SU(3)$ gauging is in fact conformal and that the $SU(3)\times U(1)$ theory is an SCFT.

Since we do not know much about the $SU(3)\times U(1)$ SCFT we cannot put to much tests on this duality. Yet it is apparent that the global symmetry agrees, both having a $U(1)$ global symmetry. Also the Higgs branch dimension calculated from the duality using $d_H = 24(c-a)$ agrees with that evaluated from the web. The dimension of the Coulomb branch also agrees as the $5d$ construction suggests the $SU(3)\times U(1)$ SCFT having a $2$ dimensional Coulomb branch. 

\begin{figure}
\center
\includegraphics[width=1.0\textwidth]{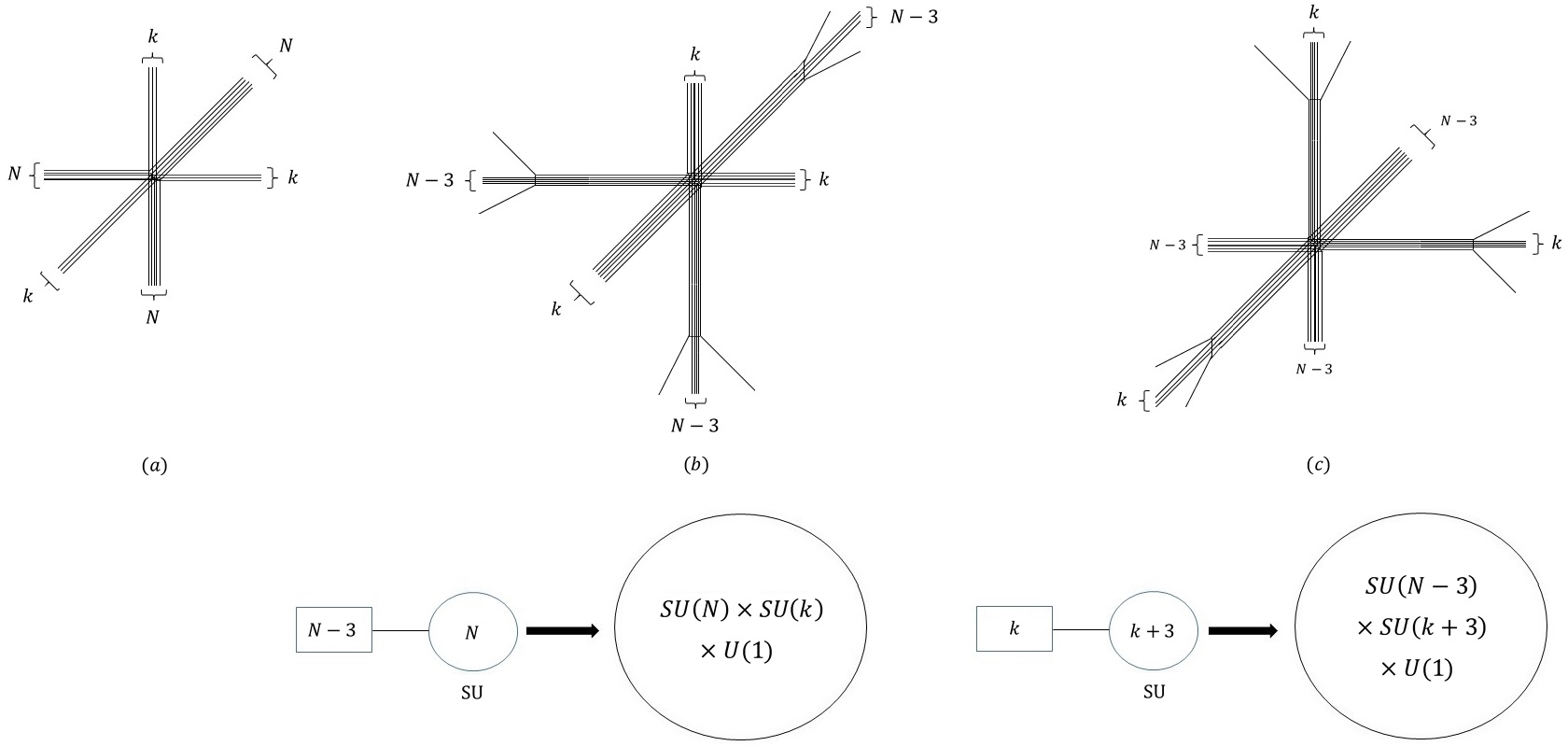} 
\caption{ (a) The brane web for a $5d$ SCFT with a $Z_3$ symmetry. When reduced to $4d$ with a $Z_3$ twist we expect it to lead to a $4d$ theory with $SU(N)\times SU(k)\times U(1)$ global symmetry. (b) The web after a $Z_3$ symmetric gauging by an $SU(N)+(N-3)F$ gauge theory of the SCFT in (a), in the limit of $g^{-2}_{SU(N)}\rightarrow \infty$. Bellow is the $4d$ theory resulting from a twisted compactification in this limit. (c) The same web but now in the $g^{-2}_{SU(4)}\rightarrow -\infty$ limit. In this limit there is a better description as an $SU(k+3)+kF$ gauging of the $5d$ SCFT in (a). Bellow is the $4d$ theory resulting from a twisted compactification in this limit.}
\label{Ils445}
\end{figure}

We can generalize this to other cases. Consider the $5d$ SCFT shown in figure \ref{Ils445} (a).  It has an $SU(N)^3\times SU(k)^3\times U(1)^3$ global symmetry as well as the $Z_3$ discrete symmetry, and reduces to the $5d$ $T_N$ theory when $k=0$. It can be generated from $k$ mass deformations of the $5d$ $T_{N+2k}$ SCFT or alternatively from $N$ mass deformations of the $5d$ $T_{k+2N}$ SCFT. When compactifyied with a $Z_3$ twist we expect it to lead to a $4d$ theory with $SU(N)\times SU(k)\times U(1)$ global symmetry.

We can now consider weakly gauging the global $SU(N)$ symmetry by an $SU(N)+(N-3)F$ gauge theory. In the $5d$ description this can by done by performing the $SU(N)+(N-3)F$ gauging in a $Z_3$ symmetric manner and consider the $g^{-2}_{SU(N)}\rightarrow \infty$ limit. This is shown in figure \ref{Ils445} (b). We can now consider taking the opposite limit $g^{-2}_{SU(N)}\rightarrow -\infty$ shown in figure \ref{Ils445} (c). When reduced to $4d$ this leads to an $SU(k+3)+kF$ gauging of the $SU(N-3)\times SU(k+3)\times U(1)$ theory. 

From the previous cases, we expect the two theories to be weakly coupled descriptions of one conformal theory on different points on its conformal manifold. Therefore we conjecture that the $SU(N)\times SU(k)\times U(1)$ theory appearing is an SCFT and the gauging is conformal. We can perform a few consistency checks. First we can compare the global symmetries and their central charges. One can see that the global symmetries matches. To compare the central charge under the flavor symmetry we use the assumption that the $SU(N)+(N-3)F$ gauging is conformal implying that $k_{SU(N)}=2N+6$. Now, due to the symmetry of the $5d$ SCFT under the interchange of $N$ and $k$, this effectively determine the central charge also for $SU(k)$. With this central charges we must have that the $SU(k+3)+kF$ gauging is conformal as well as matching of the central charges of global symmetries. This is indeed obeyed. We can also compare the conformal anomaly combination $c-a$, where we use the Higgs branch dimension evaluated from the web, and $d_H = 24(c-a)$, to determine this combination for the $SU(N)\times SU(k)\times U(1)$ theory. We indeed find they agree. 

The construction done here is quite reminiscent of the one done in section $2$ and can be considered as a generalization of it to a $Z_3$ case. Similarly to that case we can consider more general dualities like the dualities in figure \ref{Ils20} also in this case. As this is a straightforward application of the things discussed here we won't carry it. Unfortunately, unlike the previous case, we do not find a duality frame with purely known theories so we cannot use this to determine their properties. 

\subsection{$Z_2$ twist}

In this subsection we discuss compactification of the $5d$ $T_N$ theory and related theories on a circle with a $Z_2 \subset S_3$ twist. The $Z_2$ discrete symmetry we twist by is given by exchanging two of the $SU(N)$ global symmetry groups. Together with the previously discussed $Z_3$ element, these generate the group $S_3$. 

\subsubsection*{$N=2$ case}

It is again convenient to start with the $N=2$ case where the $5d$ SCFT reduces to eight free half-hypermultiplets in the $(\bold{2},\bold{2},\bold{2})$ of the $SU(2)^3$ global symmetry. It is not difficult to carry out the reduction where we find the twisted $4d$ theory to be that of six free half-hypermultiplets in the $(\bold{3},\bold{2})$ of the $SU(2)^2$ global symmetry. This is again visible from the dimension of the Higgs branch consistent with the $Z_2$ symmetry being $3$. Again this assumes that the direction given by separating junctions along the $7$-branes is not projected out. Alternatively this can be used to argue this is true which can then be applied to the higher $N$ cases. 

\subsubsection*{$N=3$ case}

Next we consider the $N=3$ case. We remind the reader that the $5d$ SCFT has an $E_6$ global symmetry and so possesses moment map operators in the adjoint of $E_6$ which decomposes to $(\bold{8},\bold{1},\bold{1})+(\bold{1},\bold{8},\bold{1})+(\bold{1},\bold{1},\bold{8})+(\bold{3},\bold{3},\bold{3})+(\bar{\bold{3}},\bar{\bold{3}},\bar{\bold{3}})$ under the $SU(3)^3$ subgroup of $E_6$. Implementing the $Z_2$ projection on these states lead us to suspect the resulting $4d$ theory possesses moment map operators in the $(\bold{8},\bold{1})+(\bold{1},\bold{8})+(\bold{6},\bold{3})+(\bar{\bold{6}},\bar{\bold{3}})$ under the visible $SU(3)^2$ global symmetry. This in fact span the adjoint of $F_4$ so we conclude that we get a rank $1$ theory with $F_4$ global symmetry. This can be also inferred as the operation exchanging two $SU(3)$ subgroups in $E_6$ is identical to its $Z_2$ outer automorphism. It is well known that the invariant part in $E_6$ under this outer automorphism is $F_4$. 

It is also interesting to examine the Higgs branch of this theory. From the brane web we can determine that it has an $8$ dimensional Higgs branch. Interestingly this is also the dimension of the $1$ instanton moduli space of localized $F_4$ instantons. Furthermore as $F_4\subset E_6$ it is naturally embedded in the localized $E_6$ $1$ instanton moduli space which is the Higgs branch of the $T_3$ theory. So it is natural to conjecture that the resulting $4d$ theory has this space as its Higgs branch. 

To our knowledge, there is no known theory possessing these properties, and so we view this as a hint for the potential existence of such a theory. It is also natural to expect it to be an SCFT as an $F_4$ global symmetry suggests strong interactions are involved and the low rank severely limits it from having additional scale dependent coupled parts. It will be interesting to further study this and see if additional evidence for the existence of such a theory can be uncovered. 

\subsubsection*{Higher $N$ and related theories}

We can also consider other theories by compactifying other $T_N$ theories or theories related to them by mass deformations or Higgs branch flows. This then leads to a large class of potential $4d$ $\mathcal{N}=2$ theories that to our knowledge are unknown. It is interesting to look for known theories among them as this can hint as to whether or not these theories exist, and if so whether they are conformal or not. 

\begin{figure}
\center
\includegraphics[width=0.25\textwidth]{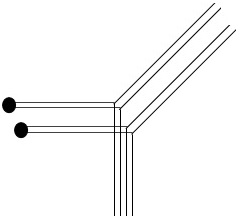} 
\caption{The brane web for the rank $1$ $E_7$ theory.}
\label{Ils382}
\end{figure}

As an example consider the $5d$ rank $1$ $E_7$ theory whose web is shown in figure \ref{Ils382}. This has a $Z_2$ symmetry exchanging the two maximal punctures, given in the web by the $4$ external $(1,1)$ and NS $5$-branes. From the field theory view point this corresponds to exchanging the two $SU(4)$ parts in the $SU(2)\times SU(4)^2$ classically visible global symmetry. We can compactify this theory with a $Z_2$ twist and inquire as to properties of the resulting $4d$ theory. 

First we ask what is the global symmetry of the theory which we try to answer by studying the moment map operators that survive the compactification. The $\bold{133}$ of $E_7$ decomposes under its $SU(2)\times SU(4)^2$ subgroup as: $(\bold{3},\bold{1},\bold{1}) + (\bold{1},\bold{15},\bold{1}) + (\bold{1},\bold{1},\bold{15}) + (\bold{1},\bold{6},\bold{6}) + (\bold{2},\bold{4},\bold{4}) + (\bold{2},\bar{\bold{4}},\bar{\bold{4}})$. Enforcing the $Z_2$ projection we get: $(\bold{3},\bold{1}) + (\bold{1},\bold{15}) + (\bold{1},\bold{20}') + (\bold{2},\bold{10}) + (\bold{2},\bar{\bold{10}})$, under the $SU(2)\times SU(4)$ global symmetry\footnote{In projecting the $(\bold{1},\bold{6},\bold{6})$ we have taken the traceless part. Like in the previous example involving $T_4$ this amounts to a constraint on the operator with no analogue in the full space.}. These span the adjoint of $E_6$. In addition one can see that its Higgs branch is $11$ dimensional, like the rank $1$ $E_6$ theory. All of these lead us to conjecture that the resulting theory is the rank $1$ $E_6$ theory. 

\section{Superconformal index}

In the previous section, we have argued that we can infer some of the operators in the Hall-Littlewood chiral ring of a $4d$ theory, resulting from the twisted compactification of a $5d$ SCFT, from information on the spectrum of operators in the $5d$ SCFT. In this section we shall try to make this more accurate by conjecturing an exact expression for the full Hall-Littlewood index for some of the $4d$ theories we considered in this article (we refer the reader to appendix A for the definition of the Hall-Littlewood index). The Hall-Littlewood index is particularly useful for this owing to the following observations:

\begin{enumerate}
  \item For the theories we consider, the Hall-Littlewood index should be identical to the Hilbert series of the Higgs branch. Furthermore, the Higgs branch is invariant under quantum corrections and so also under direct dimensional reduction. Therefore, it is conceivable that the Hall-Littlewood index of the twisted theory can be generated by twisting the $4d$ index of the direct dimensional reduction.
  \item The Hall-Littlewood index is relatively easy to compute with known expressions abundant in the literature. 
	\item Due to points 1 and 2 there are ample expressions in the literature for theories we consider giving us direct expressions to compare with. 
\end{enumerate}

We consider the $4d$ theories resulting from the $Z_2$ or $Z_3$ twisted compactification of the $5d$ $T_N$ and related theories, and the $Z_2$ twisted compactification of the $5d$ SCFTs of figure \ref{Ils14} and related SCFTs. The strategy takes from point $1$ above, that is we use the known expression for the Hall-Littlewood index of the $4d$ theory resulting from direct dimensional reduction as a basis for our conjecture. The direct dimensional reductions of the theories we consider are known to be comprised of $A$ type class S isolated SCFTs. Therefore it is useful to first review the expressions for the Hall-Littlewood index for these types of SCFTs. 

The Hall-Littlewood index for $A$ type class S isolated SCFTs was determined in \cite{GRRY,GR,GRR}. These are described by a compactification of the $A_{N-1}$ $6d$ $(2,0)$ theory on a Riemann sphere with three punctures. It is given as follows:

\be
I^{HL}_{\text{class S}} = \mathcal{N}_N \sum_{\lambda} \frac{\prod^3_{i=1} \mathcal{K}(\Lambda^{'}_i(a_i)) \psi_{\lambda} (\Lambda_i(a_i))}{\psi_{\lambda} (\tau^{1-N},\tau^{3-N},...,\tau^{N-1})} \label{classSind}
\ee 
where:

\begin{itemize}
  \item $\mathcal{N}_N$ is an overall normalization factor given by:

\be
\mathcal{N}_N = (1-\tau^2)^{N+2} \prod^N_{j=2} (1-\tau^{2j}).
\ee

  \item The sum is over all the partitions of $N$ $\lambda=(\lambda_1,...,\lambda_{N-1},0)$ corresponding to irreducible representations of $SU(N)$. The product is over the three punctures.
	\item $\mathcal{K}(\Lambda^{'}_i(a_i))$ are fugacity dependent factors associated with each puncture. The exact expression for them can be found in \cite{GR}. 
	\item $\psi_{\lambda} (x_i)$ are the Hall-Littlewood polynomials given by:

\be
\psi_{\lambda} (x_i) = \mathcal{N}_{\lambda} (\tau) \sum_{\sigma \subset S_N} x^{\lambda_1}_{\sigma(1)} ... x^{\lambda_N}_{\sigma(N)} \prod_{i<j} \frac{x_{\sigma(i)} - \tau^2 x_{\sigma(j)}}{x_{\sigma(i)} - x_{\sigma(j)}}
\ee

where $\mathcal{N}_{\lambda} (\tau)$ is a normalization factor given by:

\be
\mathcal{N}^{-2}_{\lambda} (\tau) = \prod^{\infty}_{i=0} \prod^{m(i)}_{j=1} \frac{1-\tau^{2j}}{1-\tau^2},
\ee
and $m(i)$ is the number of rows in the Young diagram $\lambda$ of length $i$.
\item $\Lambda_i(a_i)$ is a list of $N$ elements whose exact form depends on the type of puncture. The procedure for determining it in the general case can be found in \cite{GR}. 
\end{itemize}




\subsection{$Z_3$ twisted $T_N$ and related theories}

We shall first start with the $Z_3$ twisted $T_N$ type theories considered in section $3.1$. These are generated by compactifying the $5d$ $T_N$ type SCFT with a $Z_3$ twist. The untwisted dimensional reduction of these theories leads to a class S isolated SCFT corresponding to the compactification of the $A_{N-1}$ $6d$ $(2,0)$ theory on a Riemann sphere with three identical punctures. Its Hall-Littlewood index, which is identical to the Higgs branch Hilbert series of the $5d$ SCFT, is given by equation (\ref{classSind}). 

The twist project the operators down to their $Z_3$ invariants so as a minimalistic assumption it should identify the three $\mathcal{K}$ factors and project the three Hall-Littlewood polynomials, $\psi_{\lambda}$, down to one for the completely symmetric product. Thus we conjecture that the index for the twisted $4d$ theory should have the form:

\be
I^{HL}_{Z_3 \text{ twisted}} = \mathcal{N}'_N \sum_{\lambda} \frac{\mathcal{K}(\Lambda^{'}_1(a_1)) \psi_{3\lambda} (\Lambda_1(a_1))}{\mathcal{N}'_{\lambda} (\tau) \psi_{\lambda} (\tau^{1-N},\tau^{3-N},...,\tau^{N-1})} \label{RYGF}
\ee 
where $\mathcal{N}'_N$ and $\mathcal{N}'_{\lambda}$ are $\tau$ dependent normalization factors and we use $3\lambda$ to mean the partition given by $(3 \lambda_1, 3 \lambda_2, ... , 3 \lambda_{N-1},0)$. In fact we further conjecture that:
 
\be
\mathcal{N}'_N = (1-\tau^2)^{2-N} \prod^N_{j=2} (1-\tau^{2j})
\ee

\be
\mathcal{N}'_{\lambda} = \mathcal{N}^{-2}_{\lambda} = \prod^{\infty}_{i=0} \prod^{m(i)}_{j=1} \frac{1-\tau^{2j}}{1-\tau^2}
\ee

We next support our conjecture by testing this against the cases where we can identify the resulting $4d$ theory with a known theory.

\subsubsection{Example 1: the $T_2$ theory}

The simplest example to start with is the $T_2$ theory. From equation (\ref{RYGF}) we find:

\be
I^{HL}_{T_2 \text{ twisted}} = \frac{1+\tau^2}{(1-\tau^2)(1-a^2\tau^2)(1-\frac{\tau^2}{a^2})}(1 + \sum^{\infty}_{i=1} \frac{\psi_{(3i,0)} (a, \frac{1}{a})}{ (1+\tau^2)\psi_{(i,0)} (\tau, \frac{1}{\tau})}) 
\ee 
where we use $a$ for the $SU(2)$ fugacity. This can be evaluated explicitly. Using Mathematica we find:

\be
I^{HL}_{T_2 \text{ twisted}} = \frac{1}{(1-a^3\tau)(1-a\tau)(1-\frac{\tau}{a})(1-\frac{\tau}{a^3})} = PE[\tau (a^3 + a + \frac{1}{a} + \frac{1}{a^3})]
\ee

This is indeed the Hall-Littlewood index of $4$ free half-hypers in the $\bold{4}$ of $SU(2)$. 

\subsubsection{Example 2: the $T_3$ theory}

Let's now consider the $T_3$ theory. From equation (\ref{RYGF}) we find:

\be
I^{HL}_{T_3 \text{ twisted}} = \frac{(1+\tau^2)(1+\tau^2+\tau^4)}{(1-\tau^2)^2(1-\frac{r\tau^2}{s^2})(1-\frac{s^2 \tau^2}{r})(1-\frac{s\tau^2}{r^2})(1-\frac{r^2 \tau^2}{s})(1-r s\tau^2)(1-\frac{\tau^2}{s r})}\sum_{\lambda} \frac{\psi_{3\lambda} (s, \frac{1}{r}, \frac{r}{s})}{\mathcal{N}'_{\lambda} (\tau) \psi_{\lambda} (\tau^2, 1, \frac{1}{\tau^2})}
\ee 
where we span the $SU(3)$ global symmetry as $\bold{3} = s+ \frac{1}{r}+ \frac{r}{s}$.

Expanding in $\tau$ we find:

\be
I^{HL}_{T_3 \text{ twisted}} = 1 + \tau^2 (\chi_{SU(3)}[\bold{8}] + \chi_{SU(3)}[\bold{10}] + \chi_{SU(3)}[\bar{\bold{10}}]) + O(\tau^4)
\ee 

The $\tau^2$ terms give the contribution of the conserved current supermultiplets and so should be in the adjoint of the global symmetry. Indeed these form the adjoint of $SO(8)$ where only an $SU(3)$ subgroup is visible. Expanding up to order $\tau^6$, we find the index naturally forms $SO(8)$ characters where it is given by:

\be
I^{HL}_{T_3 \text{ twisted}} = 1 + \tau^2 \chi_{SO(8)}[\bold{28}] + \tau^4 \chi_{SO(8)}[\bold{300}] + \tau^6 \chi_{SO(8)}[\bold{1925}] + O(\tau^8)
\ee 

We can compare this against the known Hilbert series of the $1$-instanton moduli space of localized $SO(8)$ instantons evaluated in \cite{BHM} finding perfect agreement. Recalling that the gauge theory $SU(2)+4F$ as this space as its Higgs branch and the identity between the Hall-Littlewood index and the Hilbert series, we see that this agrees with our expectations.

We can further calculate the complete unrefined index, that is setting the $SU(3)$ fugacities to $1$. Using Mathematica we find:

\be
I^{\text{HL unrefined}}_{T_3 \text{ twisted}} = \frac{1 + 18 \tau^2 + 65 \tau^4 + 65 \tau^6 + 18 \tau^8 + \tau^{10}}{(1-\tau^2)^{10}}
\ee 

This indeed agrees with the unrefined Hilbert series of the $1$-instanton moduli space of localized $SO(8)$ instantons\cite{BHM}.  

\subsubsection{Example 3: the $T_4$ theory}

Let's now consider the $T_4$ theory. From equation (\ref{RYGF}) we find:

\bea
I^{HL}_{T_3 \text{ twisted}} = \frac{(1+\tau^2)(1+\tau^2+\tau^4)(1+\tau^2+\tau^6)}{(1-\tau^2)^3(1-\frac{\tau^2}{c^2})(1-c^2 \tau^2)(1-\frac{\tau^2}{d^2})(1-d^2 \tau^2)(1-\frac{c d\tau^2}{b^2})(1-c d b^2 \tau^2)} \\ \nonumber \frac{1}{(1-\frac{b^2\tau^2}{c d})(1-\frac{\tau^2}{c d b^2})(1-\frac{c\tau^2}{d b^2})(1-\frac{c b^2\tau^2}{d})(1-\frac{d\tau^2}{c b^2})(1-\frac{d b^2\tau^2}{c})}\sum_{\lambda} \frac{\psi_{3\lambda} (b c, \frac{b}{c}, \frac{d}{b}, \frac{1}{b d})}{\mathcal{N}'_{\lambda} (\tau) \psi_{\lambda} (\tau^3, \tau, \frac{1}{\tau}, \frac{1}{\tau^3})}
\eea
where we span the $SU(4)$ global symmetry as $\bold{4} = b(c+ \frac{1}{c})+ \frac{1}{b}(d+ \frac{1}{d})$.

Expanding this in a power series in $\tau$ we find:

\bea
I^{HL}_{T_3 \text{ twisted}} & = & 1 + \tau^2 \chi_{SU(4)}[\bold{15}] + \tau^3 (\chi_{SU(4)}[\bold{20}''] + \chi_{SU(4)}[\bar{\bold{20}}''])+ \tau^4 (\chi_{SU(4)}[\bold{84}] + \chi_{SU(4)}[\bold{50}] \nonumber \\ \nonumber & + & \chi_{SU(4)}[\bold{20}'] + \chi_{SU(4)}[\bold{15}] + 1) + \tau^5(\chi_{SU(4)}[\bold{140}] + \chi_{SU(4)}[\bar{\bold{140}}] + \chi_{SU(4)}[\bold{120}] \\ \nonumber & + & \chi_{SU(4)}[\bar{\bold{120}}] + \chi_{SU(4)}[\bold{20}''] + \chi_{SU(4)}[\bar{\bold{20}}'']) + O(\tau^6) \\ \nonumber & = & PE[\tau^2 \chi_{SU(4)}[\bold{15}] + \tau^3 (\chi_{SU(4)}[\bold{20}''] + \chi_{SU(4)}[\bar{\bold{20}}'']) + \tau^4\chi_{SU(4)}[\bold{50}] \\ & - & \tau^5(\chi_{SU(4)}[\bold{20}] + \chi_{SU(4)}[\bar{\bold{20}}])] + O(\tau^6)
\eea

This agrees with the Hall-Littlewood index for the rank $1$ $SU(4)$ SCFT computed in \cite{CDTn}.

\subsubsection{Example 4: an $A_4$ case}

Consider the $5d$ SCFT represented by the web in figure \ref{Ils381} (a). This theory describes the $T_4$ SCFT with a single free hyper\cite{Zaf2}. We can compactify this theory to $4d$ with a twist where we expect to get the rank $1$ $SU(4)$ SCFT with a free hyper. We can now use this as a further test on our index conjecture, now for a case with a non-maximal puncture.

Applying equation (\ref{RYGF}) and expanding in a power series in $\tau$ we find:

\be
I^{HL}_{A_4 \text{ twisted}} = PE[\tau(a^3 + \frac{1}{a^3})] I^{HL}_{T_4 \text{ twisted}} + O(\tau^5)
\ee
which is indeed the Hall-Littlewood index of the rank $1$ $SU(4)$ SCFT with an additional free hyper. 

\subsubsection{Example 5: an $A_5$ case}

\begin{figure}
\center
\includegraphics[width=0.5\textwidth]{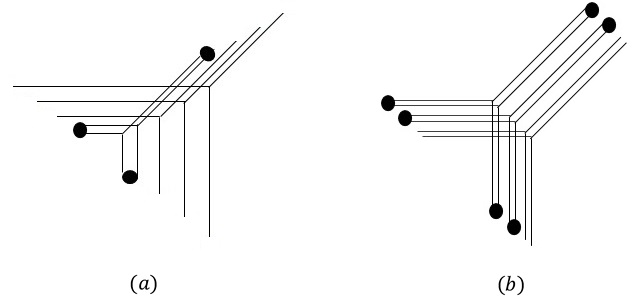} 
\caption{(a) The brane web for the $5d$ $T_4$ theory with a free hyper. (b) The brane web for a $5d$ SCFT with an $SO(8)\times SU(2)^3\times U(1)^2$ global symmetry.}
\label{Ils381}
\end{figure}

Consider the $5d$ SCFT represented by the web in figure \ref{Ils381} (b). The punctures show an $SU(2)^6\times U(1)^3$ global symmetry though the true global symmetry is $SO(8)\times SU(2)^3\times U(1)^2$ as can be inferred from the superconformal index of the associated $4d$ class S theory. The $Z_3$ discrete symmetry acts by permutating the three $SU(2)$'s, and act on the $SO(8)$ as the $Z_3$ element of its outer automorphism group. 

We can consider the $4d$ theory resulting from the $Z_3$ twisted compactification. Applying equation (\ref{RYGF}) we find:

\bea
I^{HL}_{A_5 \text{ twisted}} & = & 1 + \tau^2 (1 + \chi[\bold{3},\bold{1}]_{SU(2)^2}+\chi[\bold{1},\bold{3}]_{SU(2)^2}+q^6+\frac{1}{q^6} + (q^3+\frac{1}{q^3})\chi[\bold{4},\bold{1}]_{SU(2)^2}) \nonumber \\ & + & O(\tau^3) 
\eea
where we use $q$ for the fugacity of the $U(1)$ global symmetry. The $\tau^2$ terms show the conserved currents for the $SU(2)^2\times U(1)$ global symmetry visible from the web and the puncture, but in addition there are additional conserved currents spanning the adjoint of $G_2 \times SU(2)$. Expanding up to $\tau^4$ we indeed find that it forms characters of $G_2 \times SU(2)$ being given by:

\bea
I^{HL}_{A_5 \text{ twisted}} & = & 1 + \tau^2 (\chi[\bold{3},\bold{1}]+\chi[\bold{1},\bold{14}]) + \tau^3 (\chi[\bold{4},\bold{1}]+2\chi[\bold{2},\bold{7}])+ \tau^4 (\chi[\bold{5},\bold{1}] + \chi[\bold{3},\bold{14}] \nonumber \\ & + & \chi[\bold{1},\bold{77'}] + \chi[\bold{1},\bold{27}]+ \chi[\bold{1},\bold{14}] + 2\chi[\bold{3},\bold{7}] + 4) + O(\tau^5) \label{rtgffd}
\eea
where we have written it in characters of the $G_2 \times SU(2)$ global symmetry ordered as $\chi[SU(2),G_2]$. 

We can compactify it to $4d$ with a twist, where we expect to get a $4d$ SCFT with a $12$ dimensional Higgs branch and a $G_2 \times SU(2)$ global symmetry. We can in fact find an appropriate candidate for this theory in a known theory being theory number $19$ in \cite{CDTn}. This theory indeed has a $G_2 \times SU(2)$ global symmetry and $12$ dimensional Higgs branch agreeing with the expectation from the index and the web. As a consistency check we can see from the web that it should have a Higgs branch direction leading to the rank $1$ $SU(4)$ SCFT and a different one leading to the gauge theory $USp(4)+1AS+4F$. These indeed exist also for the theory we identified. Furthermore using the expressions in \cite{CDTn} we can compute the Hall-Littlewood index for it, finding it matches (\ref{rtgffd}) at least to the order we evaluate it. 

\subsection{$Z_2$ twisted $T_N$ and related theories} 

We next move on to the case of the $Z_2$ twisted $T_N$ type theories considered in section $3.2$. These are generated by compactifying the $5d$ $T_N$ type SCFT with a $Z_2$ twist. The untwisted dimensional reduction of these theories leads to a class S isolated SCFT corresponding to the compactification of the $A_{N-1}$ $6d$ $(2,0)$ theory on a Riemann sphere with two identical punctures and one possibly different puncture (which we conveniently choose as $i=1$). Its Hall-Littlewood index, which is identical to the Higgs branch Hilbert series of the $5d$ SCFT, is given by equation (\ref{classSind}).


The conjectured expression now takes the form:

\be
I^{HL}_{Z_2 \text{ twisted}} = \mathcal{N}''_N \sum_{\lambda} \frac{\mathcal{K}(\Lambda^{'}_1(a_1))\mathcal{K}(\Lambda^{'}_2(a_2)) \psi_{2\lambda} (a_1)\psi_{\lambda} (a_2)}{\mathcal{N}''_{\lambda} (\tau) \psi_{\lambda} (\tau^{1-N},\tau^{3-N},...,\tau^{N-1})} \label{RYG}
\ee
where $\mathcal{N}''_N$ and $\mathcal{N}''_{\lambda}$ are $\tau$ dependent normalization factors and we use $2\lambda$ to mean the partition given by $(2 \lambda_1, 2 \lambda_2, ... , 2 \lambda_{N-1},0)$. The values of $\mathcal{N}''_N$ and $\mathcal{N}''_{\lambda}$ are further given by:

\be
\mathcal{N}''_N = (1-\tau^2)^{2} \prod^N_{j=2} (1-\tau^{2j})
\ee

\be
\mathcal{N}''_{\lambda} = \mathcal{N}^{-1}_{\lambda}
\ee

We next support this conjecture by testing this against the cases where we can identify the resulting $4d$ theory with a known theory.

\subsubsection{Example 1: the $T_2$ theory}

As the simplest example let's consider the $T_2$ theory. Using equation (\ref{RYG}) we find that the index for the $Z_2$ twisted theory is:

\be
I^{HL}_{T_2 \text{ twisted}} = \frac{(1-\tau^4)}{(1-\tau^2)^2 (1-a^2\tau^2) (1-\frac{\tau^2}{a^2}) (1-b^2\tau^2) (1-\frac{\tau^2}{b^2})} \sum_{\lambda} \frac{\psi_{2\lambda} (a, \frac{1}{a})\psi_{\lambda} (b, \frac{1}{b})}{\mathcal{N}''_{\lambda} (\tau) \psi_{\lambda} (\tau, \frac{1}{\tau})} 
\ee

Using Mathematica we can perform the representation sum and find that:

\be
I^{HL}_{T_2 \text{ twisted}} = \frac{1}{(1-a^2 b\tau) (1-b\tau) (1-\frac{b\tau^2}{a^2})(1-\frac{a^2\tau}{b})(1-\frac{\tau}{b})(1-\frac{\tau}{a^2 b})} = PE[\tau (a^2 + 1 + \frac{1}{a^2})(b + \frac{1}{b})]
\ee

This is indeed the Hall-Littlewood index for $6$ half-hypers in the $(\bold{3},\bold{2})$ of $SU(2)\times SU(2)$.

\subsubsection{Example 2: the $T_3$ theory}

Let's next consider the $T_3$ theory. We have argued that this should lead to a $4d$ theory with an $F_4$ global symmetry and an $8$ dimensional Higgs branch. This can be naturally accommodated if the Higgs branch is the moduli space of localized $F_4$ $1$ instantons. We now wish to apply equation (\ref{RYG}) to this case.

Expanding the index in a power series in $\tau$, we find:

\bea
I^{HL}_{T_3 \text{ twisted}} & = & 1 + \tau^2 (\chi[\bold{8},\bold{1}] + \chi[\bold{1},\bold{8}] + \chi[\bold{6},\bold{3}] + \chi[\bar{\bold{6}},\bar{\bold{3}}]) + \tau^4 (\chi[\bold{27},\bold{8}] + \chi[\bold{24},\bold{3}] + \chi[\bar{\bold{24}},\bar{\bold{3}}] \label{HLF4} \\ \nonumber & + & \chi[\bar{\bold{15}},\bold{3}] + \chi[\bold{15},\bar{\bold{3}}] + \chi[\bold{6},\bold{15}] + \chi[\bar{\bold{6}},\bar{\bold{15}}] + \chi[\bold{6},\bar{\bold{6}}] + \chi[\bar{\bold{6}},\bold{6}] + \chi[\bold{6},\bold{3}] + \chi[\bar{\bold{6}},\bar{\bold{3}}] \\ \nonumber & + & \chi[\bold{15}',\bold{6}] + \chi[\bar{\bold{15}}',\bar{\bold{6}}] + \chi[\bold{8},\bold{8}] + \chi[\bold{27},\bold{1}] + \chi[\bold{1},\bold{27}] + \chi[\bold{8},\bold{1}] + \chi[\bold{1},\bold{8}] + 1) + O(\tau^6)
\eea
where we write the index in characters of the $SU(3)\times SU(3)$ global symmetry. As previously mentioned, the $\tau^2$ terms, which contains the contribution of the moment map operators, form the adjoint representation of $F_4$. Furthermore looking at the $\tau^4$ terms we see that they form the $\bold{1053}'$ dimensional representation of $F_4$. This is in fact the first few terms in the Hilbert series of the localized $1$ instanton moduli space of $F_4$. Furthermore, for the ubrefined index we can perform the representation summation with Mathematica finding:

\be
I^{HL}_{T_3 \text{ twisted}} = \frac{1+36\tau^2+341\tau^4+1208\tau^6+1820\tau^8+1208\tau^{10}+341\tau^{12}+36\tau^{14}+\tau^{16}}{(1-\tau^2)^{16}}
\ee 

This is indeed the unrefined Hilbert series of the localized $1$ instanton moduli space of $F_4$\cite{BHM}.

\subsubsection{Example 3: rank $1$ $E_7$ theory}

Let's consider the rank $1$ $E_7$ theory. As previously discussed compactifying this theory to $4d$ with a $Z_2$ twist, we expect to get the rank $1$ $E_6$ theory. We can use this to test equation (\ref{RYG}) also for a case with a non-maximal puncture. Expanding the index in a power series in $\tau$, we indeed find it forms characters of $E_6$, and is further given by:

\be
I^{HL}_{E_7 \text{ twisted}} = 1 + \tau^2 \chi_{E_6}[\bold{78}] + \tau^4 \chi_{E_6}[\bold{2430}] + \tau^6 \chi_{E_6}[\bold{43758}] + O(\tau^7)
\ee 

This indeed agree with the Hall-Littlewood index of the rank $1$ $E_6$ being the Hilbert series of the localized $1$ instanton moduli space of $E_6$\cite{BHM}. 

\subsection{$SU(N)\times SU(k)\times U(1)$ SCFT and related theories} 

Finally we wish to attempt to extend the conjecture for the Hall-Littlewood index also for the $Z_2$ twisted theories we originally considered in section $2$. We shall adopt a similar strategy. We first consider the index for the $4d$ theory resulting when compactifying the theory without the twist. We then use this to conjecture the form for the index with the twist. Finally we test this by comparing against cases where we can identify the $4d$ theory with a known one. 

\begin{figure}
\center
\includegraphics[width=0.5\textwidth]{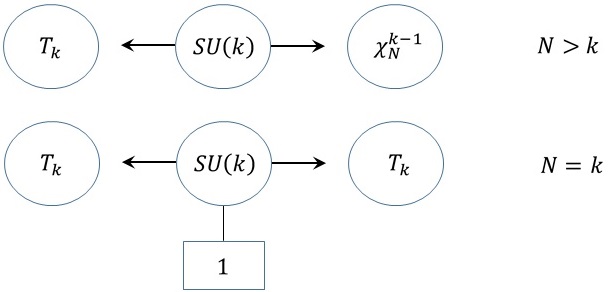} 
\caption{The $4d$ theory resulting from the compactification of the $5d$ theory of figure \ref{Ils14}. The $\chi^{k-1}_N$ is a class S theory whose definition is given in \cite{BZ}. The theory is somewhat different depending on whether $N>k$ or $N=k$. Note that due to the self duality of the theory of figure \ref{Ils14}, the $N<k$ is identical to the $N>k$ case with $N$ and $k$ replaced.}
\label{Ils383}
\end{figure}

Let's first consider the $4d$ theory resulting from compactifying the $5d$ theory of figure \ref{Ils14} with no twist. This was considered in \cite{OSTY2} and they found we get the IR free theory shown in figure \ref{Ils383}. One can see that it has two $A_{N-1}$ maximal punctures and two $A_{k-1}$ ones. These correspond to the groups of parallel $5$-branes and closing them corresponds to forcing some of the $5$-branes to end on the same $7$-brane. This essentially gives the generalization of this to other theories generated by Higgs branch flows.  

The $4d$ theory shows the $Z_2\times Z_2$ discrete symmetry of the $5d$ SCFT, given by exchanging the two pairs of punctures\footnote{When $N=k$ there is an additional discrete element given by exchanging the two $T_N$ SCFTs. This element does not commute with the $Z_2\times Z_2$ discrete symmetry and together they form the dihedral group $D_4$.}. Particularly the $Z_2$ that we twist by corresponds to exchanging both punctures simultaneously.

We can write the Hall-Littlewood index\footnote{Since the theory is IR free by index we mean the index of the two SCFTs with the gauge invariance constraint. Alternatively we can define it as the Hilbert series of the Higgs branch.} of these theories as: 

\be
I^{HL} = \int M^{SU(k)}_{Haar} PE[-\tau^2 \chi_{SU(k)}[\bold{k^2-1}]] I^{HL}_{T_k} I^{HL}_{\chi^{k-1}_N}  
\ee
for the $N>k$ case and

\be
I^{HL} = \int M^{SU(k)}_{Haar} PE[-\tau^2 \chi_{SU(k)}[\bold{k^2-1}] + \tau(m \chi_{SU(k)}[\bold{k}] + \frac{1}{m} \chi_{SU(k)}[\bar{\bold{k}}])] I^{HL}_{T_k} I^{HL}_{T_k}  
\ee
for the $N=k$ case. Here $M^{SU(k)}_{Haar}$ is the Haar measure of $SU(k)$, $m$ the fugacity for $U_F(1)$, and $I^{HL}_{T_k}$ and $I^{HL}_{\chi^{k-1}_N}$ are the Hall-Littlewood indices of the $T_k$ and $\chi^{k-1}_N$ theories. The generalization to cases with non-maximal punctures can be done by replacing the $T_k$ and $\chi^{k-1}_N$ theories by their appropriate versions. 

We can now conjecture a form for the Hall-Littlewood index of the twisted theory. As mentioned the twist act by simultaneously exchanging both pairs of maximal punctures. As suggested in the previous subsection, implementing the twist on each on the two class S theories convert it to a $4d$ theory whose Hall-Littlewood index is given by equation (\ref{RYG}). Thus the natural conjecture is to use a similar expression just with the $T_k$ and $\chi^{k-1}_N$ theories replaced with their twisted cousins:

\be
I^{HL} = \int M^{SU(k)}_{Haar} PE[-\tau^2 \chi_{SU(k)}[\bold{k^2-1}]] I^{HL}_{T_k \text{ twisted}} I^{HL}_{\chi^{k-1}_N \text{ twisted}}  \label{RY1}
\ee
for the $N>k$ case and

\be
I^{HL} = \int M^{SU(k)}_{Haar} PE[-\tau^2 \chi_{SU(k)}[\bold{k^2-1}] + \tau(m \chi_{SU(k)}[\bold{k}] + \frac{1}{m} \chi_{SU(k)}[\bar{\bold{k}}])] I^{HL}_{T_k \text{ twisted}} I^{HL}_{T_k \text{ twisted}} \label{RY2} 
\ee
for the $N=k$ case, and likewise for the cases with non-maximal punctures. In the rest of this section we test this relation by considering various examples.   

\subsubsection{Example $1$: the $N=2, k=3$ case}

As a starting example let's consider the original theory we discussed in section $2$ which is the $N=2, k=3$ case. The resulting $4d$ theory is expected to be the rank $1$ $SU(2)\times USp(6)$ theory. 
We wish to use this case to test equation (\ref{RY1}). In the case at hand, the $Z_2$ twisted $T_2$ theory is just $3$ free hypermultiplets while the $Z_2$ twisted $\chi^{1}_3$ theory is just the $F_4$ theory whose conjectured Hall-Littlewood index was given in equation (\ref{HLF4}). Thus, we conjecture the Hall-Littlewood index for the rank $1$ $SU(2)\times USp(6)$ theory to be:

\bea
I^{HL}_{SU(2)\times USp(6)} & = & \int M^{SU(2)}_{Haar} PE[\tau \chi[\bold{3}]\chi_{SU_G(2)}[\bold{2}]-\tau^2 \chi_{SU_G(2)}[\bold{3}]] I^{HL}_{F_4} \label{IndS2U6} \\ \nonumber & = & 1 + \tau^2 (\chi[\bold{3},\bold{1}] + \chi[\bold{1},\bold{21}]) + \tau^3 \chi[\bold{3},\bold{14}'] \\ \nonumber & + & \tau^4 (\chi[\bold{5},\bold{1}] + \chi[\bold{3},\bold{21}] + \chi[\bold{1},\bold{126}'] + \chi[\bold{1},\bold{90}] + 1) \\ \nonumber & + & \tau^5 (\chi[\bold{5},\bold{14}'] + \chi[\bold{3},\bold{216}] + \chi[\bold{3},\bold{14}'] ) + O(\tau^6) 
\eea

We now want to compare (\ref{IndS2U6}) with the Hall-Littlewood index evaluated from a class S construction. This theory can also be realized, though accompanied by free hypermultiplets, in class S constructions by a twisted compactification of an $A$ or $D$ type $(2,0)$ theory\cite{CDTDN,CDTAt}. We can use this to calculate the Hall-Littlewood index of this theory, though the presence of the free hypers makes a high order calculation quite consuming. Using this we indeed find (\ref{IndS2U6}), at least to the order we evaluated it.

\subsubsection{Example $2$: the $N=2, k=4$ case}

For our next example we consider the $N=2, k=4$ case which we conjecture should lead to the rank $2$ $SU(2)\times USp(8)$ theory. This theory can be constructed by a twisted compactification of a type $A$ or $D$ $(2,0)$ theory\cite{CDTDN,CDTAt}. Using the $A$ type construction we find:

\bea
I^{HL}_{SU(2)\times USp(8)} & = & 1 + \tau^2 (\chi[\bold{3},\bold{1}] + \chi[\bold{1},\bold{36}]) + \tau^4 (\chi[\bold{5},\bold{1}] + \chi[\bold{3},\bold{36}] + \chi[\bold{3},\bold{42}] \nonumber \\ & + & \chi[\bold{1},\bold{330}] + \chi[\bold{1},\bold{308}] + 1) + O(\tau^6) \label{IndS2U8}
\eea
where again we write the index in characters of the $SU(2)\times USp(8)$ global symmetry ordered as $\chi[SU(2),USp(8)]$.

We can now compare this against the conjectured expression in (\ref{RY1}). Again on one side we have the $Z_2$ twisted $T_2$ theory which is just $3$ free hypermultiplets. The other side is the $Z_2$ twisted $\chi^{1}_4$ theory which we have not previously discussed. The $Z_2$ twisted $\chi^{1}_4$ theory as rank $2$ and $SU(2)\times USp(8)$ global symmetry which we may be tempted to identify with the $SU(2)\times USp(8)$ SCFT we are considering. Furthermore the dimension of the Higgs branch also agrees. However this theory as a Higgs branch limit leading to the $F_4$ theory in contrary to the known $SU(2)\times USp(8)$ SCFT so it must be a different theory. Also using (\ref{RYG}) we find the following Hall-Littlewood index:

\bea
I^{HL}_{\chi^{1}_4 \text{ twisted}} & = & 1 + \tau^2 (\chi[\bold{3},\bold{1}] + \chi[\bold{1},\bold{36}]) + \tau^3\chi[\bold{2},\bold{42}] \label{IndS2U81} \\ \nonumber & + & \tau^4 (\chi[\bold{5},\bold{1}] + \chi[\bold{3},\bold{36}] +  \chi[\bold{1},\bold{330}] + \chi[\bold{1},\bold{308}] + 1) \\ \nonumber & + & \tau^5 (\chi[\bold{4},\bold{42}] + \chi[\bold{2},\bold{42}] + \chi[\bold{2},\bold{1155}]) + O(\tau^6) 
\eea 
which differs from (\ref{IndS2U8}).

Returning to the index computation for the $SU(2)\times USp(8)$ SCFT, we can now use the conjecture (\ref{RY1}) where we find:

\bea
I^{HL}_{SU(2)\times USp(8)} & = & \int M^{SU(2)}_{Haar} PE[\tau \chi[\bold{3}]\chi_{SU_G(2)}[\bold{2}]-\tau^2 \chi_{SU_G(2)}[\bold{3}]] I^{HL}_{SU(2)\times USp(8)} \label{IndS2U8c} \\ \nonumber & = & 1 + \tau^2 (\chi[\bold{3},\bold{1}] + \chi[\bold{1},\bold{36}]) + \tau^4 (\chi[\bold{5},\bold{1}] + \chi[\bold{3},\bold{36}] + \chi[\bold{3},\bold{42}] \\ \nonumber & + & \chi[\bold{1},\bold{330}] + \chi[\bold{1},\bold{308}] + 1) + O(\tau^6)
\eea

This matches the explicit expression (\ref{IndS2U81}) to the order it was evaluated.

\subsubsection{Example $3$: the $N=3, k=3$ case}

As a final example let us consider an $N=k$ case, particularly the $N=k=3$ case. Using the conjecture (\ref{RY1}) we find:

\bea
I^{HL}_{SU(3)\times SU(3)\times U(1)} & = & \int M^{SU(3)}_{Haar} PE[\tau (m\chi_{SU_G(3)}[\bold{3}] + \frac{1}{m}\chi_{SU_G(3)}[\bar{\bold{3}}])-\tau^2 \chi_{SU_G(3)}[\bold{8}]] I^{HL}_{F_4} I^{HL}_{F_4} \nonumber \\  & = & 1 + \tau^2 (1 + \chi[\bold{8},\bold{1}] + \chi[\bold{1},\bold{8}]) + \tau^3(m\chi[\bold{6},\bold{1}] + m\chi[\bold{1},\bold{6}] + \frac{1}{m}\chi[\bar{\bold{6}},\bold{1}] +  \frac{1}{m}\chi[\bold{1},\bar{\bold{6}}]) \nonumber \\  & + & \tau^4 (\chi[\bold{27},\bold{1}] + \chi[\bold{8},\bold{8}] + \chi[\bold{1},\bold{27}] + \chi[\bold{6},\bar{\bold{6}}] + \chi[\bar{\bold{6}},\bold{6}] + 2\chi[\bold{8},\bold{1}] + 2\chi[\bold{1},\bold{8}] + 3) \nonumber \\  & + & \tau^5 (m\chi[\bold{24},\bold{1}] + m\chi[\bold{1},\bold{24}] + \frac{1}{m}\chi[\bar{\bold{24}},\bold{1}] +  \frac{1}{m}\chi[\bold{1},\bar{\bold{24}}] + m\chi[\bar{\bold{15}},\bold{1}] + m\chi[\bold{1},\bar{\bold{15}}] \nonumber \\  & + & \frac{1}{m}\chi[\bold{15},\bold{1}] +  \frac{1}{m}\chi[\bold{1},\bold{15}] + m\chi[\bold{6},\bold{8}] + m\chi[\bold{8},\bold{6}] + \frac{1}{m}\chi[\bar{\bold{6}},\bold{8}] +  \frac{1}{m}\chi[\bold{8},\bar{\bold{6}}] \label{IndSU3Sq} \\ \nonumber & + & m\chi[\bold{6},\bold{6}] + \frac{1}{m}\chi[\bar{\bold{6}},\bar{\bold{6}}] + 2m\chi[\bold{6},\bold{1}] + 2m\chi[\bold{1},\bold{6}] + \frac{2}{m}\chi[\bar{\bold{6}},\bold{1}] +  \frac{2}{m}\chi[\bold{1},\bar{\bold{6}}]) + O(\tau^6) 
\eea
where we write the index in characters of the $SU(3)\times SU(3)$ global symmetry.

To our knowledge this theory has not been realized before so we have nothing to compare this expression to. Nevertheless, the $5d$ construction suggests that this theory obeys a duality, shown in figure \ref{Ils23} for $N=5$. We can use this as a consistency check by calculating the Hall-Littlewood index on both sides and comparing. Calculating to order $\tau^5$ we indeed find they match. 

\section{Conclusions}

In this article we have explored the dimensional reduction of $5d$ SCFTs to $4d$ with a twist in an element of their discrete global symmetry. We have concentrated on $5d$ SCFTs with a brane web representation particularly the SCFTs given by the intersection of NS and D$5$-branes and the $5d$ $T_N$ theory. We have argued that this leads to various known $4d$ isolated SCFTs as well as a wealth of potentially new ones. We then used the $5d$ description to infer various properties of these SCFTs such as their Higgs branch, mass deformations and dualities. We have also used this construction to conjecture an expression for the Hall-Littlewood index for these theories.

It is interesting to see if we can find additional evidence for the existence of the $4d$ theories we introduced. These may also teach us more about their properties. One interesting question is whether or not these can be incorporated into the known class S construction. Alternatively it is interesting if they can be constructed by alternative means such as compactification of $(1,0)$ $6d$ SCFTs or geometric engineering. For instance recently \cite{XY,CXYYZ} initiated a systematic study of $4d$ $\mathcal{N}=2$ SCFTs engineered using type IIB string theory on Calabi-Yau 3-fold singularities. It is interesting to see if the theories found in this paper can also be constructed using this method. 

This is especially true for the theories introduced in section $3.2$ particularly the rank $1$ $F_4$ theory. Recently a systematic study of rank $1$ $\mathcal{N}=2$ SCFTs was initiated in \cite{ALLM,ALLM0,ALLM1}, and it is interesting if this can support or disfavor its existence. Furthermore the existence of a rank $1$ $F_4$ SCFT was suspected from the superconformal bootstrap analysis in \cite{BLLPRR,LL}, which also calculated some of its central charges. It will be interesting if we can calculate these also for our proposed rank $1$ $F_4$ theory and compare with their results.

An additional angle is to try to generalize these constructions to more theories. One possibility is to study other $5d$ SCFTs. Another possibility is to twist by other discrete symmetries. We have seen that the $Z_3$ twist discussed in section $3.1$ can be thought of as a generalization of the $Z_2$ twist discussed in section $2$. In both of these the twisted discrete element can be represented by the action of an $SL(2,Z)$ element on the brane web, being $TS$ for the $Z_3$ case, and $-I$ for the $Z_2$ case. So a possible generalization is to consider other finite subgroups of $SL(2,Z)$, for example the $Z_4$ and $Z_6$ subgroups appearing in the construction of S-folds\cite{GER,AT}. In fact the connection with S-folds itself appear to warrant further exploration.

Yet another interesting direction is to study the compactification of these theories to $3d$. Besides the natural interest, the resulting theories are then $5d$ SCFT compactified on a torus with a twist on one of its cycles. Alternatively we can get the same theory by taking the untwisted $4d$ reduction and compactifying it to $3d$ with a twist. These should be related by a modular transformation on the torus which could potentially lead to interesting $3d$ structure.

\subsection*{Acknowledgments}

I would like to thank Oren Bergman, Shlomo S. Razamat and Leonardo Rastelli for useful comments and discussions. G.Z. is supported in part by the Israel Science Foundation under grant no. 352/13, and by the German-Israeli Foundation for Scientific Research and Development under grant no. 1156-124.7/2011.

\appendix

\section{The Hall-Littlewood index}

The Hall-Littlewood index is a special limit of the $4d$ $\mathcal{N}=2$ superconformal index. The $4d$ superconformal index is the counting of all BPS operators in the theory, annihilated by a chosen supercharge, modulo the possible merging of BPS operators to form a non BPS multiplet. It can be further refined so as to keep track of the representations of the operators under the superconformal and flavor symmetries. 

Specifically for the $4d$ $\mathcal{N}=2$ case, the bosonic part of the superconformal group is $SO(4,2)\times SU_R(2) \times U_r(1)$. The representations are then labeled by the highest weights of its $SO(4) \times SU_R(2) \times U_r(1)$ subgroup. We label the two weights of $SO(4)$ as $j_1, j_2$, that of $SU_R(2)$ as $R$ and that of $U_r(1)$ as $r$. 

The $4d$ $\mathcal{N}=2$ superconformal index is then given by the following trace formula:

\be
I = Tr (-1)^F p^{j_1 + j_2 - r} q^{j_2 - j_1 - r} \tau^{2R+2r} \prod_i a^{f_i}_i 
\ee
where $p, q$ and $\tau$ are fugacities associated with the superconformal algebra, $a_i$ are fugacities associated with the various flavor symmetries whose Cartan charges are given by $f_i$.

The Hall-Littlewood index is a special limit of the $4d$ $\mathcal{N}=2$ superconformal index given by taking $p=q=0$. This limit counts only a subsector of the full BPS operators in the theory. Alternatively it is given directly by the following trace formula:

\be
I_{HL} = Tr_{HL} (-1)^F \tau^{2E-2R} \prod_i a^{f_i}_i
\ee
where $Tr_{HL}$ denotes trace over all operators obeying: $j_1=0$, $E-2R-r=0$\cite{GRRY}.

\section{Indices for $5d$ $T_4$ theory and its mass deformations}

In this appendix we discuss the $5d$ index for the $T_4$ theory and its $Z_3$ symmetric mass deformations.

\subsection{$T_4$ theory}

The $5d$ index for the $T_4$ theory can be evaluated from its gauge theory description as $4F+SU(3)\times SU(2)+2F$. This was done in \cite{BZ} where it was indeed shown that there are additional instantonic conserved currents that enhance the classical $SU(2)^2\times U(1)^4$ global symmetry to $SU(4)^2$. The index also contains order $3$ operators in the $(\bold{4}, \bold{4}, \bold{4})$ and $(\bar{\bold{4}}, \bar{\bold{4}}, \bar{\bold{4}})$, as expected from the $4d$ Hall-Littlewood chiral ring. Furthermore we expect an order $4$ operator in the $(\bold{6}, \bold{6}, \bold{6})$. Breaking the enhanced $SU(4)^2$ group into their $SU(2)^2 \times U(1)^4$ representations, we find:

\bea
(\bold{6}, \bold{6}, \bold{6}) & = & (q_1 q^2_2 + \frac{1}{q_1 q^2_2}) \chi[\bold{6}, \bold{1}, \bold{1}] + (\frac{z^2}{b^2} + \frac{b^2}{z^2}) \chi[\bold{6}, \bold{1}, \bold{1}] +  (\frac{b^2}{q_2 z^{\frac{1}{2}}} + \frac{q_2 z^{\frac{1}{2}}}{b^2} + q_1 q_2 z^{\frac{3}{2}} + \frac{1}{q_1 q_2 z^{\frac{3}{2}}}) \chi[\bold{6}, \bold{2}, \bold{1}] \nonumber \\ & + &  (\frac{q_2 b^2}{z^{\frac{1}{2}}} + \frac{z^{\frac{1}{2}}}{q_2 b^2} + \frac{q_1 q_2}{z^{\frac{3}{2}}} + \frac{z^{\frac{3}{2}}}{q_1 q_2 }) \chi[\bold{6}, \bold{1}, \bold{2}] + (q_1 + \frac{1}{q_1} + z b^2 + \frac{1}{z b^2}) \chi[\bold{6}, \bold{2}, \bold{2}]
\eea
where we used the notation of \cite{BZ}. 

One can see that it is made of two perturbative contributions corresponding to the operators $J^{i j} \bar{Z}^{i \bar{\alpha}} \bar{Z}^{j \bar{\beta}} Q^{\alpha} Q^{\beta}$, $\epsilon^{\alpha \beta \gamma} q_i Z^{i \alpha} Q^{\beta} Q^{\gamma}$ and their conjugates, where we use $Z$ for the bifundamental field and $Q$ and $q$ for the flavors of the $SU(3)$ and $SU(2)$ gauge groups respectively. The rest are instanton charged states, and we have also verified using instanton counting methods that these exist.

\subsection{$SU(2)^3\times U(1)^3$ theory}

\begin{figure}
\center
\includegraphics[width=0.75\textwidth]{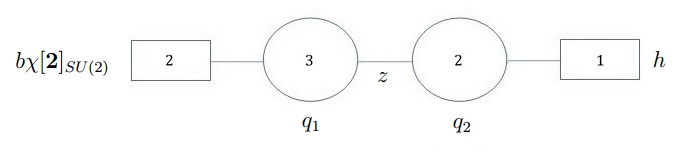} 
\caption{The the fugacity allocation for the $2F+SU(3)\times SU(2)+1F$ gauge theory.}
\label{Ils39}
\end{figure}

In this section we deal with the $5d$ $SU(2)^3\times U(1)^3$ SCFT. In particular, we calculate the index from the $2F+SU(3)\times SU(2)+1F$ gauge theory description. We use the fugacity allocation shown in figure \ref{Ils39}.

We calculate the index to order $x^3$. To that order, besides the perturbative contribution, we also get contributions from the $(1,0)$, $(0,1)$ and $(1,1)$ instantons. We find:

\bea
I & = & 1 + x^2 \left(5 + \chi[\bold{3}]_{SU(2)} + (q_2+\frac{1}{q_2})(\sqrt{h z^3}+\frac{1}{\sqrt{h z^3}}) \right) \\ \nonumber & + & x^3  \left( (y + \frac{1}{y})\left(6 + \chi[\bold{3}]_{SU(2)} + (q_2+\frac{1}{q_2})(\sqrt{h z^3}+\frac{1}{\sqrt{h z^3}}) \right) \right.  \\ \nonumber & + & \chi[\bold{2}]_{SU(2)} \left(\frac{z}{b h}+\frac{b h}{z} + (\frac{\sqrt{h}}{b\sqrt{z}}+\frac{b\sqrt{z}}{\sqrt{h}})(q_2+\frac{1}{q_2} + \sqrt{h z^3}+\frac{1}{\sqrt{h z^3}})\right)  \\ \nonumber & + &  \left. (q_1+\frac{1}{q_1})(\frac{z}{b}+\frac{b}{z}) + (q_1 q_2+\frac{1}{q_1 q_2})(b \sqrt{h z}+\frac{1}{b \sqrt{h z}})\right) + O(x^4)
\eea 

One can see from the $x^2$ terms the conserved currents of the classically visible $SU(2)\times U(1)^5$ global symmetry as well as additional conserved currents, coming from the $(0,1)$ instanton, that lead to an enhancement of $U(1)^2\rightarrow SU(2)^2$. The index can be written in characters of the $SU(2)^3\times U(1)^3$ global symmetry:

\bea
I & = & 1 + x^2 (3 + \chi[\bold{3},\bold{1},\bold{1}]^{0,0,0} + \chi[\bold{1},\bold{3},\bold{1}]^{0,0,0} + \chi[\bold{1},\bold{1},\bold{3}]^{0,0,0} ) \\ \nonumber & + & x^3  \left( \chi[\bold{2}]_{y}\left(4 + \chi[\bold{3},\bold{1},\bold{1}]^{0,0,0} + \chi[\bold{1},\bold{3},\bold{1}]^{0,0,0} + \chi[\bold{1},\bold{1},\bold{3}]^{0,0,0} \right) + \chi[\bold{2},\bold{2},\bold{2}]^{1,1,1} + \chi[\bold{2},\bold{2},\bold{2}]^{-1,-1,-1} \right.  \\ \nonumber & + & \left. \chi[\bold{2},\bold{1},\bold{1}]^{1,0,0} + \chi[\bold{2},\bold{1},\bold{1}]^{-1,0,0} + \chi[\bold{1},\bold{2},\bold{1}]^{0,1,0} + \chi[\bold{1},\bold{2},\bold{1}]^{0,-1,0} + \chi[\bold{1},\bold{1},\bold{2}]^{0,0,1} + \chi[\bold{1},\bold{1},\bold{2}]^{0,0,-1}\right)  \\ \nonumber & + & O(x^4)
\eea 
where we use the notation $\chi[d_1, d_2, d_3]^{q_1,q_2,q_3}$ for an operator with in $d_i$ dimensional representation under $SU(2)_i$ and charge $q_i$ under $U(1)_i$. In term of the fugacities these are spanned by:

\bea
 \chi[\bold{3},\bold{1},\bold{1}]^{1,0,0} &=& \frac{z}{b h} \chi[\bold{3}]_{SU(2)} \\
\chi[\bold{1},\bold{3},\bold{1}]^{0,1,0}&=& (q_2 \sqrt{h z^3} + 1 + \frac{1}{q_2 \sqrt{h z^3}})\frac{q_1 q^{\frac{1}{2}}_2 h^{\frac{1}{4}} b}{z^{\frac{1}{4}}} \\
\chi[\bold{1},\bold{1},\bold{3}]^{0,0,1}&=& (\frac{\sqrt{h z^3}}{q_2} + 1 + \frac{q_2}{\sqrt{h z^3}})\frac{h^{\frac{1}{4}} b}{q_1 q^{\frac{1}{2}}_2 z^{\frac{1}{4}}}. 
\eea


We can use this to try and guess the Hall-Littlewood chiral ring of the $4d$ SCFT where we find:

\be
I^{HL}_{4d} = \tau^2 \chi[\bold{3}]^{0} + \tau^3 (\chi[\bold{4}]^{3} + \chi[\bold{4}]^{-3} + \chi[\bold{2}]^{1} + \chi[\bold{2}]^{-1}) + O(\tau^4)
\ee

It will be interesting to see how the enhancement as well as the full index arise from the $SU(2)^3$ gauge theory description, both for this theory and the $T_4$ itself. Unfortunately, there are technical issues in performing instanton counting due to the half trifundamrntal that impedes instanton counting in these theories. Since the $U(1)^3$ SCFT, that we get by performing another $Z_3$ symmetric mass deformation, as only this description, understanding this will allow us to repeat this analysis also for this theory. We reserve this for future work.

\end{document}